\documentclass[12pt,notitlepage,a4paper]{article}
\usepackage{color,graphicx}
\usepackage{amssymb}
\usepackage{delarray,amsmath,bbm}
\usepackage[latin1]{inputenc}
\usepackage[american]{babel}
\usepackage{cite}

\def\ie{{\emph{i.e.}}}

\newcommand{\rc}{\nonumber\\}
\newcommand{\beq}{\begin{equation}}
\newcommand{\eeq}{\end{equation}}
\newcommand{\bear}{\begin{eqnarray}}
\newcommand{\eear}{\end{eqnarray}}
\numberwithin{equation}{section}

\newfont{\namefont}{cmr10}
\newfont{\addfont}{cmti7 scaled 1440}
\newfont{\boldmathfont}{cmbx10}
\newfont{\headfontb}{cmbx10 scaled 1728}
\renewcommand{\theequation}{{\rm\thesection.\arabic{equation}}}

\begin{document}
\baselineskip=15.5pt
\pagestyle{plain}
\setcounter{page}{1}

\begin{center}
\vspace{0.1in}

\renewcommand{\thefootnote}{\fnsymbol{footnote}}

\begin{center}
\Large \bf Thermodynamics of the brane\\ in Chern-Simons matter theories with flavor
\end{center}
\vskip 0.1truein
\begin{center}
\bf{Niko Jokela,${}^1$\footnote{niko.jokela@usc.es} 
Javier Mas,${}^1$\footnote{javier.mas@usc.es} 
Alfonso V. Ramallo,${}^1$\footnote{alfonso@fpaxp1.usc.es} 
and Dimitrios Zoakos${}^2$\footnote{dimitrios.zoakos@fc.up.pt}}\\
\end{center}
\vspace{0.5mm}

\begin{center}\it{
${}^1$Departamento de  F\'\i sica de Part\'\i  culas \\
Universidade de Santiago de Compostela \\
and \\
Instituto Galego de F\'\i sica de Altas Enerx\'\i as (IGFAE)\\
E-15782 Santiago de Compostela, Spain}
\end{center}

\begin{center}\it{
${}^2$Centro de F\'\i sica do Porto \\
and \\
Departamento de  F\'\i sica  e Astronomia \\
Faculdade de Ci\^encias da Universidade do Porto \\
Rua do Campo Alegre 687, 4169-007 Porto, Portugal}
\end{center}

\setcounter{footnote}{0}
\renewcommand{\thefootnote}{\arabic{footnote}}

\vspace{0.4in}

\begin{abstract}
\noindent We study the holographic dual of flavors in a Chern-Simons matter theory at non-zero temperature, realized as D6-branes in 
the type IIA black hole dual in the ABJM background geometry. We consider both massive and massless flavors. The former are treated 
in the quenched approximation, whereas the massless ones are considered as dynamical objects and their backreaction on the 
geometry is included in the black hole background. 
We compute the holographically renormalized action of the probe by imposing several physical conditions.  In the limit of massless flavors the free energy and  entropy of the probe match non-trivially the first variation of these quantities for the backreacted background when the number of flavors is increased by one unit.  We compute several thermodynamical functions for the system and analyze the meson melting phase transition between Minkowski and black hole embeddings. 
\end{abstract}

\smallskip
\end{center}

\newpage

\tableofcontents

\section{Introduction}
Recent studies of Chern-Simons matter theories 
in three dimensions by holographic techniques have provided non-trivial examples of the AdS/CFT correspondence \cite{jm,Aharony:1999ti} 
which could be of great help to shed light on the dynamics of  some strongly coupled systems in condensed matter physics. 
The paradigmatic example of these systems is the Aharony-Bergman-Jafferis-Maldacena (ABJM) theory constructed in \cite{Aharony:2008ug}, based on 
the analysis of  \cite{BL,Gustavsson:2007vu}, where the supersymmetric Chern-Simons matter theories were proposed as the low energy theories 
of multiple M2-branes. 

The ABJM theory is an ${\cal N}=6$ super Chern-Simons gauge theory in 2+1 dimensions with gauge group $U(N)_{k}\times U(N)_{-k}$  with 
opposite level numbers $k$ and $-k$. In addition to the two gauge fields, this theory contains two pairs of chiral superfields which 
transform in the $(N,\bar N)$ and $(\bar N, N)$ bifundamental representation. When $N$ and $k$ are large the theory admits a geometric 
description in terms of an $AdS_4\times {\mathbb C}{\mathbb P}^3$ with fluxes in type IIA supergravity which preserves 24 supersymmetries.  
The study of this theory and its generalizations has uncovered a very rich structure and has provided new precision tests of the AdS/CFT 
correspondence (see \cite{Klebanov:2009sg,Klose:2010ki,Marino:2011nm,Bagger:2012jb} for reviews of different aspects of the 
Chern-Simons matter theories).

The ABJM theory can be generalized in several directions. In this paper we will consider the addition of fields transforming in the fundamental representations $(N,1)$ and $(1,N)$ of the $U(N)\times U(N)$ gauge group. It was proposed in \cite{Hohenegger:2009as,Gaiotto:2009tk} that these flavors can be incorporated in the holographic dual  by considering D6-branes that fill the $AdS_4$ space and wrap an ${\mathbb R}{\mathbb P}^3$ submanifold of the internal ${\mathbb C}{\mathbb P}^3$ space.  These configurations are ${\cal N}=3$ supersymmetric. When the number $N_f$ of flavors is small one can adopt the so-called quenched approximation, in which the flavor D6-branes are considered as probes  in the $AdS_4\times {\mathbb C}{\mathbb P}^3$ geometry.  This approach has been followed in  \cite{Hikida:2009tp,Jensen:2010vx,Ammon:2009wc,Zafrir:2012yg}. 

In  \cite{Conde:2011sw} a holographic dual of ABJM with unquenched flavor was found by considering a large number $N_f$ of flavor D6-branes which are continuously distributed  in the internal space in such a way that ${\cal N}=1$ supersymmetry is preserved. To find  the unquenched solution  one has to solve the equations of motion of supergravity with brane sources, which modify the Bianchi identities of the forms and the Einstein equations. If the branes are localized,  the sources introduce Dirac $\delta$-functions in the equations, which makes the problem very difficult  to solve. For this reason we will follow the approach initiated in \cite{Bigazzi:2005md} and  study the backreaction induced by a smeared continuous distribution of  flavor branes. This procedure has been successfully   applied to  add unquenched flavor in other holographic setups \cite{CNP, conifold,D3-D7} (see \cite{Nunez:2010sf} for a review and more references). As the smeared flavor branes are not coincident, the flavor symmetry for 
$N_f$ branes is $U(1)^{N_f}$ rather than $U(N_f)$. Moreover, since we are superimposing branes with different  orientations in the internal space, the corresponding supergravity solutions are generically less supersymmetric than the ones with localized flavor. The unquenched solutions with smeared flavors are much simpler than the localized ones and, in many cases the solutions are analytic.

The unquenched solution of type IIA supergravity found in \cite{Conde:2011sw} includes  the backreaction effects due to massless flavors. The corresponding ten-dimensional geometry is of the form $AdS_4\times {\cal M}_6$, where ${\cal M}_6$ is a compact six-dimensional space whose metric is a squashed version of the unflavored  Fubini-Study metric of ${\mathbb C}{\mathbb P}^3$. In this solution the deformation introduced by the flavors is encoded in the squashing factors, which are constant and depend non-linearly on the number of flavors (although the sources of supergravity are linear in $N_f$). Notice that the backreacted metric contains an Anti-de Sitter factor. This is related to the fact that the dual Chern-Simons matter theory has conformal fixed points even when the flavors are added (see \cite{Bianchi} for a verification of this property in perturbation theory). On the gravity side this conformal behavior is responsible for the regularity of the metric at the IR, contrary to other solutions with unquenched massless flavors \cite{Nunez:2010sf}. It was checked in \cite{Conde:2011sw} that this solution captures rather well many of the effects due to loops of the fundamentals in 
several observables. In particular, it matches remarkably well with the behavior of the effective number of degrees of freedom of the flavored theory in the Veneziano limit, which was computed in the field theory side using localization in  \cite{Santamaria:2010dm}.

In sharp contrast  to what happens to other flavored backgrounds  obtained with the smearing method (see, for example, those of refs. \cite{CNP,conifold,D3-D7}), our supergravity solution has a good UV behavior and, since the metric has an Anti-de Sitter factor,  we are dealing with a geometry for which the holographic methods are firmly established and it is possible to apply a whole battery of techniques to perform a clean analysis of the different flavor screening effects. In particular, as it is shown below,  it is straightforward to add a  further temperature deformation to the flavor deformation and to construct a black hole which contains the effects of massless flavors. This is simply done by including the standard blackening factor in the Anti-de Sitter part of the metric, without modifying the internal space ${\cal M}_6$. We can then compute different thermodynamic quantities for this flavored black hole. 

When flavor branes are embedded in a black hole geometry the system undergoes a first order phase transition when the branes fall into the horizon \cite{Mateos:2006nu,Mateos:2007vn}. On the field theory side this phase transition corresponds to the melting of mesons in a deconfined plasma. The analysis of the influence of  unquenched flavor in this melting transition is clearly a very interesting problem. However, in order to have a complete understanding of this problem in the holographic setup one has  to find a black hole solution containing the full backreaction of {\it massive} flavors, which is very hard to find.  In this paper we will adopt a more modest approach and consider a small number of massive flavors and a large number of massless quarks. The latter will be included in the background, while the massive fundamentals will be treated in the quenched approximation. Accordingly, we will consider a D6-brane probe in the non-zero temperature version of the background found in \cite{Conde:2011sw} and we will study 
its thermodynamic properties, following the same methodology as the one employed in \cite{Mateos:2007vn} for the D3-D7 and D4-D6 systems. 

The action that governs the dynamics of our D6-brane probes contains a contribution from the  Dirac-Born-Infeld (DBI) and  Wess-Zumino (WZ) terms. This probe action must be renormalized holographically in order to get finite answers for the different thermodynamic functions. At zero temperature one can adopt a gauge for the RR seven-form potential $C_7$ in which the  two terms of the action cancel with each other on-shell for the kappa symmetric embeddings of the probe. At non-zero temperature the on-shell action of the probe in this gauge  is finite, and the only freedom left by the holographic renormalization is the addition of finite counterterms. These finite terms can be fixed by imposing regularity of $C_7$ at the horizon and by requiring that  all the  thermodynamic functions for the probe  vanish for infinitely massive flavors, as they can be integrated out.  

Once the action of the probe is fixed in this way, we should verify that it satisfies a non-trivial compatibility condition with the background. Indeed, let us consider a probe for a massless flavor. In this massless limit the quarks introduced by the probe are of the same type as those of the background. Thus, one can compare the thermodynamic functions of the probe with the variations of these same functions  for the background when $N_f$ is increased by one unit. For consistency, these two quantities should be equal. Actually, within the probe approximation one should assume that $N_f$ is large. Then, the variation induced in the background when $N_f\to N_f+1$ should be computed by a Taylor expansion in which only the first term is kept. We will verify that this compatibility condition is indeed satisfied in our case, which is a highly non-trivial test because the dependence of the background on $N_f$ is non-linear. After passing successfully this test,  we are ready to study systematically the 
thermodynamics of the probe brane. In general, the main objective is to  determine  the dependence of the different observables on the number of flavors of the background, as well as the departure from conformality induced on the system by the probe.

The plan of the rest of this paper is the following. In Section \ref{FlavABJM} we will present our flavored black hole background and compute some of its thermodynamic functions. In Section \ref{zeroTemp} we will analyze the flavor brane embeddings at zero temperature and extract some useful information which will be needed in the black hole case. In Section \ref{nonzeroTemb} we will study the action of the probe in the non-zero temperature geometry and we will check that the compatibility condition mentioned above is satisfied. In Section \ref{Min-BH} we shall study in detail the two types of embeddings, Minkowski and black hole, and we shall analyze the first order phase transition between them. Section \ref{Thermo} is devoted to the calculation of the different thermodynamic functions  of the probe (free energy, internal energy, entropy, and normal speed of sound). Section \ref{conclusions} contains a summary of our results and a discussion. The paper is completed with several appendices, which 
contain some explicit calculations and details not included in the main text.

\section{The flavored ABJM background}
\label{FlavABJM}

In this section we will present the non-zero temperature version of the deformed ABJM background found in \cite{Conde:2011sw}. The ten-dimensional metric, in string frame, of this supergravity solution takes the form
\beq
ds^2\,=\,L^2\,\,ds^2_{BH_4}\,+\,ds^2_{6}\,\,,
\label{flavoredBH-metric}
\eeq
where $L$ is the radius of curvature and $ds^2_{BH_4}$ is the metric of a black hole in the four-dimensional Anti-de Sitter space, given by
\beq\label{BH4-metric}
 ds^2_{BH_4} = -r^2h(r) dt^2+\frac{dr^2}{r^2h(r)}+r^2\big[(dx^1)^2+(dx^2)^2\big] \ ,
\eeq
and  $ds^2_{6}$ is the metric of the compact internal six-dimensional manifold.\footnote{Unless otherwise stated, we will use units for which $\alpha'=1$.} 
In (\ref{BH4-metric}) the blackening factor $h(r)$ is given by
\beq
h(r)\,=\,1\,-\,\frac{r_h^3}{r^3}
\,\,,
\label{blackening-factor}
\eeq
where the horizon radius $r_h$ is related to the temperature by 
\beq
T\,=\,{1\over 2\pi}\,\,
\Big[\,{1\over \sqrt{g_{rr}}}\,\,{d\over dr}\,
\Big(\,\sqrt{\,-g_{tt}}\,\Big)\,
\Big]_{r=r_h}\,=\,
{3\,r_h\over 4\pi}\,\,.
\eeq
The internal metric $ds^2_{6}$ in (\ref{flavoredBH-metric}) is a deformed version of the Fubini-Study metric of ${\mathbb C}{\mathbb P}^3$. This deformation is due to the backreaction of the massless flavors, generated by 
the D6-branes, and can be simply stated when the manifold ${\mathbb C}{\mathbb P}^3$ is represented as an
${\mathbb S}^2$-bundle over ${\mathbb S}^4$, with the fibration constructed by using the self-dual $SU(2)$ instanton on the four-sphere. Explicitly, this metric can be written as
\beq
ds^2_{6}\,=\,{L^2\over b^2}\,\,\Big[\,
q\,ds^2_{{\mathbb S}^4}\,+\,\big(d x^i\,+\, \epsilon^{ijk}\,A^j\,x^k\,\big)^2\,\Big]\,\,,
\label{internal-metric-flavored}
\eeq
where $b$ and $q$ are constant squashing factors, $ds^2_{{\mathbb S}^4}$ is the standard metric for the unit four-sphere, $x^i$ ($i=1,2,3$) are Cartesian coordinates that parameterize  the unit two-sphere ($\sum_i (x^i)^2\,=\,1$) and $A^i$ are the components of the non-Abelian one-form connection corresponding to the $SU(2)$ instanton. 

The squashing factors $q$ and $b$ in (\ref{internal-metric-flavored}) encode the effect of the massless flavors in the backreacted metric. Indeed, when $q=b=1$ the metric (\ref{internal-metric-flavored}) is just the canonical Fubini-Study metric of a  ${\mathbb C}{\mathbb P}^3$ manifold with radius $2L$ and (\ref{flavoredBH-metric}) is the metric of the unflavored ABJM model at non-zero temperature. The parameter $b$ represents the relative squashing of the ${\mathbb C}{\mathbb P}^3$ part of the metric with respect to the $AdS_4$ part due to the flavor, while $q$ parameterizes an internal deformation which preserves the ${\mathbb S}^4$-${\mathbb S}^2$ split of the twistor representation of ${\mathbb C}{\mathbb P}^3$.  The explicit expression for the coefficients $q$ and $b$ of the smeared solution of \cite{Conde:2011sw} is given below. They depend on the number of colors $N$ and flavors $N_f$, as well as on the 't Hooft coupling $\lambda\,=\,N/k$, through the combination
\beq
\hat \epsilon\,\equiv\,{3N_f\over 4k}\,=\,{3\over 4}\,\,{N_f\over N}\,\lambda\,\,,
\label{hatepsilon}
\eeq
where the factor $3/4$ is introduced for convenience. The $AdS$ radius $L$ can be also expressed in terms of $\lambda$  and the deformation parameter (\ref{hatepsilon}) (see  eqs. (\ref{flavored-AdS-radius}) and (\ref{screening-sigma})). 

\begin{figure}[ht]
\center
\includegraphics[width=0.5\textwidth]{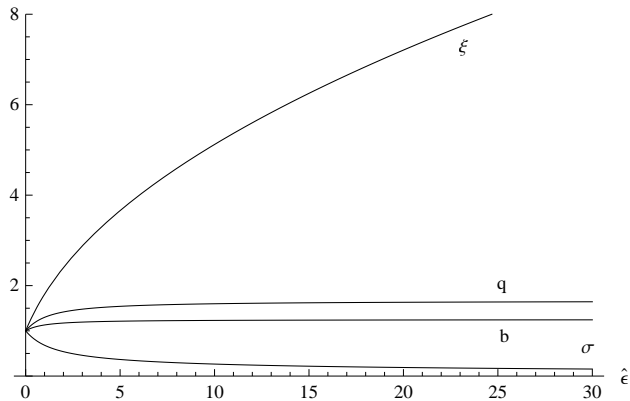}
\caption{Representation of the squashing factors $q$ and $b$, the screening function $\sigma$, and the volume 
function $\xi$ (\ref{xi}) for the background, in terms  of the deformation parameter $\hat\epsilon$.} 
\label{fig:xi}
\end{figure}

The type IIA supergravity solution found in \cite{Conde:2011sw}  contains, in  addition to the metric (\ref{flavoredBH-metric}), a constant dilaton $\phi$ and RR two- and four-forms $F_2$ and $F_4$. In order to specify the form of the latter, let us introduce  a specific system of  coordinates  to represent the metric (\ref{internal-metric-flavored}). First of all,  let $\omega^i$ ($i=1,2,3$) be the $SU(2)$ left-invariant one-forms which satisfy $d\omega^i={1\over2}\,\epsilon_{ijk}\,\omega^j\wedge\omega^k$.  Together with a new coordinate $\xi$, the $\omega^i$'s can be used to parameterize the metric of  a four-sphere ${\mathbb S}^4$ as
\beq
ds^2_{{\mathbb S}^4}\,=\,
{4\over(1+\xi^2)^2}
\left[d\xi^2+{\xi^2\over4}\left((\omega^1)^2+(\omega^2)^2+(\omega^3)^2
\right)\right]\,\,,
\label{S4metric}
\eeq
where $0\le \xi<\infty$ is a non-compact coordinate. The $SU(2)$ instanton one-forms $A^i$ can be written in these coordinates as
\beq
A^{i}\,=\,-{\xi^2\over 1+\xi^2}\,\,\omega^i\,\,. 
\label{A-instanton}
\eeq
Let us next parameterize the $x^i$ coordinates of the ${\mathbb S}^2$ by  two angles $\theta$ and $\varphi$ ($0\le\theta<\pi$, $0\le\phi<2\pi$), namely
\beq
x^1\,=\,\sin\theta\,\cos\varphi\,\,,\qquad\qquad
x^2\,=\,\sin\theta\,\sin\varphi\,\,,\qquad\qquad
x^3\,=\,\cos\theta\,\,.
\eeq
Then, one can easily prove that
\beq
\big(d x^i\,+\, \epsilon^{ijk}\,A^j\,A^k\,\big)^2\,=\,(E^1)^2\,+\,(E^2)^2\,\,,
\eeq
where  $E^1$ and $E^2$ are the following one-forms:
\bear
&&E^1=d\theta+{\xi^2\over1+\xi^2}\left(\sin\varphi\,\omega^1-\cos\varphi\,\omega^2\right) \\
\rc\rc
&&E^2=\sin\theta\left(d\varphi-{\xi^2\over1+\xi^2}\,\omega^3\right)+{\xi^2\over1+\xi^2}\,
\cos\theta\left(\cos\varphi\,\omega^1+\sin\varphi\,\omega^2\right)\,.
\label{Es}
\eear
Using these results we can represent the ten-dimensional metric (\ref{flavoredBH-metric}) as
\beq
ds^2\,=\,L^2\,\,ds^2_{BH_4}\,+\,
{L^2\over b^2}\,\Big[\,q\,ds^2_{{\mathbb S}^4}\,+\,
(E^1)^2\,+\,(E^2)^2\,\Big]\,\,.
\label{metric-AdS-flavored}
\eeq
We shall  next consider a rotated version of the forms $\omega^i$ by the  two angles $\theta$ and $\varphi$. Accordingly, we define three new one-forms  $S^i$ $(i=1,2,3)$:
\bear
&&
S^1=\sin\varphi\,\omega^1-\cos\varphi\,\omega^2 \nonumber\\
&&
S^2=\sin\theta\,\omega^3-\cos\theta\left(\cos\varphi\,\omega^1+
\sin\varphi\,\omega^2\right)\nonumber\\
&&
S^3=-\cos\theta\,\omega^3-\sin\theta\left(\cos\varphi\,\omega^1+
\sin\varphi\,\omega^2\right)\,.
\label{rotomega}
\eear 
In terms of the forms defined in (\ref{rotomega})
 the line element  of the four sphere is obtained by substituting $\omega^i\to S^i$ in (\ref{S4metric}). Let us next define the one-forms ${\cal S}^{\xi}$   and ${\cal S}^{i}$,
\beq
{\cal S}^{\xi}\,=\,{2\over 1+\xi^2}\,d\xi\,\,,\qquad\qquad
{\cal S}^{i}\,=\,{\xi\over 1+\xi^2}\,S^i \,\,,\qquad(i=1,2,3)\,\,,
\label{calS}
\eeq
in terms of which the metric of the four-sphere is 
\beq
ds^2_{{\mathbb S}^4}=({\cal S}^{\xi})^2+\sum_i({\cal S}^{i})^2\,\,.
\eeq
With these definitions,  the ansatz for $F_2$ for the flavored background written in  eq. (5.6) of ref.  \cite{Conde:2011sw} is
\beq
F_2\,=\,{k\over 2}\,\,\Big[\,\,
E^1\wedge E^2\,-\,\eta\,
\big({\cal S}^{\xi}\wedge {\cal S}^{3}\,+\,{\cal S}^1\wedge {\cal S}^{2}\big)
\,\,\Big]\,\,,
\label{F2-flavored}
\eeq
where $\eta$ is a constant squashing parameter between the ${\mathbb S}^4$ and 
${\mathbb S}^2$ components of (\ref{F2-flavored}). In the unflavored ABJM solution of \cite{Aharony:2008ug} the $F_2$ is given by (\ref{F2-flavored}) with $\eta=1$.  For a general value of $\eta$ the two-form $F_2$ is not closed. Indeed, one can easily verify that
\beq
dF_2\,=\,2\pi\,\,\Omega\,\,,
\label{dF2}
\eeq
where $\Omega$ is the following three-form
\beq
\Omega\,=\,{k\over 4\pi}\,\,
\big(\,1-\eta\,\big)\,\Big[\,
E^1\wedge ({\cal S}^{\xi}\wedge {\cal S}^{2}\,-\,{\cal S}^1\wedge {\cal S}^{3}\big)\,+\,
E^2\wedge ({\cal S}^{\xi}\wedge {\cal S}^{1}\,+\,{\cal S}^2\wedge {\cal S}^{3}\big)\,
\Big]\,\,. 
\label{Omega}
\eeq
Thus, when $\eta\not=1$  the Bianchi identity for $F_2$ is violated. This violation is due to the presence of a delocalized set of  D6-branes, whose Wess-Zumino action can be written as
\beq
S_{WZ}=\,T_{D_6}\,\,\int_{{\cal M}_{10}}\,\,
C_7\wedge \Omega\,\,,
\label{WZ-smeared}
\eeq
where $C_7$ is the RR seven-form potential and $\Omega$ is a charge distribution three-form. Clearly, the term (\ref{WZ-smeared}) induces a source for $C_7$, which modifies the Maxwell equation of $F_8=d\,C_7$. Taking into account that  $F_2=\ast F_{8}$, one easily concludes that the equation of motion for $C_7$ just takes the form of the modified Bianchi identity (\ref{dF2}). Thus, one identifies the three-form $\Omega$ written in (\ref{Omega}) with the one parametrizing the distribution of the smeared set of D6-branes.  Actually, from this identification one can relate the constant $\eta$ to the total number of flavors $N_f$. Indeed, one gets \cite{Conde:2011sw} the simple equation:
\beq
\eta\,=\,1\,+\,{3N_f\over 4k}\,\,,
\qquad\qquad
\eta\in [1,\infty)\,\,.
\label{etaNf-k}
\eeq
It is obvious from (\ref{etaNf-k}) that $\eta$ is simply related to the deformation parameter introduced in (\ref{hatepsilon}),
\beq
\eta\,=\,1+\hat \epsilon\,\,.
\label{eta-hatepsilon}
\eeq
In the solution of \cite{Conde:2011sw} the squashing parameters $b$ and $\eta$ are related by a quadratic equation, which is obtained by requiring that the background preserves ${\cal N}=1$ supersymmetry at zero temperature. This quadratic equation is 
\beq
q^2\,-\,3(1+\eta)\,q\,+5\eta\,=\,0\,\,.
\label{q-eq-flav}
\eeq
By solving this equation for $q$ and using (\ref{etaNf-k}) one can obtain $q$ as a function of the deformation parameter $\hat\epsilon$, 
\beq
q\,=\,3+{3\over 2}\,\hat \epsilon\,-\,2
\sqrt{1+ \hat \epsilon\,+\,{9\over 16}\,\hat \epsilon^2}\,\,.
\label{q-hatepsilon}
\eeq
Moreover, the solution of the BPS system of \cite{Conde:2011sw} allows to relate the parameter $b$  to the squashing factors $q$ and $\eta$:
\beq
b\,=\,{q(\eta+q)\over 2(q+\eta q-\eta)}\,\,.
\label{b-squashing}
\eeq
From this equation we get the explicit expression of $b$ in terms of the deformation parameter $\hat\epsilon$:
\beq
b\,=\,{4+{13\over 4}\,\hat\epsilon\,-\,\sqrt{1+ \hat \epsilon\,+\,{9\over 16}\,\hat \epsilon^2}
\over 3+2\hat\epsilon}\,\,.
\label{b-hatepsilon}
\eeq
By construction $\eta=q=b=1$ when $N_f=0$, whereas in the flavored solutions these coefficients are greater than one. In order to have a better idea of the behavior of $q$ and $b$ it is quite useful to expand them in powers of $N_f/k$. We get
\beq
q\,=\,
1+{3\over 8}\,{N_f\over k}-{45\over 256}\,\big({N_f\over k}\big)^2\,+\,
\cdots
\,\,,
\qquad
b\,=\,1\,+{3\over 16}\,{N_f\over k}\,-\,{63\over 512}\,\,\Big(\,{N_f\over k}\,\Big)^2\,+\,
\cdots\,\,.
\label{q-b-smallNf}
\eeq
Notice, however, that $q$ and $b$ reach a finite limiting value when the deformation parameter is very large. Indeed, one can check from (\ref{q-hatepsilon}) and (\ref{b-hatepsilon}) that
\beq
q\to {5\over 3}\,\,,\qquad\qquad\qquad
 b\to {5\over 4}\,\,,\qquad\qquad\qquad
 {\rm as}\,\,\qquad
\,{N_f\over k}\to\infty\,\,.
\label{q-b-largeNf}
\eeq
To fix completely the metric (\ref{flavoredBH-metric}) we need to know the value of the $AdS$ radius $L$. In the unflavored case $L^2$ is  proportional to the square root of the 't Hooft coupling $\lambda$. This value gets deformed by the backreaction of the flavors. Actually, we have \cite{Conde:2011sw},
\beq
L^2\,=\,\pi\,\sqrt{2\lambda}\,\,\sigma\,\,,
\label{flavored-AdS-radius}
\eeq
where $\sigma$ is defined as the following function of the deformation parameter:
\beq
\sigma\,\equiv\,\sqrt{
{2-q\over q(q+\eta q-\eta)}}\,\,b^2\,=\,
{1\over 4}\,\,
{q^{{3\over 2}}\,\,(\eta+q)^2\,(2-q)^{{1\over 2}}
\over
(q+\eta q-\eta)^{{5\over 2}}}\,\,.
\label{screening-sigma}
\eeq
It was shown in \cite{Conde:2011sw} that $\sigma$ characterizes the  corrections of the static quark-antiquark potential due to the screening produced by the flavors.
In Fig.~\ref{fig:xi} we depict $q,b,\sigma,$ and $\xi$ (\ref{xi}) as functions of the deformation parameter $\hat\epsilon$.

The solution is completed by a constant dilaton $\phi$ given by
\beq
e^{-\phi}\,=\,{b\over 4}\,{\eta+q\over 2-q}\,{k\over L}\,\,,
\label{dilaton-flavored-squashings}
\eeq
and a RR four-form $F_4$, whose expression is
\beq
F_4\,=\,{3k\over 4}\,\,\,{(\eta+q)b\over 2-q}\,\,L^2\,\,\Omega_{BH_4}\,\,,
\eeq
where $\Omega_{BH_4}$ is the volume-form of the four-dimensional black hole (\ref{BH4-metric}). The regime of validity of the type IIA  supergravity description can be obtained by requiring that $L \gg 1$ and $e^{\phi}\ll 1$. For the flavored ABJM background at zero temperature these two conditions were worked out in detail in \cite{Conde:2011sw} and will not be discussed further here. 

In the zero temperature case this background was found in \cite{Conde:2011sw}  by solving the system of first order BPS equations required to preserve ${\cal N}=1$ supersymmetry. Then, one can verify that the solution satisfies the second order equations of type IIA supergravity with sources (see appendix D of \cite{Conde:2011sw}).  In the black hole case one can easily check that these equations of motion are still satisfied after the introduction of the blackening factor $h(r)$ in the metric. 

\subsection{Thermodynamics of the background}

Let us now find the values of the different thermodynamic functions for the flavored black hole presented above. We begin by computing the entropy density $s_{back}$, which is given by:\footnote{We use the same conventions as the first paper in \cite{D3-D7}.}
\beq
s_{back}\,=\,{2\pi\over \kappa_{10}^2}\,\,{A_8\over V_2}\,\,,
\eeq
where $A_8$ is the volume at the horizon $r=r_h$ of the eight-dimensional part of the space obtained by setting $r, t={\rm constant}$ in the ten-dimensional geometry and $V_2$ is the infinite volume of the 2d space directions $x^i$.  The volume $A_8$ has to be computed with the Einstein frame metric, which in our case is obtained by changing $L$ by $e^{-\phi/4}\,L$ in (\ref{flavoredBH-metric}) and (\ref{internal-metric-flavored}). After a simple calculation one can check that $A_8/V_2$ is given by
\beq
{A_8\over V_2}\,=\,{32\pi^3\over 3}\,\,{q^2\,L^8\,e^{-2\phi}\over b^6}\,r_h^3\,\,.
\label{A8/V2}
\eeq
We can now use the values of the different factors appearing on the right-hand side of  (\ref{A8/V2}) to obtain the value of the entropy density in terms of gauge theory quantities. Taking into account that, in our units, $2\kappa_{10}^2=(2\pi)^2$, we get
\beq
s_{back}\,=\,{1\over 3}\,
\left(\frac{4\pi}{3}\right)^2 \frac{N^2}{\sqrt{2\lambda}}\,\xi\left(\frac{N_f}{k}\right) T^2\,\,,
\label{s_back}
\eeq
where
\beq
 \xi\left(\frac{N_f}{k}\right) \equiv \frac{1}{16}\frac{q^{\frac{5}{2}}(\eta+q)^4}{\sqrt{2-q}(q+\eta q-\eta)^{\frac{7}{2}}} \ .
 \label{xi}
\eeq
The quadratic dependence of the entropy with the temperature is a reflection of the conformality of the system which, in our solution,  is not affected by the massless flavors. Notice that $s_{back}$ displays the characteristic $N^{{3\over 2}}$ behavior of the  effective number of degrees of freedom of the ABJM theory in the 't Hooft limit.  The correction to this behavior introduced by the flavors is parameterized  by the function $ \xi$, which was introduced in \cite{Conde:2011sw} and shown to be very close to the function obtained by using the localization technique.  The function $\xi$ determines how the volume of the internal manifold (and, hence, the area of the horizon) changes due to the addition of flavor. 

The internal energy density can be obtained from the ADM energy,
\beq
E_{ADM}\,=\,-{1\over \kappa_{10}^2}\,
\sqrt{|G_{tt}|}\,\,\int_{{\cal M}_{t,r_{\infty}}}\sqrt{\det G_8}\,\,
(\,K_T\,-\,K_0\,)\,\,.
\label{ADM_energy}
\eeq
In (\ref{ADM_energy}) $G_8$ is the Einstein frame metric of the $t,r={\rm constant}$  hypersurface. The integral is taken over this hypersurface for a large value $r=r_{\infty}$ of the radial coordinate. The symbols $K_T$ and $K$ denote the extrinsic curvatures of the eight-dimensional subspace within the nine-dimensional (constant time) space, at finite and zero temperature, respectively. For an arbitrary hypersurface $K$ is given by
\beq
K\,=\,{1\over \sqrt{\det G_9}}\,\partial_{\mu}\,
\Big(\,\sqrt{\det G_9}\,\,n^{\mu}\,\Big)\,\,,
\eeq
with $n^{\mu}$ being a normalized vector perpendicular to the surface. For a constant $r$ hypersurface,
\beq
n^{\mu}\,=\,{1\over \sqrt{G_{rr}}}\,\delta_{r}^{\mu}\,\,,
\eeq
and one can show that $K$  for our background becomes
\beq
K\,=\,{2\,e^{{\phi\over 4}}\,\sqrt{h}\over L}\,\,.
\eeq
By using these results it is easy to find the value of the integrand in (\ref{ADM_energy}),
\beq
\sqrt{|G_{tt}|}\,\sqrt{\det G_8}\,\,(\,K_T\,-\,K_0\,)\,=\,-\,e^{-2\phi}\,L^2\,
\sqrt{\det g_6}\,\,r_{h}^3\,\,,
\eeq
where $g_6$ is the internal metric (\ref{flavoredBH-metric}). It is now immediate to obtain the internal energy density of the flavored black hole,
\beq
E_{back}\,=\,{E_{ADM}\over V_2}\,=\,{2\over 9}\,
\left(\frac{4\pi}{3}\right)^2 \frac{N^2}{\sqrt{2\lambda}}\,\xi\left(\frac{N_f}{k}\right) T^3
\,\,. 
\label{E_black}
\eeq
Again, the dependence on the temperature is just the one expected for a conformal system and the flavor dependence is determined by the function $\xi$. Moreover, the free energy density $F_{back}$ can be obtained from the thermodynamic relation 
$ F_{back}\,=\,E_{back}-T\,s_{back}$, yielding,
\beq
 F_{back} = -{1\over 9}\left(\frac{4\pi}{3}\right)^2 \frac{N^2}{\sqrt{2\lambda}}\,\xi\left(\frac{N_f}{k}\right) T^3\,\,. 
\label{free-energy-background}
\eeq
As a consistency check we notice that $s_{back}\,=\,-\partial  F_{back} /\partial T$, as it should. It is also worth pointing out that the free energy  density $ F_{back}$ can be computed directly from the regularized Euclidean action (see the first paper in \cite{D3-D7} for a similar calculation for the D3-D7 black hole). The regularization is performed by subtracting the action at zero temperature with the Euclidean time suitably rescaled. Furthermore, in the action one must include the standard Gibbons-Hawking surface term. The final result of this calculation, which will not be detailed here,  is just the same as in (\ref{free-energy-background}).

\section{D6-brane embeddings at zero temperature}
\label{zeroTemp}

One key objective of this paper is to study the properties of flavor brane probes embedded in the flavored black hole background described in Section \ref{FlavABJM}.  Before dealing with this problem in full generality, let us analyze the case in which the temperature of the background is zero, which corresponds to taking the blackening factor $h(r)$  equal to one  in the formulas  of Section \ref{FlavABJM}. 

The kappa symmetric embeddings of the flavor D6-branes that preserve the supersymmetry of the zero temperature background were studied in \cite{Conde:2011sw}. As argued in \cite{Hohenegger:2009as}, these D6-branes should extend along the three Minkowski directions $x^{\mu}$, the radial coordinate $r$, and wrap a three-dimensional submanifold  of the compact internal space. For large values of the radial coordinate the metric of this three-dimensional submanifold should approach that of a (squashed) 
${\mathbb R}\,{\mathbb P}^3\,=\,{\mathbb S}^3/{\mathbb Z}_2$.  In our 
${\mathbb S}^4-{\mathbb S}^2$ representation, it was shown in \cite{Conde:2011sw} that this internal submanifold is obtained by extending the D6-brane along the 
${\mathbb S}^4$ base in such a way that the pullback of the one-forms $\omega^1$ and $\omega^2$ vanish. Accordingly, let us consider a configuration such that 
$\hat\omega^1=\hat\omega^2=0$, where the hat denotes the pullback to the D6-brane worldvolume. Moreover, for the pullback of $\omega^3$ we just take $\hat \omega^3=d\hat\psi$, where $\hat\psi$ is an angular coordinate. We will also assume that the brane is extended along the coordinate $\varphi$ of the ${\mathbb S}^2$ fiber and that the other  ${\mathbb S}^2$  coordinate $\theta$ is a function of the radial coordinate $r$, $\theta=\theta(r)$. Therefore, 
we will choose the following set of worldvolume coordinates
\beq
\zeta^{\alpha}\,=\,(x^{\mu}, r, \xi, \hat\psi, \varphi)\,\,.
\eeq
Then,  the induced metric (at zero temperature)  on the D6-brane worldvolume becomes
\beq
d\hat s^2_7\,=\,
-L^2r^2\,dt^2+
L^2\,r^2\big[\,(dx^1)^2+(dx^2)^2\,\big]\,\,+\,
L^2\Big[\,{1\over r^2}+ {1\over b^2}\,\,\Big({d\theta\over dr}\Big)^2\,\Big]\,
dr^2+{4L^2\over b^2}\,ds^2_{3}\,\,,
\label{induced-general}
\eeq
where $ds^2_{3}$ is the following three-dimensional metric
\beq
ds^2_{3}\,=\,
{q\over (1+\xi^2)^2}\,\,d\xi^2\,+\,{q\over 4}\,\,
{\xi^2\over (1+\xi^2)^2}\,d\hat\psi^2\,+\,{1\over 4}
\sin^2\theta\,
\big(\,d\varphi\,-\,{\xi^2\over 1+\xi^2}\,d\hat\psi\,\big)^2\,\,.
\eeq
If we redefine the angular coordinates as
\beq
\xi\,=\,\tan\Big({\alpha\over 2}\Big)\,\,,\qquad\qquad
\beta\,=\,{\hat\psi\over 2}\,\,,\qquad\qquad
\psi\,=\,\varphi\,-\,{\hat\psi\over 2}\,\,,
\eeq
then the 3d   metric  $ds_{3}^2$ becomes
\beq
ds_{3}^2\,=\,{1\over 4}\,\,\Big[\, q d\alpha^2\,+\,q\,\sin^2\alpha d\beta^2\,+\,
\sin^2\theta\,
\big(\,d\psi\,+\,\cos\alpha\,d\beta\,\big)^2\,\Big]\,\,,
\label{wv-angularmetric}
\eeq
where $\theta$ is assumed to be a function of $r$. The range of the angular coordinates in (\ref{wv-angularmetric}) is,
\beq
0\le \alpha<\pi\,\,,
\qquad
0\le\beta <2\pi\,\,,
\qquad
0\le\psi< 2\pi\,\,.
\eeq

Notice that, in these coordinates,  the massless configurations whose backreaction is included in the background of Section \ref{FlavABJM}, correspond to embeddings with $\theta(r)$ being constant and equal to $\pi/2$. In order to simplify the study of all possible embeddings that satisfy the equations of motion of the probe, it is convenient to choose an isotropic system of coordinates. To find these coordinates,  let us consider the $(r,\theta)$ part of the induced metric (\ref{induced-general}), which can written as,
\beq
{L^2\over r^2}\,dr^2\,+\,{L^2\over b^2}\,d\theta^2\,=\,
{L^2\over b^2}\,\Big[\,{b^2\over r^2}\,dr^2\,+\,d\theta^2\,\Big]\,\,.
\label{r-theta-metric}
\eeq
We want to find a new radial coordinate $u$ such that the first term inside the brackets in   (\ref{r-theta-metric}) becomes $du^2/u^2$ and the whole right-hand side of 
(\ref{r-theta-metric})  is proportional to $du^2+u^2\,d\theta^2$. Clearly, we must require 
\beq
{b\,dr\over r}\,=\,{du\over u}\,\,,
\label{isotropic-DE}
\eeq
and thus (\ref{r-theta-metric}) becomes
\beq
{L^2\over b^2 u^2}\,\big[\,du^2\,+\,u^2\,d\theta^2\,\big]\,\,.
\label{isotropic-u-theta-metric}
\eeq
Eq. (\ref{isotropic-DE}) can be immediately integrated, with the result
\beq
u\,=\,r^{b}\,\,.
\eeq
Notice that the change $r\to u$ of the radial coordinate is only non-trivial in the flavored case with $b\not=1$. In terms of  this  $u$ variable, the ten-dimensional metric (\ref{flavoredBH-metric}) (for $h=1$), becomes
\beq
ds^2\,=\,L^2\,\Big[\,u^{{2\over b}}\,\,dx^2_{1,2}\,+\,{1\over b^2}\, {du^2\over u^2}\,\Big]\,+\, ds^2_6\,\,,
\eeq
where $ds^2_6$ is the metric (\ref{internal-metric-flavored}) of the squashed ${\mathbb C}{\mathbb P}^3$. 

Let us now introduce a system of Cartesian-like coordinates  $(\rho,R)$, defined as
\beq
R\,=\,u\,\cos\theta\,\,,
\qquad\qquad
\rho\,=\,u\,\sin\theta\,\,.
\label{R-rho-def}
\eeq
The inverse relation is
\beq
u^2\,=\,R^2\,+\,\rho^2\,\,,
\qquad\qquad
\tan\theta\,=\,{\rho\over R}\,\,,
\eeq
and, since $du^2\,+\,u^2\,d\theta^2=d\rho^2+dR^2$, the line element  (\ref{r-theta-metric})  becomes
\beq
{L^2\over b^2 (\rho^2+R^2)}\,\,\big[\,d\rho^2\,+\,dR^2\,\big]\,\,.
\eeq
Let us now  consider embeddings  of the D6-brane in which $R=R(\rho)$. Then, the induced metric takes the form
\bear
&&d\hat s^2_7\,=\,L^2\,\big[\,\rho^2+R^2\,\big]^{{1\over b}}\,dx_{1,2}^2\,+\,{L^2\over b^2}\,\,{1+R'^2\over \rho^2+R^2}\,d\rho^2\,+\,\rc\rc
&&\qquad\qquad
+{L^2\over b^2}\,\Big[\, q\,d\alpha^2\,+\,q\,\sin^2\alpha d\beta^2\,+\,
{\rho^2\over \rho^2+R^2}\,
\big(\,d\psi\,+\,\cos\alpha\,d\beta\,\big)^2\,\Big]\,\,,
\eear
with $R'\equiv dR/d\rho$.  The embeddings corresponding to massless flavors are the ones for which $R=0$. In the general case,  the determinant of the induced metric takes the form
\beq
\sqrt{-\det\hat  g_7}\,=\,{L^7\over b^4}\,\,q\,\sin\alpha\,\rho\,
[\rho^2+R^2]^{{3\over 2b}-1}\,\,\sqrt{1+ R'^2}\,\,.
\label{g7-zeroT}
\eeq

In order to obtain the explicit form of the embeddings, let us now study the action of the probe brane. We begin by computing the DBI action, which is given by
\beq
S_{DBI}\,=\,-T_{D6}\,\int d^7\zeta \,e^{-\phi}\,\sqrt{-\det\hat  g_7}\,\,,
\label{DBI-D6-general}
\eeq
where the tension of the D6-brane $T_{D6}\,=1/(2\pi)^6$ in our units. 
Let us use (\ref{g7-zeroT})  in (\ref{DBI-D6-general})  and integrate over the angular coordinates $\alpha$, $\beta$, and $\psi$. We define a Lagrangian density ${\cal L}_{DBI}$ as
\beq
S_{DBI}\,=\,\int d^3 x\,d\rho\,{\cal L}_{DBI}\,\,,
\eeq
where 
\beq
{\cal L}_{DBI}\,=\,-{\cal N}_0\,\rho\,
[\rho^2+R^2]^{{3\over 2b}-1}\,\,\sqrt{1+ R'^2}\,\,,
\eeq
with ${\cal N}_0$ being the following constant
\beq
{\cal N}_0\,=\,{8\pi^2\,L^7\,q\over b^4}\,T_{D6}\,e^{-\phi}\,\,.
\eeq
Next, let us compute the WZ term of the action, which becomes
\beq
S_{WZ}\,=\,T_{D6}\,\int \hat C_7\,\,,
\label{WZ-probe}
\eeq
where $C_7$ is the RR seven-form potential ($F_8=dC_7$) and, as before,  the hat denotes the pullback to the worldvolume.  In this zero temperature case the RR seven-form potential $C_7$ is naturally given in terms of the calibration seven-form ${\cal K}$ that characterizes the G-structure of the supersymmetric solution. Indeed, we can take $C_7$ as
\beq
C_7=e^{-\phi}\,{\cal K}\,\,.
\label{C7-SUSY}
\eeq
The seven-form ${\cal K}$ is naturally defined in terms of a fermion bilinear which, in turn, can be obtained from the projections satisfied by the Killing spinors of the background. This calculation was performed in \cite{Conde:2011sw} and here we will limit ourselves to recall this result. As shown in \cite{Conde:2011sw}, to represent ${\cal K}$  it is useful to define the following basis of one-forms:
\bear
&&e^{0}\,=\, L\,r\,dt\,\,,\qquad\qquad
e^{1}\,=\, L\,r\,dx\,\,,\qquad\qquad
e^{2}\,=\, L\,r\,dy\,\,,\rc\rc
&&e^{3}\,=\,{L\over r}\,\,dr\,\,,\qquad\qquad
e^{4}\,=\,{L\over b}\,\sqrt{q}\,{\cal S}^{\xi}\,\,,\rc\rc
&&e^{i}\,=\,{L\over b}\,\sqrt{q}\,{\cal S}^{i-4}\,\,,
\qquad\qquad (i=5,6,7)\,\,,\rc\rc
&&e^{j}\,=\,{L\over b}\,E^{j-7}
\,\,,\qquad\qquad (j=8,9)\,\,,
\label{framebasis}
\eear
which are a frame basis for the zero-temperature version of the metric (\ref{flavoredBH-metric}). 
In terms of the forms (\ref{framebasis}) the form ${\cal K}$ can be written as \cite{Conde:2011sw},
\beq
{\cal K}\,=\,-e^{012}\,\wedge\big(\,
e^{3458}\,-\,e^{3469}\,+\, e^{3579}\,+\,e^{3678}\,+\,e^{4567}\,+\,e^{4789}\,+\,e^{5689}
\,\big)\,\,.
\label{cal-K-explicit}
\eeq
To evaluate the WZ action we need to compute the pullback of ${\cal K}$ to the worldvolume. Let us write the pullbacks of the  frame one-forms  (\ref{framebasis}) in the $(\rho, R)$ coordinates. In this calculation it is convenient to use
\beq
d\theta\,=\,{R-\rho R'\over \rho^2\,+\,R^2}\,d\rho\,\,,
\qquad\qquad
dr\,=\,{1\over b}\,{RR'\,+\,\rho\over \big[\,\rho^2\,+\,R^2\,\big]^{1-{1\over 2b}}}
\,d\rho\,\,.
\eeq
We find
\bear
&& \hat e^{\mu}\,=\,
L\,[\rho^2+R^2]^{{1\over 2b}}\,dx^{\mu}\,\,
\,\,,\qquad\qquad
\hat e^{3}\,=\,{L\over b}\,\,{R\,R'\,+\,\rho\over \rho^2+R^2}\,
d\rho\,\,,\qquad\qquad
\hat e^{4}\,=\,{L\over b}\,\sqrt{q}\,\,d\alpha\,\,,
\qquad\rc\rc
&&\hat e^{5}\,=\,0\,\,,\qquad\qquad
\hat e^{6}\,=\,{L\sqrt{q}\over b}\,\sin\alpha\,{\rho\over \sqrt{\rho^2\,+\,R^2}}\,
\,d\beta\,\,,
\qquad
\hat e^{7}=-
{L\sqrt{q}\over b}\,\sin\alpha\,{R\over \sqrt{\rho^2\,+\,R^2}}\,
\,d\beta
\,\,,\rc\rc
&&\hat e^{8}\,=\,{L\over b}\,{R-\rho R'\over \rho^2\,+\,R^2}
\,d\rho\,\,,\qquad
\hat e^{9}\,=\,
{L\over b}\,{\rho\over \sqrt{ \rho^2\,+\,R^2}}
\,\big(\,d\psi+\cos\alpha d\beta
\,\big)\,\,.
\label{pullback-es}
\eear
By inspecting these pullbacks one readily verifies that the only non-zero contributions to
 $\hat {\cal K}$ are
 \beq
 \hat {\cal K}=\hat e^{012}\wedge \big(\,\hat e^{3469}\,-\,\hat e^{4789}\,)\,=\,
 {L^7 q\over b^4}\,\sin\alpha\,\rho\,
\big[\,\rho^2\,+\,R^2\,\big]^{{3\over 2b}-1}\,\,
d^3x\wedge  d\rho\wedge d\alpha\wedge d\beta\wedge d\psi\,\,.
\label{pullback-calK}
\eeq
Thus, after integrating over the angular variables, we can write
\beq
S_{WZ}\,=\,\int d^3 x\,d\rho\,{\cal L}_{WZ}\,\,,
\eeq
with the Lagrangian density 
\beq
{\cal L}_{WZ}\,=\,{\cal N}_0\,\rho\,
 \big[\,\rho^2\,+\,R^2\,\big]^{{3\over 2b}-1}\,\,.
\eeq
Therefore, the total Lagrangian density  is
\beq
{\cal L}\,=\,-{\cal N}_0\,\rho\,
\big[\,\rho^2\,+\,R^2\,\big]^{{3\over 2b}-1}\,
\big(\,\sqrt{1+ R'^2}\,-\,1\,\big)\,\,.
\label{calL-zeroT}
\eeq
Clearly, $R={\rm constant}$ is a solution of the equations of motion derived from ${\cal L}$ (notice that the on-shell action for this configuration vanishes). This is just the kappa symmetric solution that preserves SUSY which was found in \cite{Conde:2011sw}.\footnote{Notice that, in the angular $(r,\theta)$ parameterization of \cite{Conde:2011sw}, the $R={\rm constant}$ solution reads $\theta(r)=\arccos\big({r_0\over r}\big)^b$, where  $r_0^b=R_0$.} Let us now study the form of a general solution in the UV region of large $\rho$. In this case one can approximate $\rho^2\,+\,R^2\approx \rho^2$  in (\ref{calL-zeroT}) and take $R'$ small. At second order in  $R'$, we find that ${\cal L}$ can be approximately taken as
\beq
{\cal L}\,\approx\,-{{\cal N}_0\over 2}\,\rho^{{3\over b}\,-\,1}\,R'^2\,\,.
\eeq
The equation of motion derived from this second-order Lagrangian is simply
\beq
\partial_{\rho}\,\Big(\,\rho^{{3\over b}-1}\,\,R'\,\big)\,=\,0\,\,,
\label{eom-R-UV}
\eeq
and can be integrated trivially
\beq
R\,\sim m\,+\,{c\over \rho^{{3\over b}-2}}\,\,,
\label{R-UV}
\eeq
In (\ref{R-UV}) $m$ and $c$ are constants, which should be related to the mass of the quarks and to the vacuum expectation value of the corresponding bilinear operator $\bar\psi\,\psi$ (see below), respectively.  The power of $\rho$ of the subleading term in (\ref{R-UV}) should determine the conformal dimension of the bilinear operator. Indeed,  let us consider  a canonically normalized field $\phi$ in $AdS_4$ with conformal dimension $\Delta$. The behavior of $\phi$ near the boundary of $AdS_4$ is
\beq
\phi\sim \phi_0\,r^{\Delta-3}\,+\,{\langle{\cal O}\rangle\over r^{\Delta}}\,\,,
\label{bdy-behavior}
\eeq
where $ \phi_0$ (the boundary value of $\phi$) is identified with the source of the dual gauge theory operator ${\cal O}$ and the coefficient $\langle{\cal O}\rangle$ is identified with its VEV. In (\ref{bdy-behavior}) $\Delta$ is the dimension of ${\cal O}$ and $r$ is the canonical coordinate of $AdS_4$ (in terms of which the  $AdS_4$  metric  takes the form $r^2\,dx^2_{1,2}+dr^2/r^2$). It is clear that this canonical coordinate is just the  one in (\ref{BH4-metric}). In the UV,  $r$ and $\rho$ are related as $r\sim \rho^{1/b}$, and therefore we can rewrite (\ref{R-UV}) in terms of $r$ as
\beq
R\,\sim\,m\,+\,{c\over r^{3-2b}}\,\,. 
\label{R-r-UV}
\eeq
In order to extract the dimension of the operator dual to the scalar $R$, let us rewrite (\ref{bdy-behavior}) in such a way that the asymptotic value of the right-hand side is constant,
\beq
r^{3-\Delta}\,\phi\sim \phi_0\,+\,{\langle{\cal O}\rangle\over r^{2\Delta-3}}\,\,.
\label{bdy-behavior-const}
\eeq
Clearly, by comparing (\ref{bdy-behavior-const}) and (\ref{R-r-UV}) we find that, in our flavored ABJM case, $2\Delta-3=3-2b$, which yields
\beq
\Delta\,=\,3-b\,\,,
\eeq
in agreement with the value obtained in \cite{Conde:2011sw} for the dimension of the bilinear operator $\bar\psi\psi$.  Notice also that $\Delta_m=3-\Delta$ is the dimension of the source (the mass in our case). This dimension is just $\Delta_m=b$ in the flavored ABJM case. Thus,  the mass anomalous dimension is
\beq
\gamma_m = \Delta_m - 1 = b-1 \ .
\label{gamma_m}
\eeq
It is evident from (\ref{gamma_m}) that 
the anomalous dimension $\gamma_m$ depends on the number of flavors $N_f$ and, according to (\ref{q-b-largeNf}), it becomes maximum when $N_f\to\infty$:
\beq
\gamma_{m}^{max}\,=\,{1\over 4}\,\,.
\eeq

As it was already mentioned, the asymptotic value $m$ should be related to the quark mass $m_q$. To find the precise relation let us consider a fundamental string extended from the origin to the point with $R=R_0=m$ at $\rho=0$. The induced metric on the worldsheet of this string is
\beq
ds^2_2\,=\,-L^2 R^{{2\over b}} dt^2\,+\,{L^2\over b^2}\,\,{dR^2\over R^2}\,\,,
\eeq
whose determinant is
\beq
\sqrt{-\det g_2}\,=\,{L^2\over b}\,\,R^{{1\over b}-1}\,\,.
\eeq
The Nambu-Goto action for this string is
\beq
S_{NG}\,=\,-{1\over 2\pi}\,\int dt\,\int_{R=0}^{R=m}dR\,\sqrt{-\det g_2}\,=\,
-{L^2\over 2\pi}\,\int dt m^{{1\over b}}\,\,.
\eeq
The action per unit time should be identified with $m_q$. Thus, by using (\ref{flavored-AdS-radius}) we arrive at
\beq
m_q\propto {\sqrt{\lambda}\,\sigma\over \sqrt{\alpha'}}\,\,m^{{1\over b}}
\qquad\qquad
\Longrightarrow
\qquad\qquad
m\propto \Big(\,{m_q\sqrt{\alpha'}
\over \sqrt{\lambda}\,\sigma}\,
\Big)^{b}\,\,,
\label{m_q_zero_T}
\eeq
where $\lambda$ is the 't Hooft coupling and $\sigma$ is the function of $N_f/k$ that  has been defined in (\ref{screening-sigma}).  We have included a factor of $\sqrt{\alpha'}=l_s$ to reinstate the correct dimensions.

The constant $c$ in (\ref{R-r-UV}) should be related to the vacuum expectation value of the meson operator $\bar \psi\psi$ (the quark condensate). In order to find this relationship we should relate $c$ to the derivative of the action with respect to the mass  parameter $m$. In principle, to perform this calculation we should holographically renormalize the action to ensure its finiteness \cite{Skenderis:2002wp,Karch:2005ms}. It turns out, however, that the action corresponding to the Lagrangian density  (\ref{calL-zeroT}) is convergent and, therefore, this renormalization is not needed. Indeed, by using the asymptotic behavior (\ref{R-r-UV}) we obtain for large $\rho$,
\beq
{\cal L}\sim \rho^{1-{3\over b}}\,\,,
\eeq
and, since the maximum value of $1-{3\over b}$ is $-{7\over 5}$, the integral over $\rho$ is convergent as claimed. Notice that this convergent behavior is a consequence of the particular gauge for $C_7$ chosen. Indeed, performing a gauge transformation of the type $C_7\to C_7+d\Lambda_6$ is equivalent to adding a boundary term to the action of the probe and to choose a particular renormalization scheme. In our gauge $C_7$ is chosen to be the calibration form and, as a consequence, the action for a supersymmetric embedding $R={\rm constant}$ vanishes. For a more general embedding the WZ term introduces a subtraction of the DBI term, which renders the total action finite.

The probe configuration is obtained by solving  the equation of motion derived from the Lagrangian density (\ref{calL-zeroT}) for $R(\rho)$. In this process we have to impose boundary conditions at some value of the $\rho$ coordinate. The simplest thing is to take 
$\rho=0$ as this initial value of the coordinate and to integrate the system outwards. It is easy to verify from the limit  of the differential equation at $\rho=0$ that the only possibility to have non-singular solutions is to take $R(\rho=0)=R_0$ and $R'(\rho=0)=0$ as initial conditions. At the UV region of large $\rho$ the function $R(\rho)$ must behave as in (\ref{R-UV}), where the constants $m$ and $c$ are not independent since both should be determined by the IR value $R_0$ of $R(\rho)$. The on-shell action is obtained by evaluating the integral of ${\cal L}$ for these configurations. It can be considered as a function of the mass parameter $m$. The derivative of $S$ with respect to $m$ can be computed  as follows:
\beq
{\partial S\over \partial m}\,\sim\,\int d\rho\,\Big[\,
{\partial {\cal L}\over \partial R}\,{\partial R\over \partial m}\,+\,
{\partial {\cal L}\over \partial R'}\,{\partial R'\over \partial m}\,\Big]\,=\,
\int d\rho\,{\partial\over \partial \rho}\,\,\Big[\,{\partial {\cal L}\over \partial R'}\,{\partial R\over \partial m}\,\Big]\,\,,
\label{partialS-Rstar}
\eeq
where we have integrated by parts and used the equations of motion of $R(\rho)$. In (\ref{partialS-Rstar}) we have already integrated over the Minkowski coordinates and we have assumed that this integration gives rise to a constant factor.  The value of the right-hand side of (\ref{partialS-Rstar}) can be obtained by evaluating the ``momentum" density $\partial {\cal L}/\partial R'$ at the boundary values of the worldvolume. It is readily checked that, for regular embeddings, the IR contribution at $\rho=0$ is zero. To obtain the UV contribution at $\rho=\infty$, let us use (\ref{R-r-UV}):
\beq
{\partial {\cal L}\over \partial R'}\,\propto\,\rho\,\big[\,\rho^2+R^2\,\big]^{{3\over 2b}-1}\,\,
{R'\over \sqrt{1+R'^2}}\,\,\sim {3-2b\over b}\,c\,+\,{\rm subleading}\,\,,
\eeq
where we include different factors coming from the constant ${\cal N}_0$. 
Taking into  account that
\beq
{\partial R\over \partial m}\,=\,1\,+\,{\rm subleading}\,\,,
\eeq
we get
\beq
{\partial S\over \partial m}\sim {3-2b\over b}\,c\,\,.
\eeq
The quark condensate $\langle\bar \psi \psi\rangle$ is obtained by performing the derivative of the action with respect to the bare quark mass $\mu_q$. The latter can be obtained by taking $\sigma=b=1$ in the dressed mass $m_q$. It is clear from (\ref{m_q_zero_T}) that $\mu_q\sim m$ and, thus,
\beq
\langle\bar \psi \psi\rangle\sim {\partial S\over \partial \mu_q}\,\sim\,{3-2b\over b}\, c\,\,.
\eeq
Therefore, $c$ is indeed proportional to the quark condensate. It turns out, however, that the only regular solutions in this $T =0$ case are those for which $R={\rm constant}=m$, \ie, the kappa symmetric ones. They have $c=0$ and therefore the quark condensate vanishes in this case. Notice that the on-shell action for these solutions is zero, as expected on general grounds from their supersymmetric character (see (\ref{calL-zeroT})).

\section{Flavor brane probes at non-zero temperature}
\label{nonzeroTemb}
In this section we come back to the analysis of brane probes in the general non-zero temperature background of Section \ref{FlavABJM}. The main difference from the $T=0$ analysis of Section \ref{zeroTemp} is due to the presence of an event horizon in the metric. Thus we  will have two types of embeddings: Minkowski and black hole. In the former type the brane probe does not reach the horizon, whereas in the latter case the brane ends on the horizon. In order to describe correctly the thermodynamics of these two types of configurations and of the phase transition connecting them one has to define carefully the action of the probe. It turns out that there is a subtlety which we shall address in this section. 

As in the $T=0$ case, we will consider D6-brane probes embedded in the internal 
${\mathbb C}{\mathbb P}^3$ in such a way that the one-forms $\hat\omega^1$ and $\hat\omega^2$ vanish. We will take $(x^{\mu}, r, \alpha, \beta, \psi)$ as worldvolume coordinates and describe the embedddings by  a function $\theta=\theta(r)$. The induced metric takes the form
\bear
&&d\hat s^2_7\,=\,
-L^2r^2h(r)\,dt^2+
L^2\,r^2\big[\,(dx^1)^2+(dx^2)^2\,\big]\,\,+\,
{L^2\over r^2\,h(r)} \Big[\,1+ {r^2\,h(r)\over b^2}\,\,\dot\theta^2\,\Big]\,
dr^2+\rc\rc
&&\qquad\qquad
+\,{L^2\over b^2}\,
\Big[\, q\,d\alpha^2\,+\,q\,\sin^2\alpha \,d\beta^2\,+\,
\sin^2\theta\,\big(\,d\psi\,+\,\cos\alpha\,d\beta\,\big)^2\,\Big]\,\,,
\label{induced-r-thetal}
\eear
where the dot represent the derivative with respect to $r$. The determinant of the incuded metric is 
\beq
\sqrt{-\det g_7}\,=\,{L^7\,q\over b^3}\,r^2\,\sin\theta\,\sin\alpha\,
\sqrt{1+ {r^2\,h(r)\over b^2}\,\,\dot\theta^2}\,\,.
\eeq	
After integrating over the internal space we get the following DBI action:
\beq
S_{DBI}\,=\,{\cal N}_r\,\int\,d^3x dr \,
r^2\,\sin\theta\,
\sqrt{1+ {r^2\,h(r)\over b^2}\,\,\dot\theta^2}\,\,,
\eeq
with ${\cal N}_r$ being the following constant
\beq
{\cal N}_r\equiv {8\pi^2\,L^7\,q\over b^3}\,\,T_{D6}\,\,e^{-\phi}
\,\,.
\label{Cal_Nr}
\eeq
In terms of gauge theory quantities, we can write ${\cal N}_r$ as
\beq
 {\cal N}_r\,=\, {1\over 4\sqrt{2}\,\pi}\,{N^{{3\over 2}}\over \sqrt{k}}\,\,
 \zeta \Big({N_f\over k}\Big)\,\,,
 \label{cal_Nr_mu}
\eeq
where the function $ \zeta (N_f/k)$ contains all the dependence on $N_f$ and is given by
\beq
 \zeta \Big({N_f\over k}\Big)\,\equiv\,{1\over 2}\,\,
 {\sqrt{2-q}\,(\eta+q)\,b^4\over \sqrt{q}\,
 (q+\eta q-\eta)^{{3\over 2}}}\,=\,{1\over 32}\,
 {\sqrt{2-q}\,(\eta+q)^5\,q^{{7\over 2}}\over
  (q+\eta q-\eta)^{{11\over 2}}}
 \,\,.
 \label{mu}
 \eeq
Notice that $\zeta =1$ for $N_f=0$ and  for an arbitrary number of flavors this function 
is related to the screening function $\sigma $ defined in (\ref{screening-sigma}) by a simple equation
\beq
\sigma\,=\,{q\over b^3}\, \zeta\,\,.
\label{sigma-mu}
\eeq

Let us now focus on the WZ action, which requires some extra consideration to eventually yield consistent thermodynamics. Recall that the WZ term of the probe action is proportional to the integral of the pullback of the  RR seven-form potential $C_7$ (see (\ref{WZ-probe})). In the zero-temperature case analyzed in Section \ref{zeroTemp} we represented $C_7$ in terms of the calibration form ${\cal K}$ (eq. (\ref{C7-SUSY})). Actually,   one can easily verify that introducing the blackening factor $h(r)$ does not change the field strength
$F_8=-\ast F_2$ (the dependence on $h(r)$ cancels when one computes  the Hodge dual of $F_2$). We verified in Section \ref{zeroTemp} that this is a gauge choice that leads to an on-shell action of the probe which is finite at the UV. Since in this region the modification of the background due to the temperature vanishes asymptotically, it is clear that $C_7$ for $T\not=0$ should also contain ${\cal K}$. Moreover, in the general case we should worry about the behavior at the horizon.  Let us explore the possibility to improve the behavior of the  worldvolume action at the horizon without modifying its regular character at the UV. In general, we will write $C_7$ as
\beq
C_7=e^{-\phi}\,{\cal K}\,+\,\delta C_7\,\,,
\label{C7-improved}
\eeq
where $\delta C_7$ is  a closed seven-form which must vanish in the SUSY (zero temperature) case. To determine the improving term $\delta C_7$  in (\ref{C7-improved}), we first study the pullback of ${\cal K}$ in the black hole case. The expression of ${\cal K}$ is the one written in (\ref{cal-K-explicit}), where the $e^a$ are the one-forms defined in (\ref{framebasis}) (notice that they do not contain the blackening factor).  Recall that the angular embedding of the D6-brane is characterized by the conditions $\hat\omega^1=\hat\omega^2=0$, which imply that $\hat e^5=0$. Along this submanifold, the pullbacks of the one-forms in  (\ref{framebasis}) are
\bear
&&\hat e^{\mu}\,=\,L\,r\,dx^{\mu}\,\,,\qquad (\mu=0,1,2)\,\,,
\qquad \hat e^{\,3}\,=\,{L\over r}\,dr\,\,,
\qquad\qquad\hat e^4\,=\,{L\over b}\,\sqrt{q}\,d\alpha\,\,,
\rc\rc
&& \hat e^5\,=\,0\,\,,\qquad\qquad
\hat e^6\,=\,{L\sqrt{q}\over b}\,\sin\alpha\,\sin\theta  d\theta\,\,,\qquad\qquad
\hat e^7\,=\,-{L\over b}\,\sqrt{q} \,\sin\alpha\cos\theta\,d\beta\,\,,\rc\rc
&& \hat e^8\,=\,{L\over b}\,d\theta\,\,,\qquad\qquad
\,\,\,\,\,\,\,\,\,\,
\hat e^9\,=\,{L\over b}\,\sin\theta\,(d\psi\,+\,\cos\alpha\,d\beta)\,\,,
\eear
where the hat over the forms denotes the restriction to the angular submanifold  defined by the conditions $\hat\omega^1=\hat\omega^2=0$.  Using these results we get immediately  that the pullback of ${\cal K}$ is given by
\beq
e^{-\phi}\,\hat {\cal K}\,=\,
{L^7 q\over b^3}\,e^{-\phi}\,d^3\,x\wedge\Big[\,
{r^3\over b}\,\sin\theta\,\cos\theta\,d\theta\,+\,
r^2\,\sin^2\theta dr\,\Big]\wedge
\Xi_3\,\,,
\label{pullback-SUSY-C7}
\eeq
with $\Xi_3 $ being the following  three-form:
\beq
\Xi_3\,=\,\sin\alpha\,d\alpha\wedge d\beta\wedge d\psi\,\,.
\label{Xi_3}
\eeq
Let us now represent  the improving term $\delta C_7$ in a way similar to the right-hand side of (\ref{pullback-SUSY-C7}), 
\beq
\delta C_7\,=\,
{L^7 q\over b^3}\,e^{-\phi}\,d^3x\wedge\Big[\,
L_1(\theta)\,d\theta\,+\,L_2(r)\,dr\,\Big]\wedge \Xi_3\,\,,
\eeq
with $L_1(\theta)$ and $L_2(r)$ being two functions to be determined.  Notice that $\delta C_7$ is closed when $L_1$ is only a function of $\theta$ and $L_2$ only depends on $r$. The pullback of the total $C_7$ takes the form
\beq
\hat C_7\,=\,
{L^7 q\over b^3}\,e^{-\phi}\,d^3x\wedge\Big[
\Big({r^3\over b}\,\sin\theta\,\cos\theta\,+\,L_1(\theta)\Big)d\theta\,+\,
\Big(r^2\,\sin^2\theta\,+\,L_2(r)\Big)dr\,\Big]\wedge \Xi_3\,\,.
\eeq
As argued in \cite{Jensen:2010vx} (see also \cite{Zafrir:2012yg}), a non-zero value of $C_7$ at the horizon introduces extra sources in the theory which change the boundary conditions of the fields and should be avoided. Accordingly, we impose the condition that the angular part of $\hat C_7$ (\ie, the one that does not contain $dr$)  vanishes at the horizon $r=r_h$. This  regularity  condition determines uniquely the function $L_1(\theta)$, 
\beq
L_1(\theta)\,=\,-{r_h^3\over b}\,\sin\theta\,\cos\theta\,\,.
\label{L1-sol}
\eeq
Notice that, for this value of $L_1(\theta)$, one can recast the $d\theta$ component of
$\hat C_7$ in terms of the blackening factor, 
\beq
{r^3\over b}\,\sin\theta\,\cos\theta\,+\,L_1(\theta)\,=\,{r^3\over b}\, h(r)\,
\sin\theta\,\cos\theta\,\,.
\eeq
It is important to point out  that this term always vanishes at the bottom of the brane which is either at $r=r_h$ (for black hole embeddings) or at $\theta=0$ (for Minkowski embeddings).  Thus, the pullback of $C_7$ to the submanifold with  $\hat\omega^1=\hat\omega^2=0$ is
\beq
\hat C_7\,=\,
{L^7 q\over b^3}\,e^{-\phi}\,d^3x\wedge\Big[
\,{r^3\over b}\, h(r)\,\sin\theta\,\cos\theta\,d\theta\,+\,
\Big(r^2\,\sin^2\theta\,+\,L_2(r)\Big)dr\,\Big]\wedge \Xi_3\,\,,
\eeq
and the WZ term of the action is given by
\beq
S_{WZ}\,=\,{\cal N}_r\,\int d^3x\,dr\,r^2\,\sin\theta\,
\Big(\sin\theta+{rh(r)\over b}\,\cos\theta\,\dot\theta\Big)\,+\,
{\cal N}_r\,\int d^3\,x\,dr\,L_2(r)\,\,.
\eeq
Let us now introduce a constant $\Delta_0$, defined as
\beq
\int dr\,L_2(r)\,\equiv\,r_h^3\,\Delta_0\,\,,
\eeq
where the factor $r_h^3$ has been introduced for convenience and the definite integral is over the whole range of the radial coordinate. 
Then,
\beq
S_{WZ}\,=\,{\cal N}_r\,\int d^3x\,dr\,r^2\,\sin\theta\,
\Big(\sin\theta+{rh(r)\over b}\,\cos\theta\,\dot\theta\Big)\,+\,
{\cal N}_r\,r_h^3\,\int d^3\,x\,\Delta_0\,\,.
\label{WZ-probe-improved}
\eeq
Clearly, as the constant $\Delta_0$ does not depend on the embedding, it is a counterterm that represents a zero-point energy.\footnote{However, its contribution to the free energy is not a thermodynamic constant since it is multiplied by $T^3$, as it is clear from the $r_h^3$ factor multiplying it in (\ref{WZ-probe-improved}) (see below).}
The total action is given by
\beq
S\,=\,-{\cal N}_r\,\int d^3x\,dr\,r^2\,\sin\theta
\Big[\,\sqrt{1+ {r^2\,h(r)\over b^2}\,\,\dot\theta^2}\,-\,\sin\theta\,-\,
{rh(r)\over b}\,\cos\theta\,\dot\theta\Big]\,+\,
{\cal N}_r\,r_h^3\,\int d^3\,x\,\Delta_0\,\,.
\label{S-Delta0}
\eeq
Notice that the canonical momentum for the improved action (\ref{S-Delta0}) vanishes at the horizon, 
\beq
{\partial {\cal L}\over \partial \dot \theta}\,\Big |_{r=r_h}\,=\,0\,\,.
\label{horizon-flow}
\eeq
This means that the IR contribution to on-shell quantities like the one in (\ref{partialS-Rstar}) will vanish for black hole embeddings that end on the horizon. This is related to the fact that, due to (\ref{horizon-flow}), there is no momentum flow through the horizon and thus the latter is not a dynamical surface. This property will be important in what follows.

Let us now fix the zero-point constant $\Delta_0$ in (\ref{S-Delta0}). With this purpose we will compute the free energy of the probe and compare this result with the free energy of the flavored background that was obtained in Section \ref{FlavABJM}. In general, the free energy $F$ is obtained from the Euclidean action $S_E$ by the relation $F=T\,S_E$. In the calculation of $S_E$ one has to integrate over the Euclidean time $\tau$ in the range $0\le \tau\le 1/T$ and over the non-compact two-dimensional space. The latter gives rise to an (infinite) two-dimensional volume $V_2$. In what follows we will divide all extensive thermodynamic quantities by $V_2$ and we deal with densities. In particular, the free energy density (which we will continue to denote by $F$) corresponding to the probe action (\ref{S-Delta0}) is
\beq
F\,=\,{\cal N}_r\,\int dr\,r^2\,\sin\theta
\Big[\,\sqrt{1+ {r^2\,h(r)\over b^2}\,\,\dot\theta^2}\,-\,\sin\theta\,-\,
{rh(r)\over b}\,\cos\theta\,\dot\theta\Big]\,-\,
{\cal N}_r\,r_h^3\,\Delta_0\,\,.
\label{F-Delta0}
\eeq
In the next subsection we will determine the constant $\Delta_0$ by considering the case in which the probe brane remains very far from the horizon. This case corresponds to having quarks with very large mass which should decouple and therefore should not contribute to the free energy. As we will soon demonstrate, the condition $F(m_q\to\infty)=0$ will determine a simple value for $\Delta_0$.

\subsection{Decoupling infinitely massive flavors}

To characterize the embeddings which correspond to flavors with infinite mass it is very convenient to work in a system with isotropic (Cartesian-like) coordinates. 
Let us proceed as in the zero temperature case and find a coordinate $u$ such that the $(r,\theta)$ part of the metric is written as in (\ref{isotropic-u-theta-metric}). It is immediate to conclude that, in this black hole case, the differential equation for $u(r)$ is
\beq
{b\,dr\over r\sqrt{h}}\,=\,{du\over u}\,\,,
\label{isotropic-DE-BH}
\eeq
which again can be integrated straightforwardly
\beq
u^{{3\over 2b}}\,=\,\Big(\,{r\over r_h}\,\Big)^{{3\over 2}}\,+\,\sqrt{
\Big(\,{r\over r_h}\,\Big)^{3}\,-\,1}\,\,.
\label{u-coordinate-def}
\eeq
Notice that the horizon $r=r_h$ corresponds  to $u=1$. The inverse relation is
\beq
\Big(\,{r\over r_h}\,\Big)^{{3\over 2}}\,=\,{1\over 2}\,\Big[\,u^{{3\over 2b}}\,+\,u^{-{3\over 2b}}\,\Big]\,=\,{1\over 2}\,u^{{3\over 2b}}\,
\tilde f(u)\,\,,
\eeq
where we  defined a new function $\tilde f(u)$,
\beq
\tilde f(u)\,\equiv\,1\,+\,u^{-{3\over b}}\,\,.
\label{tilde-f-def}
\eeq
Let us next define a function $f(u)$ as
\beq
 f(u)\,\equiv\,1\,-\,u^{-{3\over b}}\,\,.
 \label{f-def}
 \eeq
One can verify that the blackening factor can be written in terms of $f$ and $\tilde f$ as follows
\beq
\sqrt{h}\,=\,{f\over \tilde f}\,\,.
\eeq
Let us next write the ten-dimensional metric of the ABJM flavored black hole in terms of the isotropic coordinate $u$. We have
\beq
ds^2\,=\,{L^2\,r_h^2\over 2^{{4\over 3}}}\,u^{{2\over b}}\,\tilde f^{{4\over 3}}\,\,
\Big[-{f^2\over \tilde f^2}\,dt^2\,+\,(dx^1)^2\,+\,(dx^2)^2\,\Big]\,+\,
{L^2\over b^2}\,\,{du^2\over u^2}\,+\,ds^2_6\,\,,
\eeq
 where $ds^2_6$ is the squashed ${\mathbb C}{\mathbb P}^3$ metric written in (\ref{internal-metric-flavored}). 

Let us now define new coordinates $R$ and $\rho$ as in (\ref{R-rho-def}) and parameterize the embedding of the probe by a function $R=R(\rho)$. Following the same steps as above we can readily obtain the action of the probe and the corresponding free energy. The details of this calculation are given in Appendix \ref{Iso}. The total action for an arbitrary value of $\Delta_0$ is written in (\ref{L_total_R}). By studying the $\rho\to\infty$ limit of the equation of motion derived from (\ref{L_total_R}) it can be easily proven that the function $R(\rho)$ has the asymptotic behavior displayed in (\ref{R-UV}) and therefore the solutions are characterized by two constants $m$ and $c$, which are related to the quark mass and condensate, respectively. Moreover, from (\ref{L_total_R}) it is immediate to obtain the expression for the free energy density $F$.  To write this result 
it is quite useful to define a new quantity  ${\cal N}$ as
\beq
{\cal N}\,\equiv\,
{{\cal N}_r \over 4b}\,r_h^3\,=\,{2\pi^2\,r_h^3\,L^7\,q\over b^4}\,\,T_{D6}\,e^{-\phi}\,\,.
\label{calN_calN_r}
\eeq
In terms of gauge theory quantities ${\cal N}$ has the following expression
\beq
{\cal N}\,=\,{2\sqrt{2}\,\pi^2\over 27}\,N\,\,\sqrt{\lambda}\,\,{\zeta\over b}\,\,T^3\,\,,
\label{Nu-gauge}
\eeq
where $\zeta$ is the function of $N_f/k$ defined in (\ref{mu}). 
Then, the  free energy density for an embedding characterized by a function $R(\rho)$ is
\beq
F={\cal N}
\Big[\int d\rho
\rho\big[\rho^2+R^2\big]^{{3\over 2b}-1}f\tilde f
\Big[\sqrt{1+ R'^2}-
1+\Big({f\over \tilde  f}-1\Big){R\over \rho^2+R^2}(\rho R'-R)\,\Big]
-4\, b\, \Delta_0\,
\Big]
\,\,,
\label{F_R}
\eeq
where $R'=dR/d\rho$. This expression simplifies greatly when we take $R=R_0={\rm constant}$. In this case we have
\beq
{F(R=R_0)\over {\cal N}}\,=\,2\,R_0^2\,
\int_{0}^{\infty}\,d\rho\,{\rho\over (\rho^2+R_0^2)^2}\,\Big[\,
1\,-\,{1\over (\rho^2+R_0^2)^{{3\over 2b}}}\,\Big]\,-\,4b\,\Delta_0\,\,.
\label{F_R_0}
\eeq
The integral on the right-hand side of (\ref{F_R_0}) can be integrated straightforwardly. The result is
\beq
{F(R=R_0)\over {\cal N}}\,=\,1\,-\,{2b\over 2b+3}\,R_0^{-{3\over b}}\,-\,4b\,\Delta_0\,\,.
\label{F_R_0_integrated}
\eeq
By looking at the equations of motion of the probe in the $(R,\rho)$ variables it is easy to convince oneself that $R=R_0={\rm constant}$ is a solution only in the case for which $R_0\to\infty$, which corresponds to the case for which the quark mass parameter $m$ is very large. In this limit (\ref{F_R_0_integrated}) becomes
\beq
\lim_{R_0\to\infty}\,{F(R=R_0)\over {\cal N}}\,=\,
1\,-\,4b\,\Delta_0\,\,.
\label{FR_0_infinty}
\eeq
As argued above, infinitely massive flavors can be integrated out and therefore their contribution to the thermodynamic functions should vanish. Thus, on physical grounds one should choose $\Delta_0$ in such a way that the right-hand side of (\ref{FR_0_infinty}) vanishes, which means that $\Delta_0$ is simply given by
\beq
\Delta_0\,=\,{1\over 4b}\,\,.
\label{Delta_0_b}
\eeq

\subsection{A highly non-trivial test}

Let us show that the value of $\Delta_0$ written in (\ref{Delta_0_b})  is precisely the one required to satisfy a non-trivial compatibility condition between the free energy density of the probe and the one obtained  from the flavored geometry. With this aim let us determine $\Delta_0$ again by considering the case of zero mass embeddings (for which $\theta$ is constant and given by $\theta=\pi/2$). One can readily verify that this configuration solves the equations of motion derived from the action (\ref{S-Delta0}) and that the only contribution to the free energy (\ref{F-Delta0}) is precisely given by the zero-point term. Thus, in this case we have
\beq
F\approx -{\cal N}_r\,r_h^3\,\Delta_0\,=\, -\Big({4\pi\over 3} \Big)^3\,{\cal N}_r\,\Delta_0\,T^3\,\,.
\label{Fprobe-largeT}
\eeq
At this point it is interesting to remember that our background includes the backreaction of $N_f$ massless flavor branes. Actually, the free energy (\ref{free-energy-background}) contains the effects of $N_f$ flavor branes at non-linear order in $N_f$.  In the limit of small mass  the free energy of the probe should match the variation, at linear order,  of the free energy of the backreacted background  (\ref{free-energy-background}) when one flavor is added. Let us compute this variation at linear order by expanding the function $\xi\left(\frac{N_f}{k}\right)$ defined in (\ref{xi})  in a Taylor series and keeping only the first order,
\beq
\xi \Big({N_f+1\over k}\Big)\,=\,\xi \Big({N_f\over k}\Big)\,+\,
\xi' \Big({N_f\over k}\Big)\,{1\over k}\,+\,\cdots
\,\,,
\label{xi_Nf+1}
\eeq
where the prime denotes derivative of $\xi$ with respect to $N_f/ k$. 
Therefore, the variation of the free energy (\ref{free-energy-background}) of the background  (at linearized level) is
\beq
\Delta  F_{back}\,=\,-\left(\frac{4\pi}{3}\right)^2\,\frac{N^2}{9\sqrt{2\lambda}}
\,{1\over k}\,\xi' \Big({N_f\over k}\Big)\,
\,T^3\,\,.
\eeq
By equating $\Delta  F_{back}$ with the right-hand side of (\ref{Fprobe-largeT}) we find the following value of $\Delta_0$
\beq
\Delta_0\,=\,{1\over 12\sqrt{2}\,\pi}\,
{N^{{3\over 2}}\over  k^{{1\over 2}}}\,
{1\over {\cal N}_r}\,\,\xi' \Big({N_f\over k}\Big)\,\,.
\label{Delta0-Cal_Nr}
\eeq
To simplify this expression of $\Delta_0$, let us rewrite ${\cal N}_r$  as in (\ref{cal_Nr_mu}). Then, we can readily check that all the dependence on $N$ and $k$ drops and the expression for the zero-point constant $\Delta_0$ is greatly simplified. We arrive at
\beq
\Delta_0\,=\,{\xi'\over 3 \zeta}\,\,.
\label{Delta0-mu}
\eeq
Remarkably, by computing explicitly the derivative with respect to the deformation parameter $N_f/k=\epsilon$  one can find a simple expression of $\xi'$ in terms of $q$, $\eta$, and $b$ for arbitrary values of the deformation parameter. This expression is
\beq
\xi'\,=\,{3\over 8}\,\,{\sqrt{2-q}\over \sqrt{q}}\,\,
{(\eta+q)\,b^3\over (q+\eta q-\eta)^{{3\over 2}}}\,=\,
\begin{cases}{3\over 4}\,\,,&
\qquad{\rm for}\,\,\, {N_f\over k}\to 0\,\,,
\cr\cr
{255\over 512}\,\sqrt{{5\over 2}}\,\,
\sqrt{{k\over N_f}}\,\,,&
\qquad{\rm for}\,\,\,  {N_f\over k}\to \infty\,\,,
\end{cases}
\label{xiprime-eta-q-b}
\eeq
where the limiting cases  match with eqs. (7.9) and (7.11) of \cite{Conde:2011sw}, respectively. 
Amazingly, this value of $\xi'$ is simply related to the function $\zeta $ that encodes the flavor dependence of the prefactor of the probe free energy. Actually, by comparing the right-hand sides of (\ref{xiprime-eta-q-b}) and (\ref{mu}) one readily concludes that
\beq
\xi'\,=\,{3\over 4b}\, \zeta\,\,,
\label{xi_prime}
\eeq
which, after taking (\ref{Delta0-mu}) into account,  means that $\Delta_0$ is just given by
(\ref{Delta_0_b}), in remarkable agreement with our calculation in the opposite $m\to\infty$ limit.

 The result just found implies that the first variation of the free energy of the flavored back hole can be written as
\beq
\Delta \,F_{back}\,=\,-{{\cal N}_r \over 4b}\,r_h^3\,\,.
\eeq
Obviously, in terms of ${\cal N}$, the first flavor variation of the free energy of the background takes the form
\beq
\Delta \,F_{back}\,=\,-{\cal N}\,\,.
\eeq
It  follows that the limiting value of the free energy for massless embeddings is
\beq
F\approx -{\cal N}\ , \,\,\,\,
\qquad {\rm as}\,\,m\to 0\,\,.
\eeq
It is interesting to formulate the matching between the action of the  probe and background  in terms of the entropy density. In the $m\to 0$ limit  the entropy density of the probe is just:
\beq
s\,=\,-{\partial F\over \partial T}\approx {3\,{\cal N}\over T}\,\,,
\qquad\qquad (m\to 0)\,\,,
\eeq
which, after using (\ref{Nu-gauge}) and (\ref{xi_prime}), can be written as:
\beq
s\,\approx\,{1\over 3}\,\,
\left(\frac{4\pi}{3}\right)^2 \frac{N^2}{\sqrt{2\lambda}}\,
{1\over k} \,\xi'\left(\frac{N_f}{k}\right)
T^2\,\,,\qquad\qquad (m\to 0)\,\,.
\label{s_probe_m0}
\eeq
Let us now calculate the total entropy of the system, \ie, the sum of (\ref{s_probe_m0}) and  the background entropy (\ref{s_back}). By linearizing  the function $\xi$ as in (\ref{xi_Nf+1}), we can write
\beq
s_{total}\,=\,s_{back}+s\approx\,{1\over 3}\,\,
\left(\frac{4\pi}{3}\right)^2 \frac{N^2}{\sqrt{2\lambda}}\,
\xi\left(\frac{N_f+1}{k}\right)\,
T^2\,\,,\qquad\qquad (m\to 0)\,\,,
\eeq
which means that $s_{total}$ is equal to the entropy of the flavored black hole in which $N_f$ is increased by one unit. Therefore, the effect of adding a probe brane with $m\to 0$ is equivalent to the increase of the area of the horizon which is produced in the geometry when $N_f\to N_f+1$ and, thus,  the effect of the probe in this limit  is very nicely encoded in the geometry of the backreacted background. 

Notice that the dependence on $N_f$ of the entropy of the background is determined by
the volume of the squashed ${\mathbb C}{\mathbb P}^3$ manifold, while that of the massless probe is related to the volume of the squashed ${\mathbb R}{\mathbb P}^3$ cycle that it wraps.  Thus, the compatibility condition just checked means that the volume of the cycle is simply related to the derivative with respect to $N_f$ of the total volume of the internal manifold. Given the fact that these volumes depend non-linearly on $N_f$, this is a remarkable property of the background which we regard  as a highly non-trivial test of the consistency of our flavored geometry.

\subsection{Summary of the RR potential  and action}
To finish this section let us summarize the result of the previous discussion. We have found that the RR seven-form potential which satisfies the requirements imposed by  the holographic renormalization and regularity at the horizon of the flavor brane must have the form:
\beq
C_7\,=\,e^{-\phi}\,{\cal K}\,+\,{{\cal N}_r\over 8\pi^2\,T_{D6}}\,d^3x\,\wedge\,\Big[
l(r)\,dr\,-\,4\sin\theta\cos\theta\,d\theta \Big]\wedge \Xi_3\,\,,
\eeq
where ${\cal K}$ is the calibration form (\ref{cal-K-explicit}), ${\cal N}$ is written in (\ref{calN_calN_r}), $\Xi_3$ is the three-form (\ref{Xi_3}) and $l(r)$ is a function whose integral over $r$ must be one in order to decouple the infinitely massive flavors. 
If the embedding of the brane is parameterized by  a function $\theta(r)$, the total action 
of the probe is given by
\beq
S\,=\,{\cal N}_r\,\int\,d^3x\,\Bigg[1-{4b\over r_h^3}\,
\int dr\,r^2\,\sin\theta
\Big[\,\sqrt{1+ {r^2\,h(r)\over b^2}\,\,\dot\theta^2}\,-\,\sin\theta\,-\,
{rh(r)\over b}\,\cos\theta\,\dot\theta\Big]\Bigg]\,\,,
\eeq
while in terms of the $(R,\rho)$ variables becomes
\beq
S=-{\cal N}\int d^3x
\Big[\int d\rho
\rho\big[\rho^2+R^2\big]^{{3\over 2b}-1}f\tilde f
\Big[\sqrt{1+ R'^2}-
1+\Big({f\over \tilde  f}-1\Big){R\over \rho^2+R^2}(\rho R'-R)\,\Big]
-1\,
\Big]
\,\,.
\label{S_total_R}
\eeq

Once the action is completely fixed it is rather straightforward to study the different solutions of the equations of motion and their corresponding thermodynamical properties. This analysis will be carried out in the next sections.

\section{Minkowski and black hole embeddings}
\label{Min-BH}
 
The action (\ref{S_total_R}) is certainly more complicated than its zero temperature counterpart (\ref{calL-zeroT}). However, in the UV region of large $\rho$ the equation that determines $R(\rho)$ is still given by (\ref{eom-R-UV}) and therefore the embedding function $R(\rho)$ behaves asymptotically as in (\ref{R-UV}). The constants $m$ and $c$ are related, respectively, to the quark mass $m_q$ and to the quark condensate $\langle{\cal O}_m\rangle$. The detailed relation between $m$ and $m_q$ is worked out in Appendix \ref{m-and-VEV}, and is given by
\beq
m_q\,=\,{2^{{1\over 3}}\pi\over 3}\,\,\sqrt{2\lambda}\,\,\sigma\,T\,
m^{{1\over b}}\,\,,
\label{m-mq}
\eeq
where $\lambda=N/k$ is the 't Hooft coupling and $\sigma$ is the screening function defined in (\ref{screening-sigma}). Notice that, according to (\ref{m-mq}), taking $m\to 0$ ($m\to\infty$) for fixed $m_q$ is equivalent to sending $T\to\infty$ ($T\to 0$). Moreover, following the same steps as in the zero temperature case, we can relate the constant $c$ to the quark condensate $\langle{\cal O}_m\rangle$.  Indeed, it is proved in Appendix \ref{m-and-VEV} that this relation is
\beq
\langle{\cal O}_m\rangle\,=\,-{2^{{2\over 3}}\,\pi\over 9}\,\,{(3-2b)\,b\over q}\,\,
\sigma
\,\,N\,T^2\,c\,\,.
\label{O-T-c}
\eeq
\begin{figure}[ht]
\begin{center}
\includegraphics[width=0.4\textwidth]{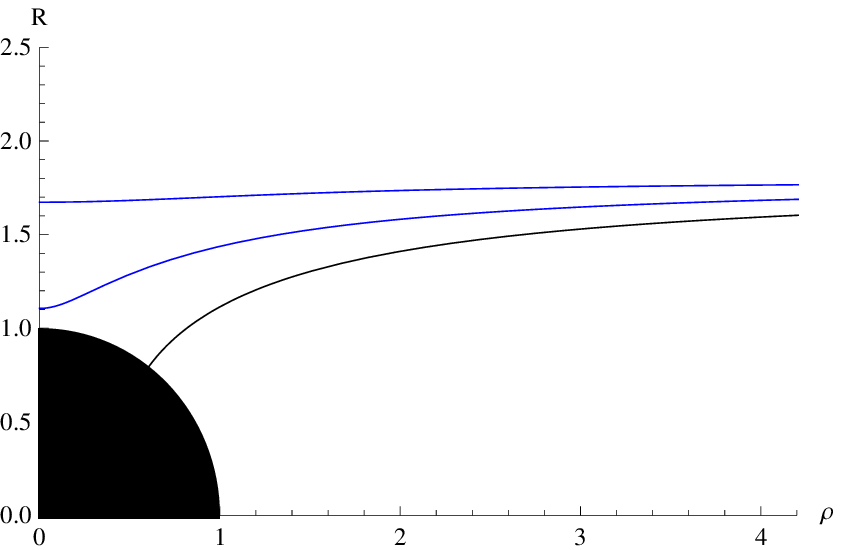}
\qquad\qquad
\includegraphics[width=0.4\textwidth]{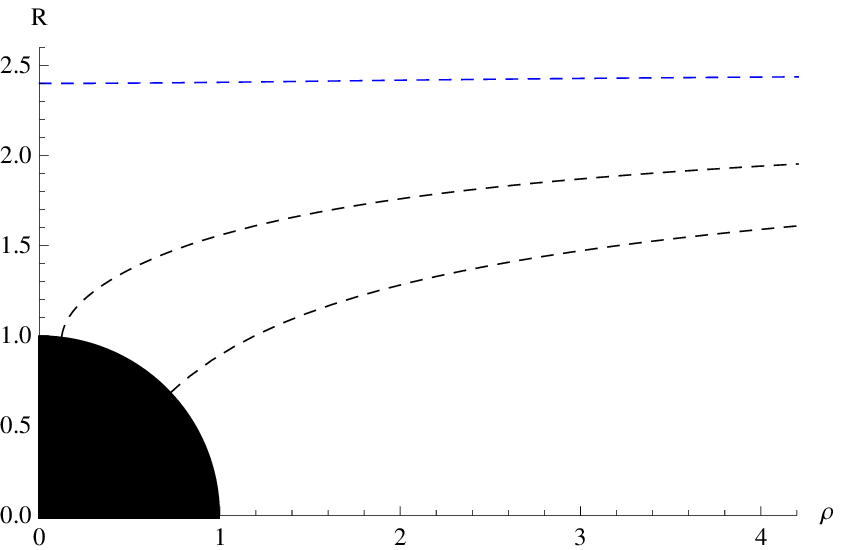}
\end{center}
\caption[embeddings]{Different embeddings in the $(R,\rho)$ plane for the unflavored background (left) with $m=1.8$ and for the flavored one with $\hat\epsilon=10$ (right) with $m=2.5$. \label{embeddings}} 
\end{figure}

\begin{figure}[ht]
\begin{center}
\includegraphics[width=0.4\textwidth]{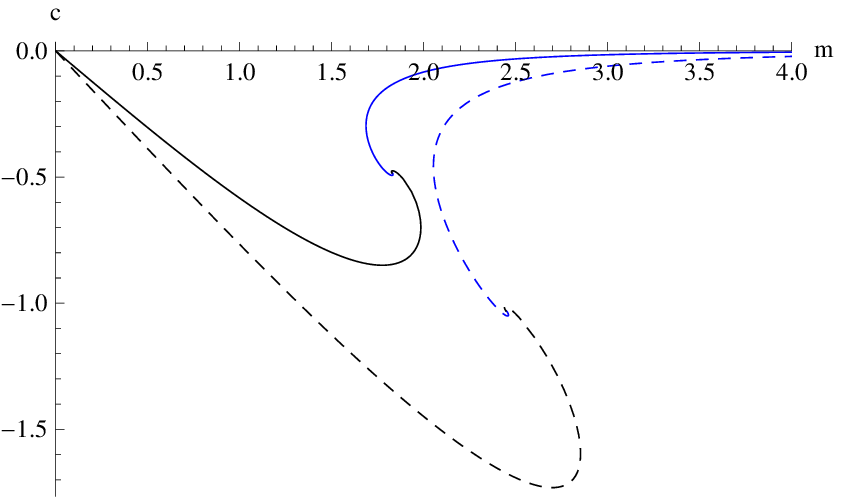}
\qquad
\includegraphics[width=0.4\textwidth]{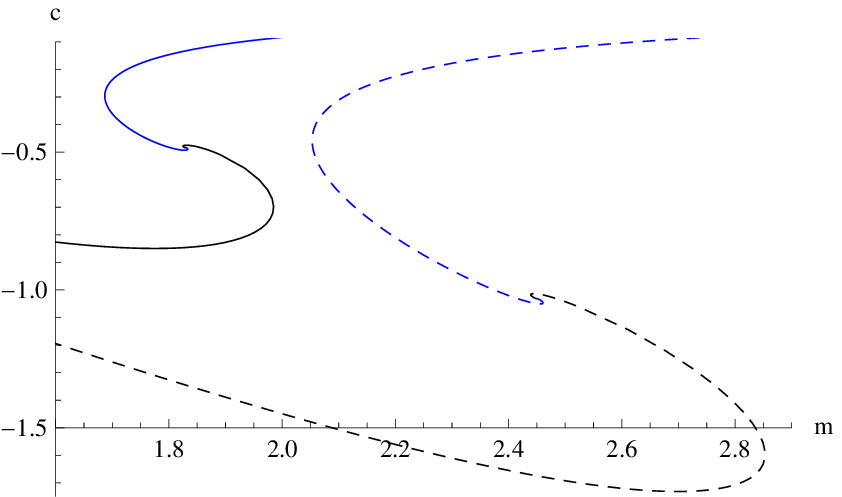}
\end{center}
\caption[c_vs_m]{On the left we plot $c$ versus $m$. The solid curve corresponds to the unflavored background while the dashed curve is for $\hat \epsilon=10$.  In both curves the black color stands for black hole embeddings, while the blue for Minkowski. On the right we present a zoom showing the spiraling behavior near the phase transition point.  \label{c_vs_m}} 
\end{figure}

At low temperature (or large mass parameter $m$) the probe brane closes off outside the horizon and one has a Minkowski embedding. In this case the brane reaches the point $\rho=0$ (or $\theta=0$) where the coordinate $R$ takes the value $R(\rho=0)=R_0$. One can  readily check that the only solutions of the equation of motion derived from the Lagrangian (\ref{L_total_R}) which are non-singular at the endpoint $\rho=0$ are those such that $R'=0$. 
By imposing these two initial conditions at $\rho=0$ one can integrate numerically the equation of motion and find the function $R(\rho)$. 
Some of these solutions for different values of $R_0$ are shown in Fig.~\ref{embeddings}. In general the value of $R_0$ determines the the asymptotic constants $m$ and $c$ and, by eliminating $R_0$, one can determine $c=c(m)$. For general values of $R_0$ this relation can only be found numerically (see Fig.~\ref{c_vs_m}). However, for large $R_0$ (or, equivalently, large $m$ or small $T$) one can establish  an approximate 
relationship. Indeed, it is shown in Appendix \ref{LowT} that
\beq
m\approx R_0\,+\,
{3\over 3+2b}\,
\Big[\,{2b\over 3-2b}\,+\,\psi \Big({3\over b}\Big)\,-\,
\psi\Big({3\over 2b}\Big)\,\Big]\
\,R_0^{1-{6\over b}}\,\,,
\qquad\,\,\,
(R_0, m\,\,\,{\rm large})\,\,,
\label{m-R0}
\eeq
where $\psi(x)\,=\,\Gamma'(x)/\Gamma(x)$ is the digamma function. Moreover, in this low $T$ regime one can also obtain the function $c(m)$ for large $m$, which is given by
\beq
c \approx {6b\over 4b^2-9}\,\,
{1\over m ^{1+{3\over b}} }\,\,,
\qquad\qquad m\gg 1
\,\,.
\label{c_m_lowT}
\eeq

When the temperature is large enough the probe brane ends at the horizon and we have a black hole embedding.  In this case it is more convenient to use  the isotropic coordinate $u$ as the holographic coordinate and to represent the profile of the brane in terms of the function $\chi(u)$, defined as
\beq
\chi(u)\,=\,\cos\theta(u)\,\,.
\eeq
The action in these variables has been obtained in Appendix \ref{Iso} (see eq. (\ref{total_L-chi})). 
The  corresponding equation of motion for $\chi(u)$ is
\begin{equation}
\partial_u \left[ {f} \, u^{\frac{3}{b}} \left( {f}\,  \chi \, + \, \frac{\tilde  f \, u \,  \dot\chi}{\sqrt{1-\chi^2+ u^2 \dot\chi^2}}\right) \right] \, 
- \,  {f} \, u^{\frac{3}{b}-1} \left( {f}\, u \, \dot \chi \, - \, \frac{\tilde f \,  \chi}{\sqrt{1-\chi^2+ u^2 \dot \chi^2}} \, + \, 2  \, \tilde f \,  \chi \right)\,=\,0\,\,,
\label{eom-chi(u)}
\end{equation}
where now the dot denotes differentiation with respect to $u$. From (\ref{eom-chi(u)}) we can infer the asymptotic behavior of $\chi(u\to\infty)$:
\begin{equation} 
\chi \,  =  \, \frac{m}{u} \, + \,  \frac{c}{u^{\frac{3}{b}-1}} \, + \, \cdots \,,
\label{asympD6} 
\end{equation} 
where $m$ and $c$ are the same constants as in (\ref{R-r-UV}).  Using the fact that 
$f(u=1)=0$ one can immediately show that the solutions of (\ref{eom-chi(u)}) which are non-singular at the horizon $u=1$ are those  which satisfy the conditions
\beq
\chi(u=1)\,=\,\chi_h\,\,,
\qquad\qquad
\dot \chi(u=1)\,=\,0\,\,.
\label{chi-bc}
\eeq
\begin{figure}[ht]
\begin{center}
\includegraphics[width=0.5\textwidth]{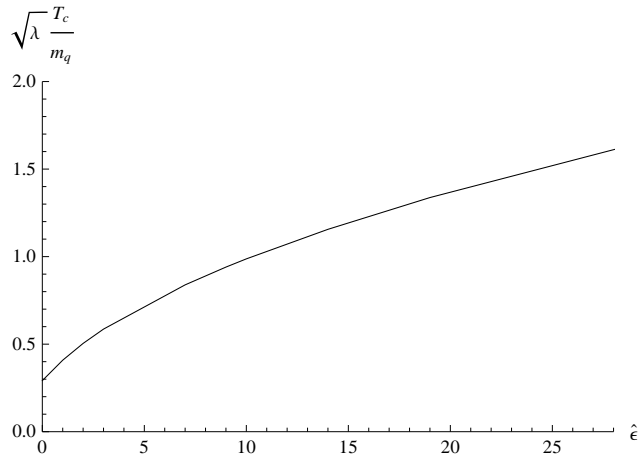}
\end{center}
\caption[xi]{Phase transition temperature  $T_c$ versus $\hat \epsilon$. \label{Tc}} 
\end{figure}
Some of the  numerical solutions of (\ref{eom-chi(u)})  with the initial conditions (\ref{chi-bc}) are shown in Fig.~\ref{embeddings}. In (\ref{chi-bc}) $\chi_h$ is an IR constant which determines the UV constants $m$ and $c$. As in the Minkowski embeddings, by eliminating  $\chi_h$ one gets $c=c(m)$, a relation which can only be obtained for all values of $m$ numerically. These results are plotted in Fig.~\ref{c_vs_m}. For high temperature (or low mass) $\chi$ remains small for all values of $u$ and one can linearize (\ref{eom-chi(u)}), which then can be solved analytically. This analysis is performed in detail in Appendix \ref{HighT}, where it is shown that, in this limit, $\chi_h$ is linearly related to $m$ by 
\beq
\chi_h\,\approx\,\sqrt{\pi}\,\,{\Gamma\big(1-{b\over 3}\big)\over 
\Gamma\big({1\over 2}-{b\over 3}\big)}\,\,m\,\,,
\qquad\qquad m\ll 1
\,\,.
\label{chi_0_m}
\eeq
Notice that the coefficient multiplying $m$ contains the dependence on the number of flavors. Similarly, one can find the function $c=c(m)$ for small $m$, which is given by the following analytic equation
\beq
c\,\approx\,-\,{\Gamma\big({1\over 2}+{b\over 3}\big)\, \Gamma\big(1-{b\over 3}\big)
\over
\Gamma\big({b\over 3}\big)\,\Gamma\big({3\over 2}-{b\over 3}\big)}\,m
\,\,,
\qquad\qquad m\ll 1
\,\,,
\label{c_m_highT}
\eeq
which implies that $c$ vanishes linearly as $m\to 0$ with a slope that depends on the deformation parameter $N_f/k$.

The temperature $T_c$ of the first order phase transition grows with the number of flavors as shown in Fig.~\ref{Tc}. This temperature is determined as the point where the curves of the  free energies of the black hole and Minkowski embeddings intercept each other.  It is important to point out that the value of $1/m$ where  the Minkowski-black hole transition  takes place does not change much with $N_f$. Indeed, it (monotonically) decreases from being $\approx 0.544$ at $N_f = 0$ down to $\approx 0.400$ as $N_f \to \infty$. However, from (\ref{m-mq}) we have ${T\over m_q}\,\sqrt{\lambda}\propto m^{-{1\over b}}\,\sigma^{-1}$, with a proportionality constant which does not depend on the number of flavors. This means that the flavor dependence of $T_c$ is dominated by the function $\sigma^{-1}$ which, for large $N_f$,  grows with the deformation parameter as $\sqrt{\hat \epsilon}$. This is precisely the behavior displayed in Fig.~\ref{Tc}.

The black hole and Minkowski embeddings are separated by a critical solution in which the brane probe just touches the horizon. This critical solution occurs for certain values $m=m_*$ and $c=c_*$ of the mass and condensate parameters. The detailed analysis of these critical embeddings is performed in Appendix \ref{critical}, where it is shown that they can be approximately represented near the horizon as $R(\rho)\approx 1+\rho$. The solutions near the critical embedding display a discrete self-similarity behavior, as it corresponds to a first order phase transition. Indeed, as shown in Appendix \ref{critical}, the mass and condensate parameters exhibit an oscillatory behavior around their critical values and, as a consequence, the quark condensate is not a single-valued function of the mass. This last fact is clearly visible in the plots of Fig.~\ref{c_vs_m}.

\section{Brane thermodynamics}
\label{Thermo}
\begin{figure}[ht]
\begin{center}
\includegraphics[width=0.45\textwidth]{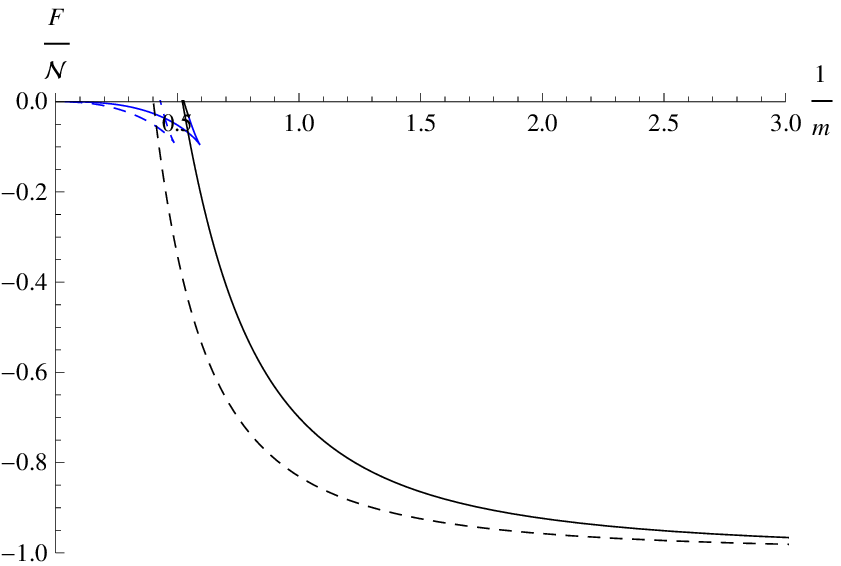}
\qquad
\includegraphics[width=0.45\textwidth]{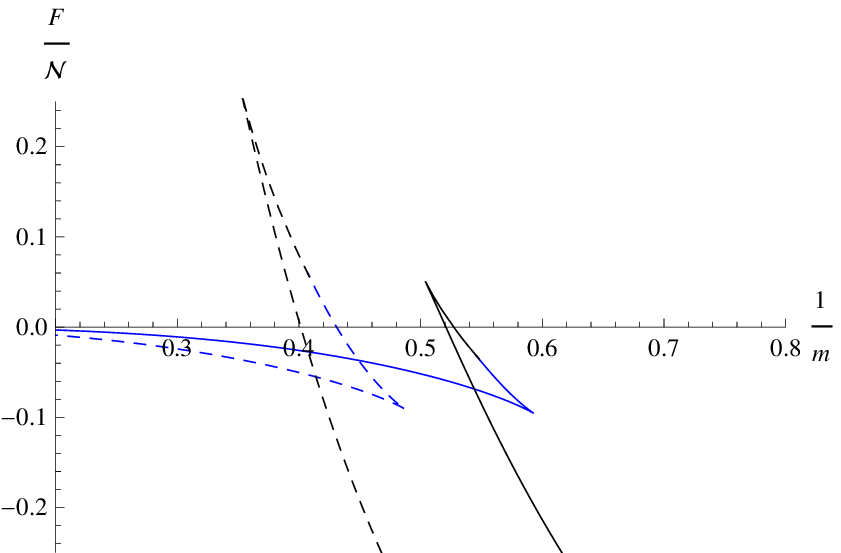}
\end{center}
\caption[F]{We plot the free energy $F/{\cal N}$ versus $1/m$ for black hole (black curves) and Minkowski (blue curves) embeddings. The solid (dashed) curves are for the unflavored (flavored with $\hat\epsilon=10$) background. On the right an amplification of the phase transition region is shown.
\label{F}} 
\end{figure}

In this section we address the main objective of this paper, the calculation of the different thermodynamic functions of the brane probe. The first of these quantities is the free energy density $F$, which can be obtained as in (\ref{F-Delta0}) from the Euclidean on-shell  action of the probe. Actually, the expression of $F$ can be easily related to the integrals of the Lagrangian density ${\cal L}$ of (\ref{L_total_R}) and (\ref{total_L-chi}). Indeed,  let $V_3$ be the value of the volume of three-dimensional Minkowski space and let us represent the on-shell Minkowski action of the brane in terms of a function ${\cal G}(m)$ by 
\beq
{1\over V_3}\,\,{S^{on-shell}\over {\cal N}}\,=\,1\,-\,{\cal G}(m)\,\,.
\label{S-calG}
\eeq
Then, the free energy  density $F$ is given by
\beq
{F\over {\cal N}}\,=\,{\cal G}(m)\,-\,1 \ .
\label{F-calG}
\eeq
The explicit expression for the function ${\cal G}(m)$  can be straightforwardly obtained from the results of Section \ref{Min-BH}. For Minkowski embeddings parameterized by a function  $R(\rho)$, it is given by
\beq
{\cal G}(m)\,\equiv\,
\int_{0}^{\infty}d\rho\,
\rho\,\big[\,\rho^2\,+\,R^2\,\big]^{{3\over 2b}-1}f\,\tilde f\,
\Big[\sqrt{1+ R'^2}\,-
1\,+\,\Big({f\over \tilde f}-1\Big)\,{R\over \rho^2+R^2}\,(\rho R'-R)\,\Big]\,\,,
\label{calG-R}
\eeq
while  for black hole embeddings  it is more convenient to represent ${\cal G}(m)$  as
\beq
{\cal G}(m)\,\equiv\,
\int_{1}^{\infty} du\,
 f   {\tilde f}  u^{\frac{3}{b}-1} 
\Big[\sqrt{1-\chi^2+ u^2 \dot\chi^2}   - 1   +  \chi^2  +   u\,  \frac{f}{\tilde f}\,  \chi \, \dot\chi\, \Big]\,\,.
\label{calG-chi}
\eeq
In (\ref{calG-R}) and (\ref{calG-chi}) it is understood that $R(\rho)$ and $\chi(u)$ are the result of integrating the equations of motion with the regular boundary conditions at the IR which correspond to the UV parameter $m$. In Fig.~\ref{F} we plot $F$ as a function of $m^{-1}$ for both types of embeddings. Notice that the curves for Minkowski and black hole embeddings cross and show the typical ``swallow tail" form, which is characteristic of first order phase transitions. It is important to point out that the improvement term (\ref{L1-sol}) that regularizes the behavior of $C_7$ at the horizon  is essential in obtaining this behavior. 

Let us now compute the entropy density $s$. We start from the definition of  $s$ as a derivative of the free energy $F$ with respect to the temperature, which we organize as follows:
\beq
s\,=\,-{\partial F\over \partial T}\,=\,-{\cal N}\,
{\partial \over \partial T} \Big({F\over {\cal N}}\Big)\,-\,
{F\over {\cal N}}\,
{\partial {\cal N} \over \partial T}\,\,.
\label{s-definition}
\eeq
Taking into account that ${\cal N}\sim T^3$, we have
\beq
{\partial {\cal N} \over \partial T} \,=\,{3\over T}\,\, {\cal N}\,\,,
\eeq
and therefore we can write (\ref{s-definition}) as
\beq
s\,=\,-{3F\over T}\,-\,{\cal N}\,\,{\partial \over \partial T} \Big({F\over {\cal N}}\Big)\,\,.
\label{s-derivatives}
\eeq
Let us now use the fact that  for fixed quark mass $m_q$ the parameter $m$ behaves as  $m\propto T^{-b}$ (see (\ref{m-mq})) and thus ${\partial m\over \partial T}\,=\,-b\,{m\over T}$.
Using the chain rule, the derivative appearing on the second term in (\ref{s-derivatives}) becomes
\beq
{\partial \over \partial T} \Big({F\over {\cal N}}\Big)\,=\,-b\,{m\over T}\,\,
{\partial \over \partial m} \Big({F\over {\cal N}}\Big)\,\,.
\label{T-derivative-F/N}
\eeq
The derivative with respect to $m$ appearing on the right-hand side of (\ref{T-derivative-F/N}) has been computed in Appendix \ref{m-and-VEV} (eq. (\ref{m-derivative-F/N})). 
By using this result, we can write
\beq
{\partial \over \partial T} \Big({F\over {\cal N}}\Big)\,=\,\,{m\over T}\,\,
(3-2b)\,c\,\,.
\eeq
Plugging the value of this derivative in   (\ref{s-derivatives}), we arrive at the following expression for the entropy $s$:
\beq
{s\over {\cal N}}\,=\,-{3\over T}\,{F\over {\cal N}}\,-\,
{m\over T}\,(3-2b)\,c\,\,.
\label{s_F_c}
\eeq
The first term on the right-hand side of (\ref{s_F_c}) is the one expected in a system with conformal invariance in three dimensions for which $F\propto T^3$. The term in  (\ref{s_F_c}) containing $m$ and $c$ represent the deviation from this conformal behavior due to the massive quarks introduced by the probe. Notice that it depends on the number $N_f$ of massless quarks of the background.

\begin{figure}[ht]
\begin{center}
\includegraphics[width=0.5\textwidth]{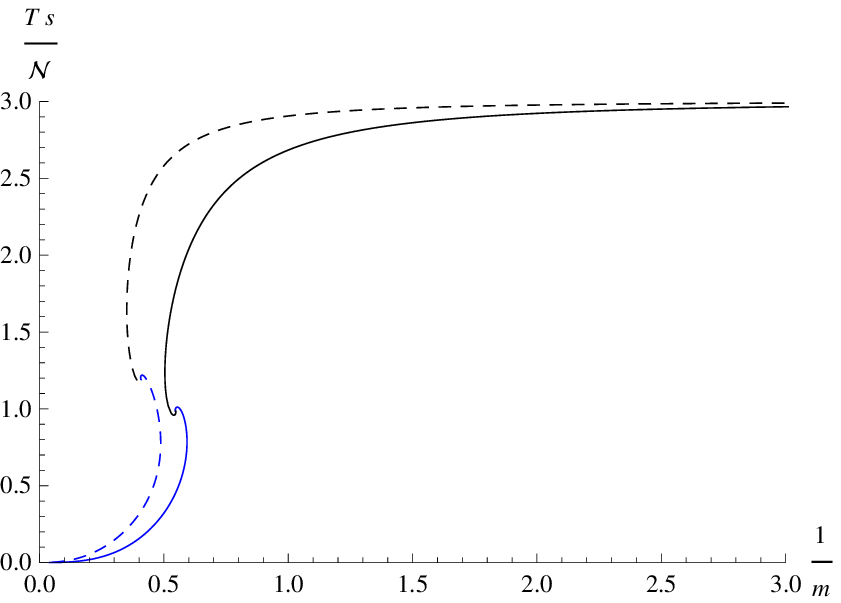}
\end{center}
\caption[S]{We plot the entropy $T s/{\cal N}$ versus $1/m$ for black hole (black curves) and Minkowski (blue curves) embeddings. 
The solid (dashed) curves are for the unflavored (flavored with $\hat\epsilon=10$) background.
 \label{S}} 
\end{figure}

By using (\ref{F-calG}) we can write $s$  in terms of the function ${\cal G}(m)$,
\beq
T\,{s\over {\cal N}}\,=\,-3\,{\cal G}(m)\,+\,3\,-\,
(3-2b)\,c\,m\,\,.
\label{s_calG_c}
\eeq
In Fig.~\ref{S} we have plotted the numerical values of the  entropy  as a function of $m^{-1}$. We notice that $s$ is always positive. As with the free energy, the regularization of $C_7$ at $r=r_h$ is essential to avoid having a pathological thermodynamic behavior for which $s<0$ for some values of $m$.

\begin{figure}[ht]
\begin{center}
\includegraphics[width=0.5\textwidth]{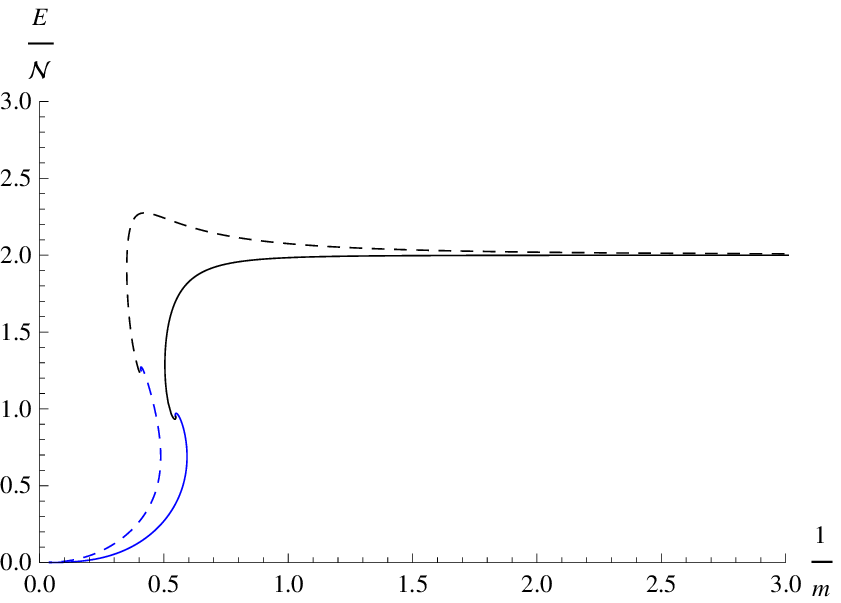}
\end{center}
\caption[E]{We plot the internal energy $E/{\cal N}$ versus $1/m$ for black hole (black curves) and Minkowski (blue curves) embeddings. 
The solid (dashed) curves are for the unflavored (flavored with $\hat\epsilon=10$) background.
\label{E}} 
\end{figure}

We can also compute the internal energy $E$ by means of the thermodynamic relation  $E=F+T\,s$. Indeed, from (\ref{s_F_c}) we get
\beq
E\,=\,-2F\,-\,\,  {\cal N}\,
(3-2b)\,c\,m\,\,.
\label{E_F_c}
\eeq
In terms of ${\cal G}(m)$, this expression can be rewritten as
\beq
{E\over {\cal N}}\,=\,-2\,{\cal G}(m)\,+\,2\,-\,
 \,(3-2b)\,c\,m\,\,.
 \label{E_calG_c}
\eeq
In Fig.~\ref{E} we plot $E$ for different values of $m^{-1}$. 

We can now use the previous expressions  and the numerical results to find the limiting values for the free energy, entropy and internal energy when $m$ is  small (or $T$ is very large). Indeed, since 
${\cal G}(m)\to 0$ as $m\to 0$, it follows that
\beq
\lim_{m\to 0}\,\,{F\over {\cal N}}\,=\,-1\,\,,
\qquad\qquad
\lim_{m\to 0}\,\,T\,{s\over {\cal N}}\,=\,3\,\,,
\qquad\qquad
\lim_{m\to 0}\,\,{E\over {\cal N}}\,=\,2\,\,,
\label{m_to_0_limit}
\eeq
which are just the values expected in this conformal limit.  Moreover, in the opposite  regime $m\to \infty$ (or $T\to 0$) one has
${\cal G}(m)\to 1$  and $c\,m\sim m^{-3/b}\to 0$. Thus,
\beq
\lim_{m\to \infty}\,\,\,{F\over {\cal N}}\,=\,
\lim_{m\to \infty}\,\,T\,{s\over {\cal N}}\,=\,
\lim_{m\to \infty}\,\,{E\over {\cal N}}\,=\,0\,\,.
\label{m_to_infty_limit}
\eeq
In the next two subsections we will refine the limits (\ref{m_to_0_limit}) and (\ref{m_to_infty_limit}) by using the results of Appendices \ref{LowT} and \ref{HighT}. 

The heat capacity density $c_v$  of the probe is defined as
\beq
c_v\,=\,{\partial E\over \partial T}\,\,.
\label{cv_def}
\eeq
By computing the derivative of $E$ as given in (\ref{E_calG_c}), one arrives at the following expression of $c_v$:
\beq
T\,{c_v\over {\cal N}}\,=\,
2\,T\,{s\over {\cal N}}\,-\,(3-2b)\,
\Big[\,3\,-\,b\,-\,b\,{\partial ( \log c)\over \partial (\log m)}\,\Big]\,c\,m\,\,.
\label{cv_general}
\eeq
We have checked numerically that $c_v$ is positive for all values of $m$  and has a finite jump discontinuity at the phase transition point. 

\subsection{Low temperature functions}
Let us now evaluate $F$, $s$, and $E$ when $T\to 0$ (or $m\to\infty$).  The on-shell action of the probe in this limit has been computed in Appendix \ref{LowT}. From this result we find that ${\cal G}(m)$ behaves as
\beq
{\cal G}(m)\,\approx\,1\,-\,{2b\over 2b+3}\,{1\over m^{{3\over b}}}\,\,.
\label{G-large_m}
\eeq
Moreover, the approximate value of $c(m)$ in this $T\to 0$ regime has been written in
(\ref{c_m_lowT}). By using this result in (\ref{F-calG}), (\ref{s_calG_c}), and (\ref{E_calG_c}), we get
\beq
{F\over {\cal N}}\,\approx\,-{2b\over 3+2b}\,\,\Big({T\over \bar M}\Big)^{3}\,\,,
\qquad\qquad
T\,{s\over {\cal N}}\, \approx\,{12b\over 2b+3}\,\Big({T\over \bar M}\Big)^{3}\,\,,
\qquad\qquad
{E\over {\cal N}}\, \approx\,{10b\over 2b+3}\,\Big({T\over \bar M}\Big)^{3}\,\,,
\label{F_s_E_lowT}
\eeq
where $\bar M$ is the constant defined in (\ref{barM}). Taking into account that ${\cal N}\sim T^3$, we find that $F$ vanishes as $T^6$ when $T\to 0$ with a coefficient which depends on the number of flavors. As a check of (\ref{F_s_E_lowT}), one can immediately verify that the coefficients of $F$ and $s$  in (\ref{F_s_E_lowT}) are such that the thermodynamic relation $s=-\partial F/\partial T$ is indeed satisfied. Furthermore, one can verify from (\ref{F_s_E_lowT}) (or directly from the general expression (\ref{cv_general})) that the specific heat $c_v$ vanishes at low temperatures as $T^5$,
\beq
T\,{c_v\over {\cal N}}\,\approx\,
{60b\over 2b+3}\,\Big({T\over \bar M}\Big)^{3}\,\,.
\eeq

\subsection{High temperature functions}
It follows from the results of Appendix \ref{HighT} that ${\cal G}(m)$ vanishes, when $m\to 0$, as
\beq
{\cal G}(m)\approx
-{3-2b\over 2b}\,\,c\,m\,\,.
\label{G_small_m}
\eeq
Moreover, the condensate parameter $c$ for small $m$ can be estimated as in (\ref{c_m_highT}). From these results we can show that $F$ can be approximated as
\beq
{F\over {\cal N}}\approx -1\,+\,{3\over b}\,\Bigg[{
\Gamma\big(1-{b\over 3}\big)\over
\Gamma\big({1\over 2}-{b\over 3}\big)}\Bigg]^2\,
\tan\big({\pi b\over 3}\big)\,
\Big({\bar M\over T}\Big)^{2b} \ ,
\label{F_highT}
\eeq
from which it follows that the deviation of $F$ from its conformal value decays as $T^{-2b}$ when $T\to\infty$.  Notice that, in this case, both the power of  the temperature  and the coefficient of this non-conformal contribution  depend on the number of flavors. 

By combining (\ref{s_calG_c}) and (\ref{G_small_m}) we can approximate  the entropy in this limit as
\beq
T\,{s\over {\cal N}}\, \approx\,3\,-\,(3-2b)\,{\cal G}(m) \ ,
\eeq
which can also be written as
\beq
T\,{s\over {\cal N}}\approx 3\,-\,
{3(3-2b)\over b}\,\Bigg[{
\Gamma\big(1-{b\over 3}\big)\over
\Gamma\big({1\over 2}-{b\over 3}\big)}\Bigg]^2\,
\tan\big({\pi b\over 3}\big)\,
\Big({\bar M\over T}\Big)^{2b} \ .
\label{s_highT}
\eeq
As a check  of (\ref{s_highT}) one can verify that $s=-\partial F/\partial T$. Moreover,  from (\ref{F_highT}) and (\ref{s_highT}) we arrive at the following high temperature expression for  the internal energy:
\beq
{E\over {\cal N}}\,\approx\,2\,+\,
{6(b-1)\over b}\,
\Bigg[{
\Gamma\big(1-{b\over 3}\big)\over
\Gamma\big({1\over 2}-{b\over 3}\big)}\Bigg]^2\,
\tan\big({\pi b\over 3}\big)\,
\Big({\bar M\over T}\Big)^{2b} \ .
\label{E_highT}
\eeq
Curiously, the $T^{-2b}$ subleading term in (\ref{E_highT}) vanishes in the unflavored case $b=1$. Finally, from (\ref{E_highT}) we can readily obtain the behavior of  the specific heat $c_v$ for large $T$,
\beq
T\,{c_v\over {\cal N}}\,\approx\,6\,\,\Big[\,1\,+\,
{(b-1)(3-2b)\over b}\,
\Bigg[{
\Gamma\big(1-{b\over 3}\big)\over
\Gamma\big({1\over 2}-{b\over 3}\big)}\Bigg]^2\,
\tan\big({\pi b\over 3}\big)\,
\Big({\bar M\over T}\Big)^{2b}\Big]\,\,.
\eeq

\subsection{Speed of sound}
\begin{figure}[ht]
\begin{center}
\includegraphics[width=0.7\textwidth]{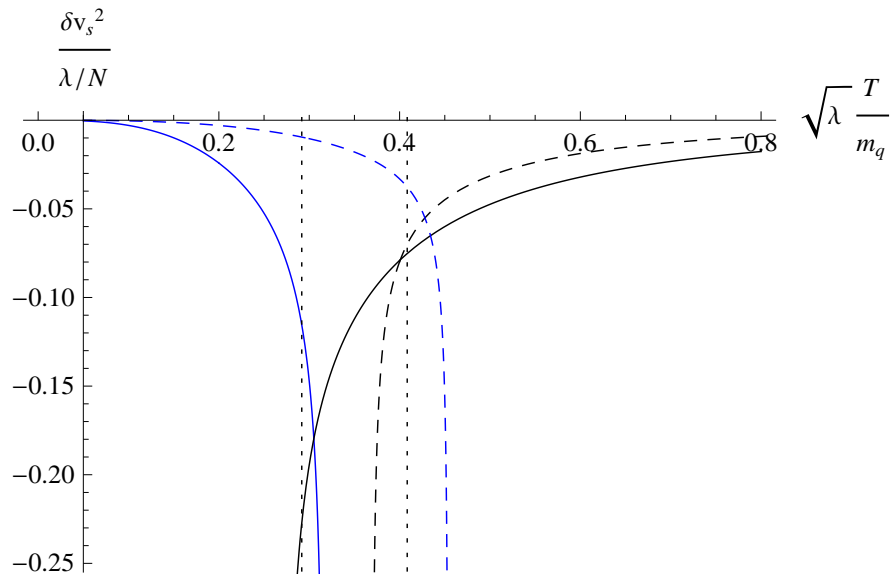}
\end{center}
\caption[xi]{We plot the speed of sound versus temperature for black hole (black curves) and Minkowski (blue curves) embeddings. 
The solid (dashed) curves are for the unflavored (flavored with $\hat\epsilon=1$) background. The dotted vertical lines correspond to the locations
of the first order phase transition.
 \label{vs}} 
\end{figure}

The speed of sound of a thermodynamic system can be obtained from the other thermal quantities by the relation
\beq
v_s^2 = \frac{\partial P}{\partial E} =- \frac{\partial F}{\partial T}
\left(\frac{\partial E}{\partial T}\right)^{-1} = \frac{s}{c_v}\,\,.
\label{vs-def}
\eeq
For a conformal system in 2+1 dimensions, as our flavored background, the formula (\ref{vs-def}) yields  $v_s^2 = 1/2$.  In this section we analyze the effect of the massive flavors introduced by the probe in the deviation from this conformal value. With this purpose we will apply (\ref{vs-def}) to the background plus probe system, \ie, we will substitute in (\ref{vs-def}) $s$ and $c_v$ by $s_{back}+ s$ and $c_{v,back}+c_{v}$, respectively, where $s$ and $c_v$ denote the entropy density and specific heat of the probe and (calculated in (\ref{s_calG_c}) and (\ref{cv_general})) and $s_{back}$ has been written in (\ref{s_back}). Hence, we get
\beq
v_s^2 = \frac{s_{back}+ s}{c_{v,back}+c_{v}}\,\,.
\eeq
The specific heat of the background is related to its entropy as $c_{v,back} = 2 s_{back}$. Moreover, in the probe approximation the D6-branes produce a small deviation from the conformal behavior. By expanding at first order, we arrive at the following result
\beq
v_s^2 \approx{1\over 2}\,
\Big[\,1\,+\,{3-2b\over 2\,s_{back}}\,
{\partial \over \partial T}\,\Big({\cal N}\,c\,m\,\Big)\,\Big]\,\,.
\label{delta_vs_first}
\eeq
Taking into account that ${\cal N}\propto T^3$  and that $m\propto T^{-b}$, we can cast 
(\ref{delta_vs_first}) as
\beq
\delta v_s^2\,\equiv\,v_s^2\,-\,{1\over 2}\,\approx\,
{3-2b\over 4}\,{{\cal N}\over T\,s_{back}}\,\,
\Big[\,3\,-\,b\,-\,b\,{\partial ( \log c)\over \partial (\log m)}\,\Big]\,c\,m\,\,.
\label{delta_vs_first_explicit}
\eeq
Moreover, from (\ref{s_back}) and (\ref{Nu-gauge}), one can verify that he ratio ${\cal N}/T \,s_{back}$ can be put as
\beq
{{\cal N}\over T\,s_{back}}\,=\,{\lambda\over 4 N b}\,{\zeta\over \xi}\,=\,
{1\over 4}\,{\lambda \over N}\,{q\over b^4}\,\sigma^2\,\,,
\label{N/Ts}
\eeq
where, in the last step, we have used (\ref{sigma-mu}) to write the result in terms of the screening function $\sigma$ defined in (\ref{screening-sigma}). Plugging (\ref{N/Ts}) into (\ref{delta_vs_first_explicit}), we arrive at the following expression for the deviation $\delta v_s^2$,
\beq
\delta v_s^2\,\approx\,{\lambda\over  N}\,
{q\,(3-2b)\,\sigma^2\over 16 \,b^4}\,\,
\Big[\,3\,-\,b\,-\,b\,{\partial ( \log c)\over \partial (\log m)}\,\Big]\,c\,m\,\,.
\label{delta_vs_general}
\eeq
We plot in Fig.~\ref{vs} the result of the numerical evaluation of $\delta v_s^2$ as a function of the temperature for different values of the flavor deformation parameter $\hat \epsilon$.   We see that in all cases $\delta v_s^2$ is negative, which implies that the massive flavors reduce the speed of sound.  This effect is larger as we approach the temperature where the phase transition takes place. Generically, $\delta v_s^2$ decreases as the number of massless flavors (and thus of the deformation parameter $\hat \epsilon$) is increased.  This is simply a consequence of the fact that we are considering only one D6-brane probe and therefore its effect is more and more diluted as $N_f\to\infty$. In order to have a better understanding of this behavior let us estimate $\delta v_s^2$ in the low and high temperature regimes.  At low $T$ we can use (\ref{c_m_lowT}) to compute the right-hand side of (\ref{delta_vs_general}). We get
\beq
\delta v_s^2\,\approx\,-{9\over 4}\,{\lambda\over N}\,
{q\,\sigma^2\over (2b+3)\,b^3}\,\,\Big({T\over \bar M}\Big)^3\,\,,
\qquad\qquad (T\to 0)
\,\,.
\label{delta_lowT}
\eeq
Thus, we find that the temperature dependence of the deviation from conformality  at low $T$ (\ie, $\delta v_s^2\sim T^3$) does not depend on the number of massless flavors. However, the coefficient multiplying $T^3$ in (\ref{delta_lowT}) does depend on $\hat \epsilon$ and approaches zero as $\hat\epsilon$ becomes large. To illustrate this fact let us evaluate the leading term on the right-hand-side of (\ref{delta_lowT})  when $\hat\epsilon\to\infty$ for fixed 't Hooft coupling. We get
\beq
\delta v_s^2\,\sim\,-
{1\over N}\,{1\over \hat \epsilon^{{5\over 2}}}\,
\Big({T\over m_q}\Big)^3\,\,,
\qquad\qquad (T\to 0,\,\, \hat \epsilon\to\infty)
\,\,.
\label{delta_lowT_epsilon}
\eeq
Similarly,  for large $T$ we can use (\ref{c_m_highT}) to evaluate (\ref{delta_vs_general}), 
\beq
\delta v_s^2\,\approx\,-{3\over 8}\,{\lambda\over N}\,\,
{q\,(3-2b)\,\sigma^2\over b^4}\,\,
\Bigg[{
\Gamma\big(1-{b\over 3}\big)\over
\Gamma\big({1\over 2}-{b\over 3}\big)}\Bigg]^2\,
\tan\big({\pi b\over 3}\big)\,
\Big({\bar M\over T}\Big)^{2b}\,\,,
\qquad\qquad (T\to \infty)
\,\,.
\label{delta_largeT}
\eeq
Therefore, $\delta v_s^2$ vanishes for $T\to\infty$ as a power law that depends on the parameter $b$ ($\delta v_s^2\sim T^{-2b}$). In this case the addition of massless flavor produces a faster decrease of  $\delta v_s^2$ with the temperature. However, the coefficient of this power law increases with $\hat\epsilon$. Actually, one can easily verify  from  (\ref{delta_largeT}) that for large $\hat\epsilon$ and $T$, $\delta v_s^2$  behaves as
\beq
\delta v_s^2\,\sim\,-{1\over N}\,\hat\epsilon^{\,{1\over 4}}\,
\Big({m_q\over T}\Big)^{{5\over 2}}
\,\,,
\qquad\qquad (T\to \infty,\,\, \hat \epsilon\to\infty)
\,\,.
\label{delta_largeT_epsilon}
\eeq

\section{Summary and conclusions}
\label{conclusions}
In this paper we studied the thermodynamics of flavor D6-branes in 
the gravity dual of Chern-Simons matter theory in three dimensions. The background geometry is a black hole of type IIA supergravity with delocalized sources which includes the backreaction due to massless flavors. The corresponding metric and forms are just the straightforward  $T\not= 0$ generalization of the $AdS_4\times {\cal M}_6$  solution  found in  \cite{Conde:2011sw}, in which the deformation due to the massless flavors is encoded in the constant squashing factors of the different pieces of the metric.  We added to this background an additional D6-brane probe, representing a massive flavor, and determined its holographically renormalized action, which passed several non-trivial tests. We then studied the thermodynamics of this probe in the flavored black hole geometry.

At low temperature the probe brane does not intercept the horizon of the black hole and we have a Minkowski embedding while, on the contrary, at high temperatures the brane falls into the horizon. At some intermediate temperature the system undergoes a first order phase transition which can be interpreted as a meson melting transition. We studied the different thermodynamic functions for both types of embeddings, as well as the corresponding phase transition. All the results depend both on the temperature and on the flavor deformation parameter  $\hat \epsilon\propto N_f/k$. The dependence on the latter is encoded in the different functions ($b$, $q$, $\sigma$, and $\xi$) of the background.

It is important to understand the different scales of our system. We notice that the background has only one independent scale, namely  the temperature $T$.  A massive flavor introduces a new scale in the problem, which is precisely the mass $m_q$ of the quarks. This new scale is better characterized by the mass gap of the quark-antiquark bound states, defined as the mass of the lightest meson at zero temperature. Up to numerical factors this mass gap is the quantity $\bar M$ defined in (\ref{barM}). Notice that $\bar M$ depends on the screening function $\sigma$, which was to be expected since $\sigma$ parameterizes the flavor screening corrections to the quark-antiquark Coulomb force. Given these two mass scales of the problem it is very natural to consider their ratio. As $\bar M/T=m^{{1\over b}}$, this dimensionless quantity is related to the mass parameter $m$ which we  used as the independent variable of our thermodynamic functions.

One of the main targets of the present paper was the study of the impact of the density of smeared massless flavors in various observables of the massive probe.
The feasibility of such investigation relied on the analytic dependence on the deformation parameter $\hat\epsilon$ of the background. Only a limited number of such observables had thus far been analyzed. One such example is the location of the first order phase transition between Minkowski and black hole embeddings. As shown in Fig.~\ref{Tc} the location increases with $T/m_q$. This has a dual interpretation depending on which variable one chooses to keep fixed, either $T$ or $m_q$. The increasing behavior is also observed in $3+1$ dimensions in the particular case of D3-D7-brane system which acts as a model for the quark gluon plasma (see the last paper in \cite{D3-D7}). Also, the (absolute value of the) condensate $|c|$ increases for any value of $m$ (see Fig.~\ref{c_vs_m}). Notice, however, that the relation between $c$ and $\langle {\cal O}_m\rangle$ 
involves the functions $b$, $q$, and $\sigma$ (see (\ref{O-T-c})). The first two functions reach a constant value when $\hat \epsilon\to\infty$, while the screening function $\sigma$ decreases as $1/\sqrt{\hat\epsilon}$ in this limit (see Fig.~\ref{fig:xi}), which implies that 
$\langle {\cal O}_m\rangle\to 0$ with infinitely many flavors.  The $c$ versus $m$ curve enjoys the self-similarity properties in the neighborhood of the transition point. We also examined the deviation of the speed of sound $\delta v_s^2$ away from the conformal result due to the massive probe and found that it decreases as a function of $\hat\epsilon$ (Fig.~\ref{vs}).

The work presented here can be continued in several directions. First of all, we could study the fluctuations of the probe brane and obtain the meson mass spectrum at non-zero temperature. This study would allow to characterize more precisely the meson melting transition. Secondly, it is quite natural to analyze the thermodynamics of the D6-brane probe at non-zero baryon density and chemical potential which, as in  \cite{Kobayashi:2006sb},  can be introduced by switching on a non-vanishing worldvolume gauge field. A related project would be the study of the thermodynamics of the holographic systems introduced in \cite{Ammon:2012mu}, which contain self-dual configurations of the worldvolume gauge fields that represent   D2-branes dissolved in  the D6-brane.

Another possible future direction of the present work could be the
addition of a magnetic field in order to
study the phenomenon of the magnetic catalysis of ``chiral symmetry
breaking". At weak coupling this has
been studied with conventional perturbative field theory techniques
while at strong coupling a
holographic study has been performed using flavored ${\cal N}$=4
Yang--Mills theory \cite{Filev:2007gb}.
As a warm-up analysis, the magnetic field could only couple with
the probe flavor brane while
a more elaborate approach would correspond to a coupling of the
magnetic field with the
backreacted flavors of the background
(for the similar analysis in the D3-D7 case see \cite{Filev:2011mt,
Erdmenger:2011bw}).

A combination of the charge density and the magnetic field (with non-vanishing NSNS $B$-field in the background and/or supplementary internal flux on the worldvolume) would uncover many interesting
phenomena with potential applications to condensed matter physics. For example, reduced supersymmetry ${\cal N} = 3 \to 1$ due to smeared backreacted flavor branes may help bypassing the arguments in \cite{Zafrir:2012yg} and would thereby allow for the study of the quantum Hall effect, along the lines of \cite{Bergman:2010gm}. It would also be important to study how does the flavor deformation parameter enter into the physics of the holographic zero sound \cite{Karch:2008fa} (for $T\ne 0$ generalization, see \cite{Bergman:2011rf, Davison:2011ek}), and, allowing a non-vanishing Chern-Simons term as assumed above, the properties of the magneto-roton excitation \cite{Jokela:2010nu} and the subsequent formation of the striped phase away from the quantum Hall phase \cite{Bergman:2011rf,Jokela}.

\section*{Acknowledgments}

We are grateful to P. Benincasa,  E. Conde, A. Cotrone, V. Filev,  K. Jensen, L. Mazzanti, C. N\'u\~nez, and J. Tarrio for useful discussions. 
The  work of N.~J., J.~M., and A.~V.~R. is funded in part by the Spanish grant 
FPA2011-22594,  by Xunta de Galicia (Conseller{\'i}a de Educaci\'on, grant 
INCITE09 206 121 PR and grant PGIDIT10PXIB206075PR),  by the 
Consolider-Ingenio 2010 Programme CPAN (CSD2007-00042), and by FEDER. N.~J. is supported as well by the Ministerio de Ciencia e 
Innovaci{\'on} through the Juan de la Cierva program. D.~Z. is funded by the FCT fellowship SFRH/BPD/62888/2009.
Centro de F\'{i}sica do Porto is partially funded by FCT through the projects
PTDC/FIS/099293/2008 \& CERN/FP/116358/2010.
\appendix
\vskip 1cm
\renewcommand{\theequation}{\rm{A}.\arabic{equation}}
\setcounter{equation}{0}
\medskip

\section{Probe action in isotropic coordinates}
\label{Iso}

Let us consider the coordinates $R$ and $\rho$, defined in (\ref{R-rho-def}), where one should understand that $u$ is the isotropic radial coordinate at non-zero temperature introduced in (\ref{u-coordinate-def}). We will first study embeddings of the D6-brane probes that are parameterized by a function $R(\rho)$. The induced metric takes the form
\bear
&&d\hat s^2_7\,=\,
{L^2\,r_h^2\over 2^{{4\over 3}}}\,
\,\big[\,\rho^2+R^2\,\big]^{{1\over b}}\,
\tilde f^{{4\over 3}}\,\,
\Big[-{f^2\over \tilde f^2}\,dt^2\,+\,(dx^1)^2\,+\,(dx^2)^2\,\Big]\,+\,
{L^2\over b^2}\,\,{1+R'^{\,2}\over \rho^2+R^2}\,\,d\rho^2\,+\,\rc\rc
&&\qquad\qquad
+{L^2\over b^2}\,\Big[\, q\,d\alpha^2\,+\,q\,\sin^2\alpha \,d\beta^2\,+\,
{\rho^2\over \rho^2+R^2}\,
\big(\,d\psi\,+\,\cos\alpha\,d\beta\,\big)^2\,\Big]\,\,.
\eear
The determinant of this induced metric is 
\beq
\sqrt{-\det\hat  g_7}\,=\,{L^7r_h^3\over 4 b^4}\,\,q\,\sin\alpha\,\rho\,
[\rho^2+R^2]^{{3\over 2b}-1}\,\,f\,\tilde f\,\sqrt{1+ R'^{\,2}}\,\,,
\eeq
where $f$ and $\tilde f$ are given in (\ref{f-def}) and (\ref{tilde-f-def}) and  it is understood that  $u=\sqrt{\rho^2+R^2}$.  By using these results  the Lagrangian density for the DBI part of the probe action  can be written as
\beq
{\cal L}_{DBI}\,=\,
-{\cal N}\,\rho\,
\big[\,\rho^2\,+\,R^2\,\big]^{{3\over 2b}-1}\,f\,\tilde f\,
\,\sqrt{1+ R'^{\,2}}\,\,,
\label{L_DBI_R_rho}
\eeq
where ${\cal N}$  is  the constant which has been defined in (\ref{calN_calN_r}).
  
Let us next calculate the WZ term of the action. We first compute the pullback of  $ {\cal K}$  in terms of the $R$ and $\rho$ coordinates for an embedding parameterized by a function $R(\rho)$. By using:
\beq
d\theta\,=\,{R-\rho R'\over \rho^2\,+\,R^2}\,d\rho\,\,,
\qquad\qquad
{dr\over r}\,=\,{1\over b}\,{f\over \tilde f}\,\,
{RR'\,+\,\rho\over \rho^2\,+\,R^2}
\,d\rho\,\,,
\label{dtheta-dr}
\eeq
the pullbacks of $e^{\mu}$ ($\mu=0,1,2$) and $e^3$ become
\beq
\hat e^{\mu}\,=\,
{L\,r_h\over 2^{{2\over 3}}}\,\,\tilde f^{{2\over 3}}\,
\,[\rho^2+R^2]^{{1\over 2b}}\,dx^{\mu}\,\,
\,\,,\qquad\qquad
\hat e^{3}\,=\,{L\over b}\,\,{ f\over \tilde f}\,
{R\,R'\,+\,\rho\over \rho^2+R^2}\,
d\rho\,\,,
\eeq
while the pullbacks of the other one-forms $e^a$ are the same as  in (\ref{pullback-es}). Therefore, $\hat {\cal K}$ is given by
\bear
&&e^{-\phi}\,\hat {\cal K}\,=\,{L^7\,q\,r_h^3\over 4\, b^4}\,e^{-\phi}
\,\rho\,\big[\,\rho^2\,+\,R^2\,\big]^{{3\over 2b}-1}\,f\,\tilde f\,\times
\rc\rc
&&
\times\Big[1\,+\,\Big(1-{\tilde f\over f}\Big)\,{R\over \rho^2+R^2}\,(\rho R'-R)\,\Big]
d^3x\wedge  d\rho\wedge \Xi_3\,\,,
\label{hat-calK-R-rho}
\eear
where $\Xi_3 $ is the three-form defined in (\ref{Omega}). To evaluate the WZ term we need to compute the improving term $\delta C_7$. From (\ref{L1-sol}) and (\ref{dtheta-dr}) we get
\beq
L_1(\theta)\,d\theta\,=\,{r_h^3\over b}\,\,{R\,\rho\over (\rho^2+R^2)^2}\,\,
(\rho R'-R)\,d\rho\,\,.
\label{L1dtheta-rho}
\eeq
Moreover, by using the identity,
\beq
{1\over 4}\,\big[\,\rho^2\,+\,R^2\,\big]^{{3\over 2b}}\,
f\,\tilde f\,\Big(\,{f\over \tilde f}-{\tilde f\over f}\Big)\,=\,1\,\,,
\eeq
we can insert  the unity in (\ref{L1dtheta-rho})  and rewrite this last equation as
\bear
&&{L^7\,q\,r_h^3\over b^3}\,e^{-\phi}\,L_1(\theta)\,d\theta\,=\,
{L^7\,q\,r_h^3\over 4 b^4}\,e^{-\phi}\,\rho\,\big[\,\rho^2\,+\,R^2\,\big]^{{3\over 2b}-1}\,f\,\tilde f\,\times\rc\rc
&&
\qquad\qquad\qquad
\times\Big({\tilde f\over f}\,-\,{f\over \tilde f}\,\Big)\,{R\over \rho^2+R^2}\,(\rho R'-R)\,
d\rho\,\,.
\label{L1_R_rho}
\eear
It is clear by comparing (\ref{hat-calK-R-rho}) and (\ref{L1_R_rho}) that the effect of the improving term $L_1$ is to change $\tilde f/f $ to $f/\tilde f$. By including the zero-point  energy term, which is given by $4\, b\, \Delta_0\,{\cal N}$, we arrive at the following  WZ action
\beq
S_{WZ}\,=\,{\cal N}\,\int d^3x
\Bigg[\,\int d\rho\,
\rho\,\big[\,\rho^2\,+\,R^2\,\big]^{{3\over 2b}-1}\,f\,\tilde f\,
\Big[1\,+\,\Big(1-{f\over \tilde  f}\Big)\,{R\over \rho^2+R^2}\,(\rho R'-R)\,\Big]\,+\,4\, b\, \Delta_0
\Bigg]
\,\,.
\label{L_WZ_R_rho}
\eeq
By adding (\ref{L_DBI_R_rho}) and (\ref{L_WZ_R_rho})  we obtain the
total action $S$  in the $(\rho, R)$ variables
\beq
S=-{\cal N}\int d^3x
\Big[\int d\rho
\rho\big[\rho^2+R^2\big]^{{3\over 2b}-1}f\tilde f
\Big[\sqrt{1+ R'^{\,2}}-
1+\Big({f\over \tilde  f}-1\Big){R\over \rho^2+R^2}(\rho R'-R)\,\Big]
-4\, b\, \Delta_0\,
\Big]
\,\,.
\label{L_total_R}
\eeq
The free energy density $F$ written in (\ref{F_R}) follows immediately from (\ref{L_total_R}). Moreover, by taking $4\, b\, \Delta_0=1$ the action (\ref{L_total_R}) coincides with the one written in (\ref{S_total_R}). 

Let us next take the isotropic  coordinate $u$  as the independent variable and let us represent the configuration of the probe by the function 
$\chi(u)=\cos \theta(u)$. In order to find 
the Lagrangian density in these variables we notice that
the induced metric on the worldvolume now takes the form
\bear
&&d\hat s^2_7\,=\,
{L^2\,r_h^2\over 2^{{4\over 3}}}\,
\,u^{{2\over b}}\,
\tilde f^{{4\over 3}}\,\,
\Big[-{f^2\over \tilde f^2}\,dt^2\,+\,(dx^1)^2\,+\,(dx^2)^2\,\Big]\,+\,
{L^2\over b^2 u^2}\,\,
{1-\chi^2+u^2\,\dot\chi ^2\over 1-\chi^2}
\,\,du^2\,+\,\rc\rc
&&\qquad\qquad
+{L^2\over b^2}\,\Big[\, q\,d\alpha^2\,+\,q\,\sin^2\alpha \,d\beta^2\,+\,
(1-\chi^2)\,
\big(\,d\psi\,+\,\cos\alpha\,d\beta\,\big)^2\,\Big]\,\,,
\eear
where $\dot\chi=d\chi/du$. The determinant of this metric is
\beq
\sqrt{-\det\hat  g_7}\,=\,{L^7r_h^3\over 4 b^4}\,\,q\,\sin\alpha\,
u^{{3\over b}-1}\,\,f\,\tilde f\,
\sqrt{1-\chi^2+u^2\,\dot\chi ^2}\,\,.
\eeq
Therefore, the DBI term of the Lagrangian density is
\beq
{{\cal L}_{DBI}\over {\cal N}}\,=\,-
u^{{3\over b}-1}\,\,f\,\tilde f\,
\sqrt{1-\chi^2+u^2\,\dot\chi ^2}\,\,,
\eeq
where ${\cal N}$ is the constant defined in (\ref{calN_calN_r}). 
In order to compute the WZ part we have to calculate  first the pullback of the calibration form ${\cal K}$. By using 
\beq
d\theta\,=\,-{\dot \chi\over \sqrt{1-\chi^2}}\,du\,\,,
\qquad\qquad
{dr\over r}\,=\,{f\over \tilde f}\,{du\over u}\,\,,
\eeq
we find the different $\hat e^{a}$'s in the $(u,\chi)$ variables
\bear
&& \hat e^{\mu}\,=\,
{L r_h\over 2^{{2\over 3}}}\,\,
u^{{1\over b}}\, \tilde f ^{{2\over 3}}\,dx^{\mu}\,\,
\,\,,\qquad\qquad
\hat e^{3}\,=\,{L\over b}\,\,{f\over \tilde f}\,
{du\over u}\,\,,\qquad\qquad
\hat e^{4}\,=\,{L\over b}\,\sqrt{q}\,\,d\alpha\,\,,
\qquad\rc\rc
&&\hat e^{5}\,=\,0\,\,,\qquad\qquad
\hat e^{6}\,=\,{L\sqrt{q}\over b}\,\sin\alpha\,\sqrt{1-\chi^2}
\,d\beta\,\,,
\qquad
\hat e^{7}=-
{L\sqrt{q}\over b}\,\sin\alpha\,\chi\,
\,d\beta
\,\,,\rc\rc
&&\hat e^{8}\,=\,-{L\over b}\,
{\dot \chi\over \sqrt{1-\chi^2}}
\,du\,\,,\qquad\qquad
\hat e^{9}\,=\,
{L\over b}\, \sqrt{1-\chi^2}\,
\,\big(\,d\psi+\cos\alpha\, d\beta
\,\big)\,\,.
\label{pullback-es-chi-u}
\eear
From these expressions we can show  that  the pullback of ${\cal K}$ is
\beq
\hat {\cal K}\,=\,{L^7\,q\,r_h^3\over 4\, b^4}\,
\,u^{{3\over b}-1}\,f\,\tilde f\,
\Big[1\,-\chi^2-u\,{\tilde f\over f}\,\chi\dot\chi\,\Big]
d^3x\wedge  d\rho\wedge \Xi_3\,\,.
\eeq
Let us next compute the contribution to the action of the term of $\delta C_7$ containing the function $L_1$. We get
\beq
L_1(\theta)\,d\theta\,=\,{r_h^3\over b}\,\chi \dot\chi\,du\,=\,
{r_h^3\over 4b}\,u^{3\over b}\,f\,\tilde f\,
\Big({\tilde f\over f}\,-\,{f\over \tilde f}\,\Big)\,\,
\chi \dot\chi\,du\,\,,
\eeq
and we again  see that  the effect of adding $L_1$ is equivalent to changing 
$\tilde f/ f$ by $f/\tilde f$ in $\hat {\cal K}$. Taking into account the zero-point energy, and using the value of the constant $\Delta_0$ written in (\ref{Delta_0_b}),  the
WZ action becomes
\beq
S_{WZ}\,=\, {\cal N}\,\int d^3x\,\Bigg[\int du
u^{{3\over b}-1}\,\,f\,\tilde f\,
\Big[1\,-\chi^2-u\,{f\over \tilde  f}\,\chi\dot\chi\,\Big]\,+\,1\,\Bigg]
\,\,.
\eeq
Therefore, the total action has the following expression
\beq
S\,=\,-{\cal N}\,\int d^3x
\Bigg[\,  
\int du\,u^{\frac{3}{b}-1} \,f   {\tilde f} \,
\Big[\sqrt{1-\chi^2+ u^2 \dot\chi^2}   - 1   +  \chi^2  +   u\,  \frac{f}{\tilde f}\,  \chi \, \dot\chi\, \Big]\,-\,1\,\Bigg]
\,\,\,.
\label{total_L-chi}
\eeq

\vskip 1cm
\renewcommand{\theequation}{\rm{B}.\arabic{equation}}
\setcounter{equation}{0}
\medskip

\section{Low temperature (Minkowski embeddings)}
\label{LowT}
In this appendix we will study, following closely the appendix A.2 of \cite{Mateos:2007vn}, the Minkowski embeddings for high mass (or low temperature), in which the D6-brane probe remains very far from the horizon.  In this case we have embeddings which are nearly flat (with $R(\rho)$ almost constant). Accordingly, we write $R(\rho)$ as:
\beq
R(\rho)\,=\,R_0\,+\,\delta R(\rho)\,\,,
\eeq
with $R_0$ being constant and large compared with $\delta R(\rho)$. Let us write the approximate  Euler-Lagrange equation. First of all, we represent 
$\partial {\cal L}/\partial R'$ as:
\beq
{1\over {\cal N}}\,\, {\partial {\cal L}\over \partial R'}\,\approx\,-\, f_1(\rho)
\partial_{\rho}\,\delta R\,-\,f_2(\rho)\,\,,
\label{dLdRprime}
\eeq
where $f_1(\rho)$ and $f_2(\rho)$ are given by
\bear
&&f_1(\rho)\,=\,\rho\big[\rho^2+R_0^2\big]^{{3\over 2b}-1}\,\,
\Big[1\,-\,\big[\rho^2+R_0^2\big]^{-{3\over b}}\Big]\,\,,\rc\rc
&&f_2(\rho)\,=\,-{2\rho^2\,R_0\over (\rho^2+R_0^2)^2}\,\,
\Big[1\,-\,\big[\rho^2+R_0^2\big]^{-{3\over 2b}}\Big]\,\,.
\label{f12}
\eear
In (\ref{dLdRprime}) and (\ref{f12}) we  substituted $R(\rho)$ by $R_0$ after computing the derivative of ${\cal L}$ with respect to $R'$. Let us next calculate
$\partial {\cal L}/\partial R$. As in \cite{Mateos:2007vn}, after computing the derivative we will neglect the terms with $R'$ and we will substitute $R(\rho)$ by $R_0$. After some rearrangement, we get
\beq
{1\over {\cal N}}\,\, {\partial {\cal L}\over \partial R}\,\approx-
{4\rho\,R_0\over (\rho^2+R_0^2)^3}\,\,
\Big[\rho^2\,-\,R_0^2\,-\,
{\rho^2-\big(1+{3\over 2b}\big)\,R_0^2\over 
\big[\rho^2+R_0^2\big]^{{3\over 2b}}}\,\,\Big]\,\,.
\eeq
Let us next  define a new function $f_3(\rho)$ as
\beq
f_3(\rho)\,\equiv\,-\int_0^{\rho}\,d\rho\,{1\over {\cal N}}\,\,{\partial {\cal L}\over \partial R}\,\,.
\eeq
This integral can be computed explicitly
\beq
f_3(\rho)\,=\,-{2\rho^2\,R_0\over (\rho^2+R_0^2)^2}\,
\Big[1\,-\,{2b\over 3+2b}\,\,(\rho^2+R_0^2)^{-{3\over 2b}}\,\Big]\,-\,
{6R_0\over 3+2b}\Bigg[
{R_0^2\over (\rho^2+R_0^2)^{2+{3\over 2b}}}\,-\,{1\over R_0^{2+{3\over b}}}
\Bigg]\,\,.
\eeq
The Euler-Lagrange equation of motion for $\delta R(\rho)$ can be integrated once 
\beq
f_1(\rho)\,\partial_{\rho}\delta R\,+\,f_2(\rho)\,=\,f_3(\rho)\,\,,
\label{first-integral}
\eeq
where we have imposed the boundary condition $R'(\rho=0)=\partial_{\rho} \delta R(\rho=0)=0$.  From this equation we get
\beq
\partial_{\rho}\delta R\,=\,{f_3-f_2\over f_1}\,=\,
-{6R_0\over 3+2b}\,\,{1\over \rho \big[(\rho^2+R_0^2)^{{3\over b}}\,-\,1\big]}
\Bigg[\,1\,-\,\Big(1+{\rho^2\over R_0^2}\Big)^{1+{3\over 2b}}\,\Bigg]\,\,.
\eeq
Then, the asymptotic value of $\delta R$ is given by
\beq
\delta R(\rho\to\infty)\,=\,-{6 R_0^{1-{6\over b}}\over 3+2b}\,\,F(R_0)\,\,,
\label{deltaR-asymp}
\eeq
where $F(R_0)$ is,
\beq
F(R_0)\,\equiv\,\int_0^{\infty}{d\varrho\over \varrho}\,\,
{1\,-\,\big(1+\varrho^2\big)^{1+{3\over 2b}}\over
\,\big(1+\varrho^2\big)^{{3\over b}}\,-\,R_0^{-{6\over b}}}\,\,.
\label{FR0}
\eeq
Notice that the integrand in (\ref{FR0}) behaves as $\varrho^{1-{3\over b}}$ for large $\varrho$ and therefore the integral only converges if $b<3/2$ (which is always true because the maximum value of $b$ is 5/4).

At leading order in $R_0$ we can substitute $F(R_0)$ by $F(\infty)$ in (\ref{deltaR-asymp}). Thus,
\beq
\delta R(\rho\to\infty)\,\approx\,-{6 R_0^{1-{6\over b}}\over 3+2b}\,\,F(\infty)\,\,.
\eeq
We find the following value for $F(\infty)$:
\beq
F(\infty)\,=\,-{1\over 2}\,\,\Big[\,{2b\over 3-2b}\,+\,\psi \Big({3\over b}\Big)\,-\,
\psi\Big({3\over 2b}\Big)\,\Big]\,\,,
\eeq
where $\psi(x)=\Gamma'(x)/\Gamma(x)$ is the digamma function.  For $b\approx 1$ one can represent $F(\infty)$ in powers of $b-1$ as
\beq
F(\infty)\,=\, -{3\over 4}\,-\,\log(2)\,-\,{15+\pi^2\over 8}\,(b-1)\,+\,\cdots\,\,.
\eeq
The approximate asymptotic value $m $ of $R(\rho)$  at $\rho=\infty$ can be related to $R_0$ as
\beq
m\approx R_0\,+\,a(b)\,R_0^{1-{6\over b}}\,\,,
\label{R*-R0}
\eeq
where 
\beq
a(b)\,=\,-{6\over 3+2b}\,F(\infty)\,=\,
{3\over 3+2b}\,
\Big[\,{2b\over 3-2b}\,+\,\psi \Big({3\over b}\Big)\,-\,
\psi\Big({3\over 2b}\Big)\,\Big]\,\,.
\label{a(b)}
\eeq
The relation (\ref{R*-R0}) can be easily inverted at leading order. We find
\beq
R_0\approx m\,-\,a(b)\,m^{1-{6\over b}}\,\,.
\label{R0-R*}
\eeq
In the particular case $b=1$ the previous formula gives rise to the following relation between $R_0$ and $m$
\beq
R_0\approx m\,-\,{3\over 10}\,(3+4\log 2)\,{1\over m^5}\,\,,
\qquad\qquad\qquad (b=1)\,\,.
\eeq
which should be compared with eq. (A.12) of \cite{Mateos:2007vn}.  Let us next study the large $\rho$ dependence  of $R(\rho)$. In this limit (\ref{deltaR-asymp}) reduces to
\beq
\partial_{\rho}\delta R\,\approx\, {6\over 3+2b}R_0^{-1-{3\over b}}\,\,
\rho^{1-{3\over b}}\,\,.
\eeq
This equation can be integrated immediately
\beq
\delta R\,\approx\, {\rm constant}\,+\,{6b\over 4b^2-9}R_0^{-1-{3\over b}} \,\,\rho^{2-{3\over b}}\,\,.
\label{deltaR-rho}
\eeq
From (\ref{deltaR-rho}) we read the value of the condensate  constant $c$ as a function of $R_0$,
\beq
c\approx {6b\over 4b^2-9}R_0^{-1-{3\over b}} \,\,.
\eeq
In particular, in the unflavored background  $b=1$  we get the following relation between $c$ and $m$:
\beq
c\approx -{6\over 5}\,{1\over m^4}\,\,,
\qquad\qquad\qquad (b=1)\,\,.
\eeq

\subsection{On-shell action}

Let us use the previous results to evaluate the on-shell action for the Lagrangian density (\ref{L_total_R}). To compute the leading order result at high mass (or low temperature) it is enough to take $R=R_0$ in the action. Let us express the result in terms of the function ${\cal G}(m)$ defined in (\ref{S-calG}). By taking $R=R_0$ in  (\ref{calG-R}), we get
\beq
{\cal G}(m)\,\approx\,2\,R_0^2\,\int_{0}^{\infty}\,d\rho\,
\rho\,\big[\,\rho^2+R_0^2\,\big]^{-{3\over 2b}\,-\,2}\,\,
\Big[\,\big[\,\rho^2+R_0^2\,\big]^{{3\over 2b}}\,-\,1\,\Big]\,\,.
\label{G-R0}
\eeq
The integral on the right-hand side of (\ref{G-R0}) can be easily performed, 
\beq
{\cal G}(m)\,=\,1\,-\,{2b\over 2b+3}\,R_0^{-{3\over b}}\,\,.
\label{G-R0-explicit}
\eeq
At leading order we can take $R_0=m$ in (\ref{G-R0-explicit}) and we get the estimate  (\ref{G-large_m}).

\vskip 1cm
\renewcommand{\theequation}{\rm{C}.\arabic{equation}}
\setcounter{equation}{0}
\medskip

\section{High temperature limit (black hole embeddings)}
\label{HighT}

Let us  now consider the limit of high temperature or, equivalently, low quark mass.  Note that the D6-brane embedding with $\chi=0$  is an exact solution to the equation of motion.  In order to study solutions for which $\chi$ remains small,  we expand the D6-brane action to quadratic order in $\chi$ and obtain the following equation of motion
\begin{equation}
\partial_u \Bigg[ u^{-\frac{3}{b}} \left[\left(1 - u^{\frac{3}{b}}\right)^2 \chi  - u 
\left(1 -  u^{\frac{6}{b}}\right)
 \dot\chi \right]\Bigg] 
+  u^{-1- \frac{3}{b}} \,\left(1- u^{\frac{3}{b}}\right) 
\Bigg[  \left(1+ u^{\frac{3}{b}} \right)  \chi  -  u  
\left(1 - u^{\frac{3}{b}}\right)
 \dot\chi \Bigg] =0\,.
\label{linear-chi-eom}
\end{equation}
The general solution of (\ref{linear-chi-eom}) is
\beq
\chi(u)\,=\,c_1\,u^{{3\over b}-1}\,\,F\Big({1\over 2}\,,\,1-{b\over 3};{3\over 2}-{b\over 3};
u^{{6\over b}}\Big)\,+\,
c_2\,u\,F\Big({1\over 2}\,,\,{b\over 3};{1\over 2}+{b\over 3};
u^{{6\over b}}\Big)\,\,,
\label{general-sol-chi}
\eeq
where $c_1$ and $c_2$ are two constants to be determined. Let us focus on  the $u=1$ behavior of $\chi(u)$. In general, we have
\beq
F\Big(\alpha, \beta;\alpha+\beta;z\Big)\approx -{\Gamma(\alpha+\beta)\over \Gamma(\alpha)\Gamma(\beta)}\,\,\log (1-z)\,\,,
\qquad\qquad {\rm as}\,\,z\to 1^{-}\,\,.
\eeq
Therefore, near $u=1$  the general solution (\ref{general-sol-chi}) behaves as
\beq
\chi(u)\,\approx\,{1\over \sqrt{\pi}}\,\,
\Bigg[\,
{\Gamma\big({3\over 2}-{b\over 3}\big)\over \Gamma\big(1-{b\over 3}\big)}\,\,c_1
\,+\,
{\Gamma\big({1\over 2}+{b\over 3}\big)\over \Gamma\big({b\over 3}\big)}\,\,c_2
\,\Bigg]\,\log(1-u^{{6\over b}})\,\,.
\label{chi-near-u1}
\eeq
Thus, the solution (\ref{general-sol-chi}) is generically singular at the horizon $u=1$.  To avoid this singularity we must impose that the coefficient of the logarithm in (\ref{chi-near-u1}) vanishes, which leads to
\beq
{c_1\over c_2}\,=\,-{
\Gamma\big({1\over 2}+{b\over 3}\big)\, \Gamma\big(1-{b\over 3}\big)
\over
\Gamma\big({b\over 3}\big)\,\Gamma\big({3\over 2}-{b\over 3}\big)}\,\,.
\label{c1c2ratio}
\eeq
Interestingly, this condition is equivalent to requiring $\dot\chi (u=1)=0$, as in (\ref{chi-bc}).

Let us now look at the behavior at $u=\infty$. To find the asymptotic limit of $F(\alpha,\beta;\gamma; z)$ at large $z$ we make use of the following relation
\bear
&&F(\alpha,\beta;\gamma; z)\,=\,
{\Gamma(\gamma)\Gamma(\beta-\alpha)\over 
\Gamma(\beta)\Gamma(\gamma-\alpha)}\,\,(-1)^{\alpha}\,z^{-\alpha}\,\,
F(\alpha,\alpha+1-\gamma;\alpha+1-\beta; {1\over z})\,+\,\rc\rc
&&
\qquad\qquad\qquad\qquad
+\,
{\Gamma(\gamma)\Gamma(\alpha-\beta)\over \Gamma(\alpha)\Gamma(\gamma-\beta)}
\,\,(-1)^{\beta}\,z^{-\beta}\,\,
F(\beta,\beta+1-\gamma;\beta+1-\alpha; {1\over z})\,\,.
\qquad
\eear
For the particular case of $\gamma=\alpha+\beta$ this formula leads to the following asymptotic behavior for large $z$:
\beq
F(\alpha,\beta;\alpha+\beta; z)\,\approx\,
\Gamma(\alpha+\beta)\,\Bigg[\,
(-1)^{\alpha}\,{\Gamma(\beta-\alpha)\over \Gamma^2(\beta)}\,z^{-\alpha}\,+\,
(-1)^{\beta}\,{\Gamma(\alpha-\beta)\over \Gamma^2(\alpha)}\,z^{-\beta}
\,\Bigg]\,\,.
\eeq
It remains to determine the values of $(-1)^{\alpha}$ and $(-1)^{\beta}$, which are in general multivalued. By comparing with the numerical results  when $\alpha$ and $\beta$ are as in (\ref{general-sol-chi}) one concludes that one should take $-1=e^{-i\pi}$ and thus $(-1)^{\alpha}=e^{-i\pi\alpha}$ (and similarly for 
$(-1)^{\beta}$). Thus, for the two hypergeometric functions in (\ref{general-sol-chi}), we can write at large $u$
\bear
&&u^{{3\over b}-1}F\Big({1\over 2}\,,\,1-{b\over 3};{3\over 2}-{b\over 3};
u^{{6\over b}}\Big)\approx
 -\Gamma\Big({3\over 2}-{b\over 3}\Big)\,
\Bigg[i\,{\Gamma\big({1\over 2}-{b\over 3}\big)\over 
\Gamma^2\big(1-{b\over 3}\big)}\,{1\over u}+
{\Gamma\big(-{1\over 2}+{b\over 3}\big)\over \pi}\,e^{{i\pi b\over 3}}\,
{1\over u^{{3\over  b}-1}}\,\Bigg]\,\,,\rc\rc
&&u\,F\Big({1\over 2}\,{b\over 3};{1\over 2}+{b\over 3};
u^{{6\over b}}\Big)\approx
\Gamma\Big({1\over 2}+{b\over 3}\Big)\,
\Bigg[
{\Gamma\big({1\over 2}-{b\over 3}\big)\over \pi}\,e^{-{i\pi b\over 3}}\,
{1\over u}\,-\,i
{\Gamma\big(-{1\over 2}+{b\over 3}\big)\over 
\Gamma^2\big({b\over 3}\big)}\,
{1\over u^{{3\over  b}-1}}\,\Bigg]\,\,.
\eear
Using these equations we see that the coefficient of $1/u$ in the asymptotic expansion of $\chi(u)$ is
\beq
-i\,{\Gamma\big({3\over 2}-{b\over 3}\big)\,
\Gamma\big({1\over 2}-{b\over 3}\big)\over 
\Gamma^2\big(1-{b\over 3}\big)}\,\,c_1\,+\,
{\Gamma\big({1\over 2}+{b\over 3}\big)\,\Gamma\big({1\over 2}-{b\over 3}\big)
\over \pi}\,e^{-{i\pi b\over 3}}\,\,c_2\,\,.
\label{leading-chi-asymp}
\eeq
The imaginary part of (\ref{leading-chi-asymp}) should be zero (otherwise the mass would be complex). This condition leads to
\beq
{c_1\over c_2}\,=\,-{\sin\big({\pi b\over 3}\big)\over \pi}\,\,
{\Gamma\big({1\over 2}+{b\over 3}\big)\,\Gamma^2\big(1-{b\over 3}\big)
\over \Gamma\big({3\over 2}-{b\over 3}\big)}\,\,,
\eeq
which can be shown to be equivalent to (\ref{c1c2ratio}) by taking $z=b/3$ in the reflection formula for the Gamma function, namely
\beq
\Gamma(z)\,\Gamma(1-z)\,=\,{\pi\over \sin (\pi z)}\,\,.
\label{reflection}
\eeq
The only contribution to the real part of (\ref{leading-chi-asymp}) comes from the second term, and is given by
\beq
\cos\big({\pi b\over 3}\big)\,
{\Gamma\big({1\over 2}+{b\over 3}\big)\,\Gamma\big({1\over 2}-{b\over 3}\big)\over \pi}
\,\,c_2\,\,.
\eeq
One can check that the coefficient multiplying $c_2$ in this last expression is one by using again the reflection formula (\ref{reflection}). Thus, we can identify $c_2$ with the mass parameter $m$.   Let us next study the subleading terms. The coefficient of $u^{-3/b +1}$ is
\beq
-{\Gamma\big({3\over 2}-{b\over 3}\big)\Gamma\big(-{1\over 2}+{b\over 3}\big)\over \pi}\,e^{{i\pi b\over 3}}\,c_1\,-\,
i
{\Gamma\big({1\over 2}+{b\over 3}\big)\,\Gamma\big(-{1\over 2}+{b\over 3}\big)\over 
\Gamma^2\big({b\over 3}\big)}\,c_2\,\,.
\label{subleading-chi-asymp}
\eeq
By requiring the imaginary part of (\ref{subleading-chi-asymp}) to vanish we get again an expression for $c_1/c_2$, which can be shown to be equivalent to (\ref{c1c2ratio}) by using (\ref{reflection}).  Moreover, the real part of (\ref{subleading-chi-asymp}) is
\beq
-\cos\big({\pi b\over 3}\big)\,
{\Gamma\big({3\over 2}-{b\over 3}\big)\,\Gamma\big(-{1\over 2}+{b\over 3}\big)\over \pi}
\,\,c_1\,\,,
\eeq
which can be shown to be equal to $c_1$ by using again (\ref{reflection}). Thus, we can identify $c_1$ with the condensate $c$ in (\ref {asympD6}). From these identifications of $c_1$ and $c_2$ and their relation (\ref{c1c2ratio}) it follows that,  in this low-mass regime,  the condensate $c$ is linear in the mass $m$ and is given by
\beq
c\,\approx\,-\,{\Gamma\big({1\over 2}+{b\over 3}\big)\, \Gamma\big(1-{b\over 3}\big)
\over
\Gamma\big({b\over 3}\big)\,\Gamma\big({3\over 2}-{b\over 3}\big)}\,m\,\,,
\label{c_small_m}
\eeq
which is just the expression written in (\ref{c_m_highT}). 
Let us now find the relation between the value of $\chi$ at the horizon ($\chi_h\equiv \chi(u=1)$) and the mass $m$. We find that  $\chi_h$ can be simply written as
\beq
\chi_h\,\approx\,\sqrt{\pi}\,\,{\Gamma\big(1-{b\over 3}\big)\over 
\Gamma\big({1\over 2}-{b\over 3}\big)}\,\,m\,\,,
\eeq
which coincides with (\ref{chi_0_m}).

\subsection{On-shell action}

We can now use the approximate analytic solution found in the previous subsection to compute the on-shell action which is needed to evaluate the free energy at high temperature. Instead of applying a brute force method let us use the fact that the on-shell quadratic  action can be computed as the integral of a total derivative (\ie,  by taking the appropriate limits without the need of performing the integral).  Let us consider first the generic case of an action of the type
\beq
{\cal S}\,=\,\int_{u_0}^{\infty}\,du\,\Big[\,F_1(u)\,\dot\chi^2\,+\,F_2(u)\,\chi^2\,+\,F_3(u)\,\chi\dot \chi\,\Big]\,\,,
\eeq
where the $F_i$'s are known functions of the radial variable $u$. The equation of motion derived from ${\cal S}$ is
\beq
{d\over du}\,\Big[\,F_1\,{d\chi\over du}\,\Big]\,-\,F_2\,\chi\,=\,-{1\over 2}\,
{d\over du}\big[\,F_3\,\chi\,\big]\,+\,{1\over 2}\,F_3\,{d\chi\over du}\,\,.
\label{eom-Fs}
\eeq	
If we rewrite the action as
\beq
{\cal S}\,=\,\int_{u_0}^{\infty}\,du\,\Bigg[
{d\over du}\,\Big[\,F_1\,\chi\,{d\chi\over du}\,\Big]\,-\,
\chi\,\Big[\,{d\over du}\,\Big(\,F_1\,{d\chi\over du}\,\Big)\,-\,
F_2\,\chi\,\Big]\,+\,F_3\,\chi\,{d\chi\over du}\,\Bigg]\,\,,
\eeq
then, after using the equation of motion (\ref{eom-Fs}), the on-shell action can be written as
\beq
{\cal G}\equiv{\cal S}^{on-shell}\,=\,
\int_{u_0}^{\infty}\,du\,\Bigg[
{d\over du}\,\Big[\,F_1\,\chi\,{d\chi\over du}\,\Big]\,+\,
{1\over 2}\,{d\over du}\big[\,F_3\,\chi^2\,\big]\,\Bigg]\,\,.
\label{calG-calS_onshell}
\eeq
Equivalently, we can write ${\cal G}$ in terms of boundary values at $u=u_0$ and at $u=\infty$ 
\beq
{\cal G}\,=\,\chi\,\Big(\,F_1\,{d\chi\over du}\,+\,
{1\over 2}\,F_3\,\chi\,\Big)\Bigg|^{u=\infty}_{u=u_0}\,\,.
\label{calG_UV_IR}
\eeq
We will apply this method to compute the function ${\cal G}(m)$ defined in (\ref{S-calG}) in the small $m$ regime. Thus, we will take $u_0=1$ and we will identify the function ${\cal G}$ of (\ref{calG-calS_onshell}) with ${\cal G}(m)$. By expanding the right-hand side of (\ref{calG-chi}) to quadratic order in $\chi$ we find,
\bear
&&F_1(u)\,=\,{1\over 2}\,u^{{3\over b}+1}\,f\,\tilde f\,=\,
{1\over 2}\,u^{{3\over b}+1}\,-\,{1\over 2}\,u^{-{3\over b}+1}\,\,,\rc\rc
&&F_2(u)\,=\,{1\over 2}\,u^{{3\over b}-1}\,f\,\tilde f\,=\,
{1\over 2}\,u^{{3\over b}-1}\,-\,{1\over 2}\,u^{-{3\over b}-1}\,\,,\rc\rc
&&F_3(u)\,=\,\,u^{{3\over b}}\, f^2\,=\,
u^{{3\over b}}\,-\,u^{-{3\over b}}\,-\,2\,
\,\,.
\eear
Notice that $F_1(u=1)=F_3(u=1)=0$ and therefore there is no contribution from the horizon to  the right-hand side of (\ref{calG_UV_IR}). Moreover, from the UV asymptotic behavior (\ref{asympD6}), we get for large $u$
\beq
F_1\,{d\chi\over du}\,+\,
{1\over 2}\,F_3\,\chi\,=\,-{c\over 2b}\,(3-2b)\,u\,+\,\cdots\,\,,
\eeq
where the dots represent terms which vanish when $u\to\infty$. Therefore
\beq
 \lim_{u\to\infty}\,\chi\,\Big(\,F_1\,{d\chi\over du}\,+\,
{1\over 2}\,F_3\,\chi\,\Big)\,=\,-{3-2b\over 2b}\,\,c\,m\,\,,
\eeq
and ${\cal G}(m)$ can be approximated in this large temperature regime as
\beq
{\cal G}(m)\approx -{3-2b\over 2b}\,\,c\,m\,\,,
\eeq
which is just the expression used in the main text (eq. (\ref{G_small_m})). 
Let us rewrite this equation in a more explicit way. By using (\ref{c_small_m}) and  the reflection formula (\ref{reflection}), we can write
\beq
{3-2b\over 2b}\,\,c\,m\,=\,-{3\over b}\,
{\Gamma\big({1\over 2}+{b\over 3}\big)\,
\Gamma\big(1-{b\over 3}\big)\over
\Gamma\big({b\over 3}\big)\,
\Gamma\big({1\over 2}-{b\over 3}\big)}\,\,m^2\,=\,
-{3\over b}\,\Bigg[{
\Gamma\big(1-{b\over 3}\big)\over
\Gamma\big({1\over 2}-{b\over 3}\big)}\Bigg]^2\,
\tan\big({\pi b\over 3}\big)\,m^2\,\,.
\eeq
Therefore, finally we arrive at
\beq
{\cal G}(m)\,\approx\,
{3\over b}\,\Bigg[{
\Gamma\big(1-{b\over 3}\big)\over
\Gamma\big({1\over 2}-{b\over 3}\big)}\Bigg]^2\,
\tan\big({\pi b\over 3}\big)\,m^2
\,\,.
\eeq

\vskip 1cm
\renewcommand{\theequation}{\rm{D}.\arabic{equation}}
\setcounter{equation}{0}
\medskip

\section{Mass and condensate}
\label{m-and-VEV}

In this appendix we study in detail the relation between the parameters $m$ and $c$,  the quark mass $m_q$, and the condensate $\langle{\cal O}_m\rangle$. The quark mass $m_q$ can be obtained by computing the Nambu-Goto action of a fundamental string hanging from the boundary to the horizon. The relation that is found in this way is
\beq
m_q\,=\,{1\over 2\pi}\,\,{L^2\,r_h\over 2^{{2\over 3}}}\,m^{{1\over b}}\,\,.
\label{mq-m}
\eeq
It is easy  to write the right-hand side of  (\ref{mq-m}) in terms of gauge theory quantities. First of all, we recall that $r_h=4\pi T/3$. Moreover,
the $AdS$ radius $L$ for the flavored background is given by $
L^2\,=\,\pi\,\sqrt{2\lambda}\,\,\sigma$, where $\lambda=N/k$ is the 't Hooft coupling and  $\sigma$  is the screening function  
defined in (\ref{screening-sigma}).  By using these equations we can rewrite (\ref{mq-m}) as in (\ref{m-mq}).  This expression can be inverted,
\beq
m\,=\,\Big(\,{3 m_q\over 2^{{1\over 3}}\pi \,\,\sqrt{2\lambda}\,\,\sigma}\,\,
{1\over T}\,\Big)^{b}\,=\,
\Big(\,{\bar M\over T}\,\Big)^b\,\,,
\label{m-mq-T}
\eeq
where, in the last step, we introduced the quantity
\beq
\bar M\,\equiv\, {3 m_q\over 2^{{1\over 3}}\pi \,\,\sqrt{2\lambda}\,\,\sigma}\,\,.
\label{barM}
\eeq
It follows from (\ref{m-mq-T}) that for fixed quark mass $m_q$ and 't Hooft coupling $\lambda$, $m$ depends on $T$ as $m\propto T^{-b}$.

Let us now turn ourselves to the calculation of the condensate, which can be obtained from the derivative of the free energy with respect to the  bare quark mass $\mu_q$,
\beq
\langle{\cal O}_m\rangle\,=\,{\partial F\over \partial \mu_q}\,\,.
\label{VEV-derivative}
\eeq
In order to compute the derivative on the right-hand side of (\ref{VEV-derivative}) we should find the relation between the bare mass $\mu_q$ and the mass parameter $m$. 
Notice that the quark mass written in (\ref{m-mq}) contains the screening effects due to quark loops, which should not be included in the bare mass. These effects are encoded in the functions $b$ and $\sigma$. By taking $b=\sigma=1$ in (\ref{m-mq}) we switch off the dressing due to dynamical flavors. Accordingly, our prescription for the bare mass $\mu_q$ is
\beq
\mu_q\,=\,{2^{{1\over 3}}\pi\over 3}\,\,\sqrt{2\lambda}\,\,\,T\, m\,\,.
\label{bare_mass}
\eeq
By using the chain rule we can relate the derivative with respect to $\mu_q$ to the derivative  with respect to $m$. Actually, it follows from (\ref{bare_mass}) that
\beq
{\partial m\over \partial \mu_q}\,=\,{m\over \mu_q}\,\,,
\eeq
and therefore the condensate is given by
\beq
\langle{\cal O}_m\rangle\,=\,{ m\over \mu_q}\,\,{\partial F\over \partial m}\,=\,
{m\,{\cal N}\over \mu_q}\,\,{\partial\over \partial m}\,\Big({F\over {\cal N}}\Big)
\,\,,
\label{Om_derivative_F/N}
\eeq
where, in the last step, we multiplied and divided by ${\cal N}$, which was defined in (\ref{calN_calN_r}) and does not depend on $m$.  The calculation of the derivative on the right-hand side of (\ref{Om_derivative_F/N}) is very similar to the one performed at the end of Section  \ref{zeroTemp} in the zero temperature background.  As in (\ref{F-calG}), we will represent $F/ {\cal N}$ by means of the integral  ${\cal G}(m)$. 
We will  work with the $(u,\chi)$ variables and parameterize these quantities in terms of a density ${\cal F} (u,\chi,\dot\chi)$ as follows
\beq
{F\over {\cal N}}\,=\,{\cal G}(m)-1\,=\,
\int_{u_0}^{\infty}du\,\,
{\cal F} (u,\chi,\dot\chi)
\,-\,1\,\,.
\eeq
The explicit expression of ${\cal F}$ can be read off from the right-hand side of (\ref{calG-chi}). Notice that ${\cal G}(m)$ depends on $m$ implicitly through the embedding function 
$\chi(u)$. Indeed, changing $m$ is equivalent to modifying the boundary conditions for the embedding, which in turn gives rise to a new solution of  the equations of motion of the probe. The variation with respect to $m$  of 
$F/ {\cal N}_u$ for a function $\chi$ that satisfies the equations of motion can be obtained from the asymptotic behavior of the derivatives of ${\cal F}$. Following the same steps as in (\ref{partialS-Rstar}) we arrive at
\beq
{\partial \over \partial m} \Big({F\over {\cal N}}\Big)\,=\,
{\partial {\cal F}\over \partial \dot\chi}\,\,
{\partial \chi\over \partial m}
\,\Bigg|^{u=\infty}_{u=u_0}\,\,.
\label{variation-m-F}
\eeq
Let us consider from now on a black hole embedding for which $u_0=1$ (a similar result can be obtained for the Minkowski embeddings by working  in the $(\rho,R)$ variables). The derivative of 
${\cal F}$ appearing in (\ref{variation-m-F}) is
\beq
{\partial {\cal F}\over \partial \dot\chi}\,=\,
f\,\tilde f\,u^{{3\over b}+1}\,\,
{\dot\chi\over 
\sqrt{1-\chi^2\,+u^2 \dot\chi^2}}\,+\,
u^{{3\over b}}\,\,f^2\,\chi\,\,.
\eeq
As $f(u=1)\,=\,0$ (see (\ref{f-def})), we have
\beq
{\partial {\cal F}\over \partial \dot\chi}\,\Bigg|_{u=1}=0\,\,,
\eeq
and the contribution at the lower limit of (\ref{variation-m-F}) vanish. 
In order to evaluate the asymptotic value at $u\to\infty$, let us remember that
$\chi$ and $\dot\chi$ behave as
\beq
\chi\sim {m\over u}\,+\,{c\over u^{{3\over b}-1}}\,+\,\cdots\,\,,
\qquad\qquad
\dot\chi\sim -
{m\over u^2}\,+\,\big(1-{3\over b}\big)\,{c\over u^{{3\over b}}}\,+\,\cdots\,\,.
\eeq
Thus, it follows that
\beq
u^{{3\over b}+1}\,\dot\chi\sim -m u^{{3\over b}-1}\,+\,
\big(1-{3\over b}\big)\,c\,u\,+\,\cdots\,\,,
\qquad\qquad
u^{{3\over b}}\,\chi\sim m u^{{3\over b}-1}\,+\,c\,u\,+\,\cdots\,\,.
\eeq
Then, for large $u$,  we get 
\beq
{\partial {\cal F}\over \partial \dot\chi}\sim {2b-3\over b}\,c\,u\,+\,\cdots\,\,.
\eeq
Taking into account that
\beq
{\partial \chi\over \partial m}\sim {1\over u}\,+\,\cdots\,\,,
\eeq
we finally  arrive at
\beq
{\partial \over \partial m} \Big({F\over {\cal N}}\Big)\,=\,
{\partial {\cal F}\over \partial \dot\chi}\,\,
{\partial \chi\over \partial m}
\,\Bigg|_{u=\infty}\,=\,{2b-3\over b}\,c\,\,.
\label{partialF-m}
\eeq
From (\ref{partialF-m}) we readily get
\beq
{\partial F\over \partial m}\,=\,-{3-2b\over b}\,c\,{\cal N}\,\,.
\label{m-derivative-F/N}
\eeq
By using (\ref{m-derivative-F/N})  to evaluate the right-hand side of (\ref{Om_derivative_F/N}), we obtain the relation between $\langle{\cal O}_m\rangle$ and $c$ that we were looking for
\beq
\langle{\cal O}_m\rangle\,=\,-{3-2b\over b}\,{m \,{\cal N}\over \mu_q}\,c\,\,.
\label{O-Nu-c}
\eeq
Therefore, $c$ is proportional to the condensate $\langle{\cal O}_m\rangle$ as expected. Let us now write this result in terms of gauge theory quantities. By using (\ref{Nu-gauge}) and (\ref{bare_mass}) we can write
\beq
{m \,{\cal N}\over \mu_q}\,=\,{2^{{2\over 3}}\,\pi\over 9}\,\,
{\zeta\over b}\,\,N\,\,T^2\,\,.
\eeq
Plugging this result into (\ref{O-Nu-c}) and using (\ref{sigma-mu}) to eliminate $\sigma$, we arrive at the formula written in (\ref{O-T-c})  for the condensate $\langle{\cal O}_m\rangle$. 

Let us now determine the high and low temperature behavior of $\langle{\cal O}_m
\rangle$. We start by considering the behavior for large $T$ (or small $m$) and fixed $m_q$. In this case $c\sim m\sim T^{-b}$  (see (\ref{c_small_m}) and (\ref{mq-m})) and thus,
\beq
\langle{\cal O}_m\rangle\,\sim\, T^{2-b}\,\sim\,T^{1-\gamma_m}\,\,,
\qquad\qquad (T\to\infty)\,\,,
\eeq
where in the last step we wrote the result in terms of the mass anomalous dimension $\gamma_{m}=b-1$. Thus, the  dependence  on $T$ of the condensate  in this high  $T$ regime varies with the number of flavors. Actually, it is determined by the mass anomalous dimension $\gamma_m$. Clearly $\langle{\cal O}_m\rangle$ grows linearly with $T$ for the unflavored background, whereas, since $\gamma_{m}=1/4$ for $N_f\to\infty$, the condensate only grows as $T^{3/4}$ when the number of flavors is very large. 

At low $T$ we found in (\ref{c_m_lowT})  that $c$ behaves as
\beq
c\sim m^{-1-{3\over b}}\,\sim\, T^{b+3}\,\,,
\qquad\qquad (T\to 0)\,\,.
\eeq
By using this result in (\ref{O-T-c}) we conclude that the dependence of $\langle{\cal O}_m\rangle$ on the temperature for low $T$ is given by
\beq
\langle{\cal O}_m\rangle\,\sim\, T^{5+b}\,\,,
\qquad\qquad (T\to 0)\,\,.
\eeq

\vskip 1cm
\renewcommand{\theequation}{\rm{E}.\arabic{equation}}
\setcounter{equation}{0}
\medskip

\section{Thermal screening}
\label{thermal}
In this appendix we will analyze two quantities that characterize the screening of quarks in the thermal medium of our flavored black hole. We will start by studying the quark-antiquark potential, following the approach of \cite{Maldacena:1998im,Rey:1998ik} (see also \cite{Brandhuber:1998bs,Rey:1998bq}), in which one considers a fundamental string hanging from the UV and penetrating into the bulk. If $r_0$ denotes the minimal value of the radial coordinate reached by the string, one can show that the quark-antiquark distance $d$ on the boundary is given by
\beq
 d  =  \frac{2\sqrt {h_0}}{r_0}\int_1^\infty\frac{dy}{\sqrt{y^4-y(1-h_0)}\sqrt{y^4-y(1-h_0)-h_0}}\,\,,
 \label{dqq_exact}
 \eeq
 where $h_0$ denotes
 \beq
 h_0\,\equiv\,h(r=r_0)\,=\,1\,-\,\Big({r_h\over r_0}\Big)^3\,\,.
 \eeq
 Moreover, we can  also compute the energy of the quark-antiquark pair  by evaluating the on-shell action of the string. This quantity must be regulated by subtracting the energy of two straight strings stretching from the UV to the horizon $r=r_h$. The final result of this calculation yields
 \beq
  E_{q\bar q}  =  \sqrt{2\lambda}\,\sigma\,
  \left\{ r_0\int_1^\infty dy\left[\frac{\sqrt{y^4-y(1-h_0)}}{\sqrt{y^4-y(1-h_0)-h_0}}-1\right]-r_0+r_h  \right\}\,\,.
\label{Eqq_exact}
\eeq
Notice that, as in the $T=0$ case, the screening effect due to the dynamical quarks is given by the function $\sigma$ multiplying the square root of the 't Hooft coupling $\lambda$ in (\ref{Eqq_exact}). In order to investigate the departure from the Coulomb  behavior due to the finite temperature, let us expand (\ref{dqq_exact})  and (\ref{Eqq_exact}) in powers of $T$ (or, equivalently of $r_h$) and keep the first non-trivial contribution. For the $q\bar q$ distance $d$ we get
\beq
 d  =  \frac{2}{r_0}\frac{\sqrt 2 \pi^{3/2}}{\Gamma\left(\frac{1}{4}\right)^2}-
 {r_h^3\over r_0^4}\,\,{\cal J}
 \,+\,\cdots\,\,,
 \label{dqq_lowT}
  \eeq
 where $ {\cal J}$ is the following integral:
  \beq
 {\cal J}\,\equiv\,\int_1^\infty\frac{dy}{y^2\sqrt{y^4-1}}\left\{1-\frac{1+y+y^2+2y^3}{y^3(1+y+y^2+y^3)}\right\}\approx 0.093\,\,.
 \eeq
The relation (\ref{dqq_lowT}) can be easily inverted to obtain $r_0$ as a function of $d$, 
\beq
 r_0 =  \frac{2\,\sqrt 2 \pi^{3/2}}{\Gamma\left(\frac{1}{4}\right)^2}\,\frac{1}{d}\,-\,\left(\frac{\Gamma\left(\frac{1}{4}\right)^2}{2\,\sqrt 2\,\pi^{3/2}}\right)^3 \,r_h^3\,{\cal J}\,d^2
 +\cdots \ .
 \label{r0_d_lowT}
\eeq
Similarly, $E_{q\bar q}$ for low $T$ can be expanded as
\beq
 E_{q\bar q}= -\sqrt{\lambda}\,\sigma\,\,\Big[\,
\frac{2\pi^{3/2}}{\Gamma\left(\frac{1}{4}\right)^2}\,\,r_0\,+\,
{\tilde {\cal J}\over\sqrt{2}\, r_0^2}\,\,r_h^3\,-\,\sqrt{2}\,r_h\,+\,\cdots\,\Big]\,\,,
 \label{Eqq_lowT}
\eeq
where $\tilde {\cal J}$ is defined as the following integral:
\beq
\tilde {\cal J}\,\equiv\,\int_1^\infty dy\,\,\,
 \frac{1+y+y^2}{y (1+y+y^2+y^3)\sqrt{y^4-1}}
\approx 0.485\,\,.
\eeq
Plugging the value of $r_0$ given by (\ref{r0_d_lowT}) in (\ref{Eqq_lowT}) we obtain the quark-antiquark energy as a function of $d$ at low temperature, 
\beq
 E_{q\bar q}(d,T)\,-\,E_0(T)
 = -\sqrt{2\lambda}\,\sigma\,
 \Big[\frac{4\pi^{3}}{\Gamma\left(\frac{1}{4}\right)^4}\,{1\over d}\,+\,{\pi\over 54}\,
 \Gamma\left(\frac{1}{4}\right)^4\,T^3\,d^2\,+\,\cdots\,\Big]\,\,,
 \label{Eqq_lowT_d}
 \eeq
where $E_0(T)\,=\,{4\pi\over 3}\,\,\sqrt{2\lambda}\,\sigma\,T$ is the zero-point  thermal energy introduced by our regularization and we have used $\tilde{\cal J}-{\cal J}=\pi/8$. The behavior displayed in (\ref{Eqq_lowT_d}) for low  $T$ corresponds to the one expected for a quark-antiquark  pair screened by a thermal bath. Indeed, we readily conclude that the first temperature correction in (\ref{Eqq_lowT_d}) makes the force between the $q$ and the $\bar q$ less attractive.  Actually, as in \cite{Brandhuber:1998bs,Rey:1998bq}, one can evaluate numerically the exact expressions (\ref{dqq_exact}) and (\ref{Eqq_exact}). One finds that the $q\bar q$ distance $d$ reaches a maximum and, actually, the Coulomb-like behavior $ E_{q\bar q}\sim 1/d$, valid at low temperatures, ceases to exist at high temperatures and the quarks become free due to the thermal screening.

The second observable measuring the thermal screening of quarks that we will analyze  is the constituent quark mass $M_c$ below and near the critical temperature. According to the standard holographic dictionary, $M_c$ is obtained by evaluating the action of a fundamental string hanging from the flavor D6-brane down to the horizon. Thus, following a similar calculation in \cite{Mateos:2007vn}, let us consider a  Minkowski embedding in  the $(R,\rho)$ variables. The induced metric on the worldsheet of a  fundamental string extended in $t, R$ at $\rho=0$ is
\beq
ds^2_2\,=\,-{L^2\,r_0^2\over 2^{{4\over 3}}}\,R^{{2\over b}}\,f^2\,\tilde f^{-{2\over 3}}\,dt^2\,+\,{L^2\over b^2}\,{dR^2\over R^2}\,\,.
\eeq
Taking into account that in the above metric, $f=1-R^{-{3\over b}}$ and 
$\tilde f=1+R^{-{3\over b}}$, we get the following value for the determinant of the induced metric:
\beq
\sqrt{-\det g_2}\,=\,{L^2\,r_h\over 2^{{2\over 3}}\,b}\,
R^{{1\over b}-1}\,\big(1-R^{-{3\over b}}\big)\,
\big(1+R^{-{3\over b}}\big)^{-{1\over 3}}\,\,.
\eeq
The constituent quark mass $M_c$ is minus the action per unit time of the Nambu-Goto action, 
\beq
M_c\,=\,{1\over 2\pi}\,\,\int_1^{R_0}\,\,
\sqrt{-\det g_2}\,=\, {1\over 2\pi}\,
{L^2\,r_h\over 2^{{2\over 3}}}\,
\Big[\,R_0^{{1\over b}}\,\Big(\,1\,+\,{1\over R_0^{{3\over b}}}\,\Big)^{{2\over 3}}\,-\,\sqrt[3]{4}\,
\Big]\,\,,
\eeq
where we have taken $\alpha'=1$ and  $R_0$ is the minimum value of the coordinate $R$ reached by the brane. When $R_0\to \infty$, $R_0\approx m$, and the constituent quark mass $M_c$ becomes equal  to the  quark mass $m_q$ and we recover the relation (\ref{m-mq}) between $m_q$ and $m$.  It follows that  the ratio between $M_c$   and  $m_q$ is given by
\beq
{M_c\over m_q}\,=\,{1\over m^{{1\over b}}}\,\,
\Big[\,R_0^{{1\over b}}\,\Big(\,1\,+\,{1\over R_0^{{3\over b}}}\,\Big)^{{2\over 3}}\,-\,\sqrt[3]{4}\,
\Big]\,\,.
\label{Mc_R0}
\eeq
Notice that  in (\ref{Mc_R0}) $M_c\to 0$ as we approach the critical solution with $R_0=1$.  Actually, one can show that $M_c$ decreases monotonically with $T$ as we approach the temperature of the phase transition. This is the expected physical behavior for a free quark in a plasma. Moreover, at low temperature we can use the analytic result (\ref{R0-R*}) to obtain the first screening corrections. We find
\beq
{M_c\over m_q}\,=\,1\,-\,\sqrt[3]{4}\,\,{T\over\bar M}\,+\,{2\over 3}\,
\Big(\,{T\over \bar M}\,\Big)^3-
\Big(\,{a(b)\over b}+{1\over 9}\,\Big)
\Big(\,{T\over \bar M}\,\Big)^6\,+\,\cdots\,\,,
\eeq
where $\bar M$ has been defined in (\ref{barM}) and $a(b)$ in (\ref{a(b)}).

\vskip 1cm
\renewcommand{\theequation}{\rm{F}.\arabic{equation}}
\setcounter{equation}{0}
\medskip

\section{Critical embeddings}
\label{critical}
In this appendix we study the critical behavior of the D6-brane probe in the flavored ABJM black hole. Following closely \cite{Mateos:2007vn,Frolov:1998td,Christensen:1998hg,Frolov:2006tc} we analyze the brane embeddings near the horizon. It is quite convenient to choose a new system of coordinates in which the induced metric near the horizon has the form of a Rindler space. In order to find these coordinates,   let us expand the radial coordinate $r$  in the near-horizon region as follows
\beq
r\,=\,r_h\,+\,C\,z^{\alpha}\,\,,
\eeq
where $z$ is a new coordinate, which is assumed to be small in the near-horizon region, and $C$ and $\alpha$ are constants that will be determined by looking at the $dr^2$ part of the metric (\ref{flavoredBH-metric}). We notice that the expansion of the blackening factor $h(r)$ is
\beq
h(r)\,\approx\,{3C\over r_h}\,z^{\alpha}\,+\,\cdots\,\,.
\eeq
Then, the $dr^2$ part of the metric is
\beq
{dr^2\over r^2 \,h(r)}\,=\,{\alpha^2\,C\over 3 r_h}\,z^{\alpha-2}\,dz^2\,+\,\cdots\,\,.
\eeq
We fix the constants $C$ and $\alpha$ by requiring that $z$ is a Rindler coordinate and that the right-hand side of this equation is just $dz^2$, which is achieved when $C$ and $\alpha$ are given by
\beq
\alpha\,=\,2\,\,,\qquad\qquad
C\,=\,{3 r_h\over 4}\,=\,\pi\,T\,\,.
\eeq
We will also want to explore the region in which $\theta$ is small. Accordingly, let us represent $\theta$ in terms of a new coordinate $y$ given by,
\beq
y\,=\,{\theta\over b}\,\,,
\label{y-theta}
\eeq
and  approximate  $\sin\theta\approx \theta=b\,y$. Written in the coordinates $z$ and $y$, the metric takes the form
\bear
&&{ds^2\over L^2}\,=\,-(2\pi T)^2\,z^2\,dt^2\,+\,r_h^2\,
\Big[\,(dx^1)^2\,+\,(dx^2)^2\,\Big]\,+\,dz^2\,+\,dy^2\,+\,\rc\rc
&&\qquad\qquad+
{q\over b^2}\,\Big[\,d\alpha^2\,+\,\sin^2\alpha\,d\beta^2\,\Big]\,+\,
y^2\,(\,d\psi+\cos\alpha \,d\beta\,)^2\,+\,\cdots\,\,,
\label{near-horizon-metric}
\eear
where we keep the leading terms and the dots represent terms that do not contribute to the embedding of the D6-brane.  Notice from (\ref{near-horizon-metric})  that only one of the three internal directions wrapped by the probe collapses at the tip of the brane $y=0$. 

Let us take an embedding characterized by a function $y=y(z)$, which is a appropriate  to describe a black hole embedding. To compute the induced metric we just substitute  $dz^2\,+\,dy^2\,=\,(1+\dot y^2)\,dz^2$ in the metric (\ref{near-horizon-metric}), where the dot denotes derivative with respect to $z$. The DBI Lagrangian density becomes
\beq
{\cal L}_{DBI}\,\propto\,
z\,y\,\sqrt{1+\dot y^2}\,\,,
\label{critical-DBI-y}
\eeq
where the proportionality constant is $2\pi T\,r_h^2\,{\cal N}_r$.  Let us now consider the WZ term of the action. We will expand the WZ Lagrangian by taking into account that $z$ and $y$ are of first order and $\dot y$ is of order zero. With these assignments  ${\cal L}_{DBI}$ is of second order. By inspecting (\ref{pullback-SUSY-C7}) we notice that there are two terms to look at. First of all we consider
\beq
r\,\sin\theta\,\cos\theta\,d\theta\,\approx \,b^2\,r_h\,y\,\dot y\,dz\,+\,
{3\,b^2\,r_h\over 4}\,z^2\,y\,\dot y\,dz\,+\,\cdots\,\,.
\label{first-term-WZ}
\eeq
The first term in this equation gives a contribution at first order to the Lagrangian which, however, is a total  derivative and does not contribute to the equation of motion. Thus we neglect this contribution. The second term in (\ref{first-term-WZ}) is of order three and will be neglected. The second term in (\ref{pullback-SUSY-C7}) to be considered is
\beq
r^2\,\sin^2\theta\,dr\,=\,
{3\,b^2\,r_h^3\over 2}\,\,z\,y^2\,dz\,+\,\cdots\,\,,
\eeq
which is also of order three. Therefore, the total Lagrangian density  at leading order is just the DBI  one written in (\ref{critical-DBI-y}). The corresponding equation of motion is
\beq
z\,y\,\ddot y\,+\,(\,y\,\dot y\,-\,z\,)(\,1+\dot y^2\,)\,=\,0\,\,.
\label{eom-nh-y}
\eeq
The parametrization $y=y(z)$ is appropriate to study the black hole embeddings. In this case the differential equation (\ref{eom-nh-y}) must be solved by imposing the following boundary conditions:
\beq
y(z=0)\,=\,y_h\,\,,
\qquad\qquad
\dot y(z=0)\,=\,0\,\,,
\label{nh-bc-bh}
\eeq
where $y_h$ characterizes the angle at which the brane reaches the horizon. 

In the case of Minkowski embeddings the appropriate parametrization is $z=z(y)$. The brane does not reach the horizon and ends at $y=0$ at a point whose distance from the horizon is determined by the value of $z(y)$ and $y=0$. Therefore, the differential equation to integrate is just obtained from (\ref{eom-nh-y}) by exchanging $y\leftrightarrow z$, namely
\beq
y\,z\,\ddot z\,+\,(\, z\,\dot z\,-\,y\,)(\,1+\dot z ^2\,)\,=\,0\,\,,
\label{eom-nh-z}
\eeq
with the boundary conditions
\beq
z(y=0)\,=\,z_h\,\,,
\qquad\qquad
\dot z(y=0)\,=\,0\,\,.
\label{nh-bc-Min}
\eeq
Notice that the equation of motion (\ref{eom-nh-y}) and the Lagrangian (\ref{critical-DBI-y}) are the same as the general expressions in \cite{Mateos:2007vn} with $n=1$, which is consistent with the fact that only one the internal directions of the ${\mathbb R}{\mathbb P}^3$ cycle wrapped by the brane collapses at the tip.

Clearly, the system is symmetric under the interchange of $y$ and $z$
\beq
y\,\leftrightarrow\, z\,\,.
\label{y-z-exchange}
\eeq
This symmetry exchanges black hole and Minkowski embeddings. This means that for any black hole solution $y=f(z)$ there exists a Minkowski solution $z=f(y)$ with the same function $f$.

The critical solution is the following particular solution of (\ref{eom-nh-y}):
\beq
y=z\,\,.
\label{critical-sol}
\eeq
In this solution the brane just ends at the horizon $z=0$ at the point $y=0$ (\ie, with $\theta=0$). Therefore, the critical solution (\ref{critical-sol}) is the limiting case of both Minkowski and black hole embeddings. Notice that it is invariant under the exchange (\ref{y-z-exchange}).

Let us write the critical solution in terms of the isotropic coordinate $u$. First, we recall the relation between $u$ and $r$, 
\beq
u^{{3\over 2b}}\,=\,\Big({r\over r_0}\Big)^{{3\over 2}}\,(1+\sqrt{h})\,\,.
\eeq
Taking into account that $\sqrt{h}\approx3z/4$, we have
\beq
u^{{3\over 2b}}\,\approx 1+{3\over 2}\,z\,\,
\qquad
\to
\qquad
u\approx 1+b\,z\,\,.
\eeq
It follows that $R=u\cos\theta$ and $\rho=u\sin\theta$ can be expressed in terms of $z$ and $y$ as
\beq
R\approx 1+b\,z\,\,,
\qquad\qquad
\rho\approx b\,y\,\,.
\label{R-rho-z-y}
\eeq
Then, the critical embedding $y=z$ in the near-horizon region is given by the following linear relation between $R$ and $\rho$:
\beq
R\,=\,1\,+\,\rho\,\,.
\eeq
Thus, $dR/d\rho\,=\,1$ and the incidence angle of the critical embedding in the $(R,\rho)$ plane is $\pi/4$ for all values of $b$. 

Let us next analyze the near critical black hole solutions (the corresponding analysis for the Minkowski embeddings can be obtained by exchanging $y$ with $z$ in what follows). We represent $y(z)$ as,
\beq
y\,=\,z\,+\,\xi(z)\,\,,
\eeq
with $\xi(z)$ being a small function of $z$. At first order in $\xi$, the equation of motion
(\ref{eom-nh-y}) reads,
\beq
z^2\,\ddot \xi\,+\,2\,z\,\dot\xi\,+\,2\,\xi\,=\,0\,\,.
\label{eom-xi}
\eeq
This equation can be solved by a power law $\xi=z^{\nu}$ where  the exponent $\nu$ satisfies the quadratic equation $\nu^2+ \nu +2=0$, whose two solutions are
\beq
\nu_{\pm}\,=\,-{1\over 2}\,\pm \alpha\,i\,\,,
\eeq
with $\alpha$ being
\beq
\alpha\,=\,{\sqrt{7}\over 2}\,\,.
\eeq
The two independent solutions for $ \xi$ can be taken to be
\beq
{T^{-{3\over 2}}\over \sqrt{z}}\,\,\sin\big[\,\alpha\log(T\,z)\,\big]\,\,,
\qquad\qquad
{T^{-{3\over 2}}\over \sqrt{z}}\,\,\cos\big[\,\alpha\log(T\,z)\,\big]\,\,,
\eeq
where we have introduced the temperature $T$ in order to deal with dimensionless quantities. Therefore, we can write,
\beq
y\,=\,z\,+\,{T^{-{3\over 2}}\over \sqrt{z}}\,\,
\Big[\,A\,\sin\big[\,\alpha\log(T\,z)\,\big]\,+\,B\,
\cos\big[\,\alpha\log(T\,z)\,\big]\,\Big]\,\,,
\eeq
with $A$ and $B$ being two coefficients. Notice that $A=B=0$ for the critical embeddings and therefore the coefficients $A$ and $B$ measure the deviation from the solution (\ref{critical-sol}). Let us denote by $m_*$ and $c_*$ the values of the mass and condensate parameters which correspond to the critical embedding, respectively. Clearly, $A$ and $B$ depend on the differences $m-m_*$ and $c-c_*$. Actually, it was suggested in \cite{Frolov:1998td,Christensen:1998hg,Frolov:2006tc} that $A$ and $B$ depend linearly on $m-m_*$ and $c-c_*$.

The differential equation (\ref{eom-nh-y}) satisfies the following property. 
If $y(z)\,=\,f(z)$ is a solution of the differential equation (\ref{eom-nh-y}), then $\bar y(z)$ defined as:
\beq
\bar y(z)\,=\,{f(\mu z)\over \mu}\,\,,
\label{mu-transformation}
\eeq
with $\mu\in \mathbb{R}$ being an arbitrary real number, is also a solution of (\ref{eom-nh-y}) with the initial condition,
\beq
\bar y_h\,\equiv\, \bar y (z=0)\,=\,{ y_h\over \mu}\,\,.
\label{y0-mu}
\eeq
Clearly, any two solutions of (\ref{eom-nh-y}) with the conditions (\ref{nh-bc-bh}) are related by this symmetry. Thus, we can reconstruct all black hole solutions from a given (fiducial) one. Let us see how (\ref{mu-transformation}) is realized in the coefficients $A$ and $B$. First, we define the  rotation matrix ${\cal M}(\mu)$ as
\beq
{\cal M}(\mu)\,\equiv\,
\left( \begin{array}{ccc}
\cos\big[\,\alpha\log(\mu)\big] && \sin\big[\,\alpha\log(\mu)\big] \\\\
 -\sin\big[\,\alpha\log(\mu)\big]   && \cos\big[\,\alpha\log(\mu)\big]
\end{array}
\right)\,\,.
\label{calM-def}
\eeq
Then, if we  denote by $\bar A$ and $\bar B$ 	the coefficients for the transformed solution $\bar y(z)$,   one finds that they are related to the initial coefficients $A$ and $B$ by the following  combined scaling and rotation:  
\beq
\left( \begin{array}{c}
\bar A\\\\ \bar B
\end{array}
\right)\,=\,{1\over \mu^{{3\over 2}}}\,{\cal M}(\mu)\,
\left( \begin{array}{c}
A\\\\ B
\end{array}
\right)\,\,.
\label{cd-trans}
\eeq
It is very illustrative to rewrite this result in terms of the $z=0$ values of $y(z)$ ($y_h$ and $\bar y_h$). From (\ref{y0-mu}) it follows that
\beq
\mu\,=\,{y_h\over \bar y_h}\,\,.
\eeq
Moreover, one can check that the matrix ${\cal M}$ satisfies,
\beq
{\cal M}(\mu)\,=\,{\cal M}(y_h)\,{\cal M}^{-1}(\bar y_h)\,\,.
\eeq
By using this result in the transformation law (\ref{cd-trans}) we can rewrite this last equation as
\beq
{{\cal M}(y_h)\over y_h^{{3\over 2}}}\,
\left( \begin{array}{c}
A\\\\ B
\end{array}
\right)\,=\,
{{\cal M}(\bar y_h)\over \bar y_h^{{3\over 2}}}\,
\left( \begin{array}{c}
\bar  A\\\\ \bar B
\end{array}
\right)\,=\,v\,\,,
\label{constant-v}
\eeq
where $v$ is a constant vector (it is the same for all the embeddings). Therefore, the quantity on the left-hand side of (\ref{constant-v}) is the same for all black hole solutions. Let us rewrite this property as
\beq
y_h^{-{3\over 2}}\,\left( \begin{array}{c}
A\\\\ B
\end{array}
\right)\,=\,
{\cal M}^{-1}(y_h)\,v\,\,.
\label{cd-periodic}
\eeq
\begin{figure}[ht]
\begin{center}
\includegraphics[width=0.45\textwidth]{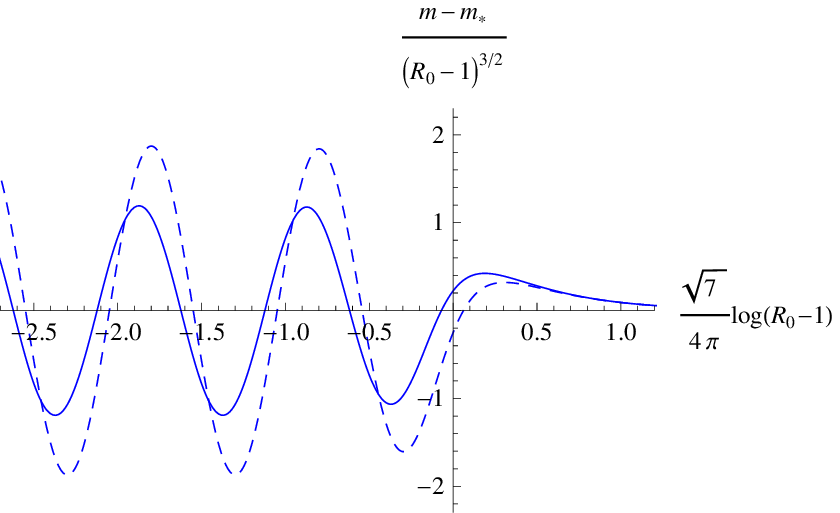}
\qquad
\includegraphics[width=0.45\textwidth]{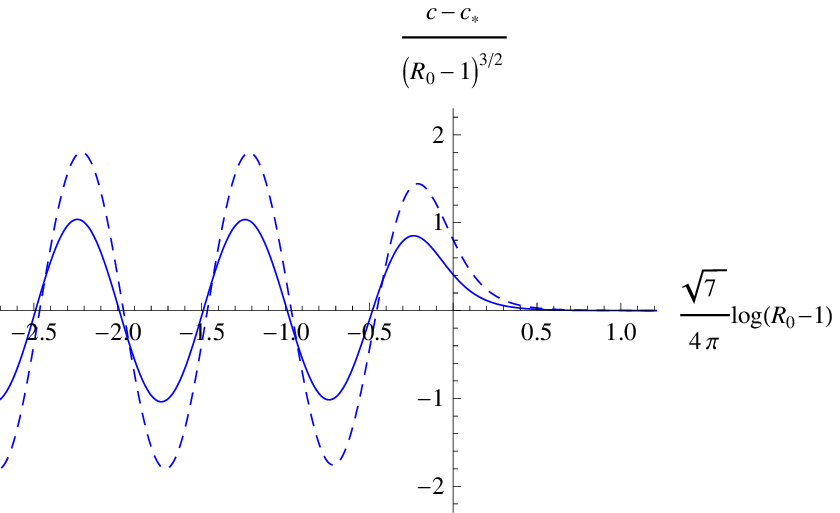}
\end{center}
\caption[criticalMN]{Values of $m$ and $c$ around the critical point for Minkowski embeddings. The solid (dashed) curves correspond to $\hat\epsilon=0$ ($\hat\epsilon=10$). 
\label{criticalMN}} 
\end{figure}
Next, we notice that ${\cal M}(y_h)$ (and its inverse) is a periodic function of $\log y_h$. Actually, it follows from (\ref{calM-def}) that ${\cal M}(y_h)$  does not change when ${\alpha\over 2\pi}\,\log y_h$ is shifted by one. Therefore, we get from (\ref{cd-periodic}) that $y_h^{-{3\over 2}}\,A$ and
$y_h^{-{3\over 2}}\,B$ are periodic functions of ${\alpha\over 2\pi}\,\log y_h$ with period one.  Since the coefficients $A$ and $B$ are linearly related to $m-m_*$ and 
$c-c_*$, 
\beq
{m-m_*\over y_h^{{3\over 2}}}\,=\,F_m\Big({\sqrt{7}\over 4\pi}\,\log y_h\Big)\,\,,
\qquad\qquad
{c-c_*\over y_h^{{3\over 2}}}\,=\,F_c\Big({\sqrt{7}\over 4\pi}\,\log y_h\Big)\,\,,
\label{BH-periodicity-y0}
\eeq
where $F_m(x)$ and $F_c(x)$ are periodic functions of $x$ with period one. 

A similar result can be found for Minkowski embeddings by exchanging $y_0\leftrightarrow z_0$. Then, we can write,
\beq
{m-m_*\over z_h^{{3\over 2}}}\,=\,G_m\Big({\sqrt{7}\over 4\pi}\,\log z_h\Big)\,\,,
\qquad\qquad
{c-c_*\over z_h^{{3\over 2}}}\,=\,G_c\Big({\sqrt{7}\over 4\pi}\,\log z_h\Big)\,\,,
\label{Min-periodicity-z0}
\eeq
with $G_m(x)$ and $G_c(x)$ being periodic in $x$ with unit period. 

Let us recast the previous results in terms of our physical variables. We consider first the case of the Minkowski embeddings, which are characterized by the value $R_0$ of the $R$ coordinate at $\rho=0$. From the relation (\ref{R-rho-z-y}) between $R$ and $z$, it follows that $z_0$ and $R_0$ are related as
\beq
z_h\,=\,{R_0-1\over b}\,\,.
\eeq
By using this result in (\ref{Min-periodicity-z0}), we can write
\beq
{m-m_*\over (R_0-1)^{{3\over 2}}}\,=\,\tilde G_m\Big({\sqrt{7}\over 4\pi}\,\log  (R_0-1)\Big)\,\,,
\qquad\qquad
{c-c_*\over (R_0-1)^{{3\over 2}}}\,=\,\tilde G_c\Big({\sqrt{7}\over 4\pi}\,\log (R_0-1)\Big)\,\,,
\label{Min-periodicity-R0}
\eeq
where the new functions $\tilde G_{m,c}(x)$ are defined as
\beq
\tilde G_{m,c}(x)\,\equiv\,b^{-{3\over 2}}\,\,G_{m,c}(x-{\alpha\over 2\pi}\,\log b)\,\,.
\eeq
It follows from this definition that $\tilde G_{m,c}(x)$ are also periodic functions of $x$ with unit period. The numerical results for the functions written on the left-hand side of  (\ref{Min-periodicity-R0}) are plotted in Fig.~\ref{criticalMN}. They confirm this periodicity behavior. 

\begin{figure}[ht]
\begin{center}
\includegraphics[width=0.45\textwidth]{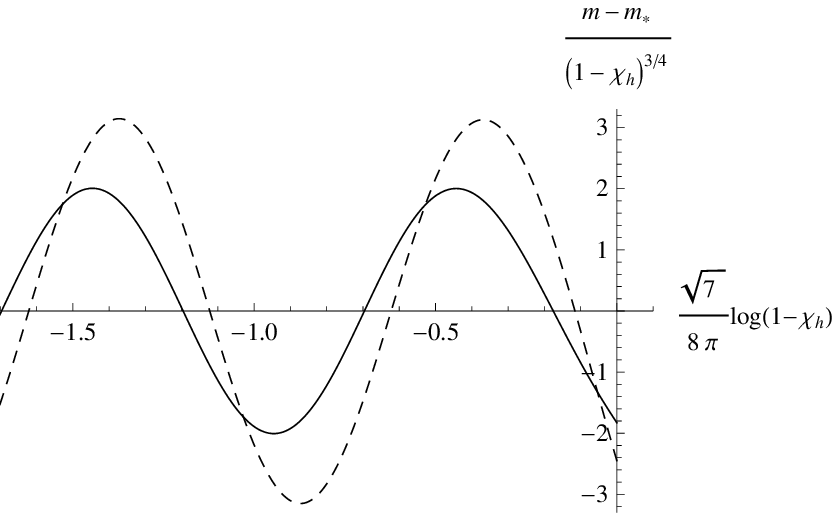}
\qquad
\includegraphics[width=0.45\textwidth]{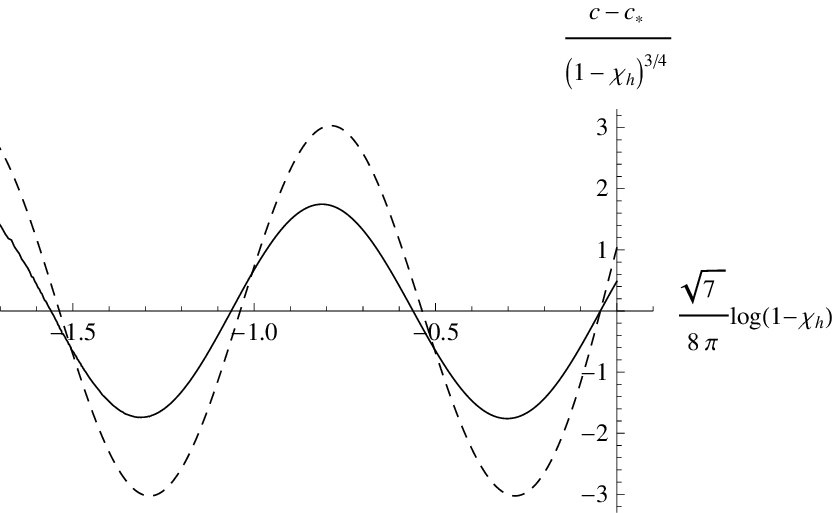}
\end{center}
\caption[criticalBH]{Values of $m$ and $c$ around the critical point for black hole embeddings. The solid (dashed) curves correspond to $\hat\epsilon=0$ ($\hat\epsilon=10$). 
\label{criticalBH}} 
\end{figure}

Similarly, we can deal with the case of black hole embeddings. In this case the solutions are characterized by the value $\chi_h$  of $\chi=\cos\theta$ at the horizon. For near-critical solutions $\chi_h\approx 1-\theta_h^2/2$ and, by using (\ref{y-theta}), we get that the relation between $\chi_h$ and $y_h$ is
\beq
y_h\,=\,{\sqrt{2}\over b}\,\,\big(1-\chi_h)^{{1\over 2}}\,\,.
\eeq
Plugging this result in (\ref{BH-periodicity-y0}) we find
\beq
{m-m_*\over (1-\chi_h)^{{3\over 4}}}\,=\,\tilde F_m\Big({\sqrt{7}\over 4\pi}\,\log   (1-\chi_h)\Big)\,\,,
\qquad\qquad
{c-c_*\over (1-\chi_h)^{{3\over 4}}}\,=\,\tilde F_c\Big({\sqrt{7}\over 4\pi}\,\log (1-\chi_h)\Big)\,\,,
\label{BH-periodicity-chi0}
\eeq
where  the new functions $\tilde F_{m,c}(x)$ are defined as:
\beq
\tilde F_{m,c}(x)\,\equiv\,{2^{{3\over 4}}\over b^{{3\over 2}}}
\,\,F_{m,c}\Big({x\over 2}+{\alpha\over 2\pi}\,\log {\sqrt{2}\over b}\Big)\,\,.
\eeq
Clearly,  $\tilde F_{m,c}(x)$ are periodic functions of $x$ with period two.  The numerical values of $(m-m_*)(1-\chi_h)^{-{3\over 4}}$ and $(c-c_*)(1-\chi_h)^{-{3\over 4}}$ are displayed in Fig.~\ref{criticalBH} and agree with the predicted periodic behavior.



\begin{thebibliography}{99}

 \bibitem{jm} J.~M.~Maldacena, ``The large $N$ limit of superconformal field
theories and supergravity'', {\it Adv.\ Theor.\ Math.\ Phys.}\  {\bf 
2} (1998) 231,
{\rm hep-th/9711200}.

 
 
\bibitem{Aharony:1999ti}
  O.~Aharony, S.~S.~Gubser, J.~M.~Maldacena, H.~Ooguri and Y.~Oz,
  ``Large N field theories, string theory and gravity,''
  Phys.\ Rept.\  {\bf 323}, 183 (2000)
  [arXiv:hep-th/9905111].



\bibitem{Aharony:2008ug}
  O.~Aharony, O.~Bergman, D.~L.~Jafferis and J. M. Maldacena,
``N=6 superconformal Chern-Simons-matter theories, M2-branes and their gravity duals,'' JHEP {\bf 0810}, 091 (2008) [arXiv:0806.1218 [hep-th]].
  



 
\bibitem{BL}
  J.~Bagger, N.~Lambert,
 ``Modeling Multiple M2's,''
  Phys.\ Rev.\  {\bf D75}, 045020 (2007)
  [hep-th/0611108];
  J.~Bagger, N.~Lambert,
 ``Gauge symmetry and supersymmetry of multiple M2-branes,''
  Phys.\ Rev.\  {\bf D77 } (2008)  065008
  [arXiv:0711.0955 [hep-th]];
  J.~Bagger, N.~Lambert,
 ``Comments on multiple M2-branes,''
  JHEP {\bf 0802}, 105 (2008)
  [arXiv:0712.3738 [hep-th]].
 
\bibitem{Gustavsson:2007vu}
  A.~Gustavsson,
  ``Algebraic structures on parallel M2-branes,''
  Nucl.\ Phys.\  {\bf B811}, 66-76 (2009)
  [arXiv:0709.1260 [hep-th]].
 
 
 
\bibitem{Klebanov:2009sg} 
  I.~R.~Klebanov and G.~Torri,
  ``M2-branes and AdS/CFT,''
  Int.\ J.\ Mod.\ Phys.\ A {\bf 25}, 332 (2010)
  [arXiv:0909.1580 [hep-th]].
 
 
\bibitem{Klose:2010ki} 
  T.~Klose,
  ``Review of AdS/CFT Integrability, Chapter IV.3: N=6 Chern-Simons and Strings on AdS4xCP3,''
  Lett.\ Math.\ Phys.\  {\bf 99}, 401 (2012)
  [arXiv:1012.3999 [hep-th]].
 
\bibitem{Marino:2011nm} 
  M.~Mari\~no,
  ``Lectures on localization and matrix models in supersymmetric Chern-Simons-matter theories,''
  J.\ Phys.\ A {\bf 44}, 463001 (2011)
  [arXiv:1104.0783 [hep-th]].
 
 
 
 
\bibitem{Bagger:2012jb} 
  J.~Bagger, N.~Lambert, S.~Mukhi and C.~Papageorgakis,
  ``Multiple Membranes in M-theory,''
  arXiv:1203.3546 [hep-th].
 
 
\bibitem{Hohenegger:2009as}
  S.~Hohenegger, I.~Kirsch,
  ``A Note on the holography of Chern-Simons matter theories with flavour,''
  JHEP {\bf 0904}, 129 (2009)
  [arXiv:0903.1730 [hep-th]].
  
  
\bibitem{Gaiotto:2009tk}
  D.~Gaiotto, D.~L.~Jafferis,
``Notes on adding D6 branes wrapping RP**3 in AdS(4) x CP**3,''
   [arXiv:0903.2175 [hep-th]].

 
\bibitem{Hikida:2009tp}
  Y.~Hikida, W.~Li, T.~Takayanagi,
``ABJM with Flavors and FQHE,''
  JHEP {\bf 0907}, 065 (2009)
  [arXiv:0903.2194 [hep-th]].

\bibitem{Jensen:2010vx}
  K.~Jensen,
  ``More Holographic Berezinskii-Kosterlitz-Thouless Transitions,''
  Phys.\ Rev.\  {\bf D82}, 046005 (2010)
  [arXiv:1006.3066 [hep-th]].


\bibitem{Ammon:2009wc}
  M.~Ammon, J.~Erdmenger, R.~Meyer {\it et al.},
 ``Adding Flavor to AdS(4)/CFT(3),''
  JHEP {\bf 0911}, 125 (2009)
  [arXiv:0909.3845 [hep-th]].




\bibitem{Zafrir:2012yg}
  G.~Zafrir,
  ``Embedding massive flavor in ABJM,''
  JHEP {\bf 1210} (2012) 056
  [arXiv:1202.4295 [hep-th]].

   
\bibitem{Conde:2011sw}
  E.~Conde and A.~V.~Ramallo,
 ``On the gravity dual of Chern-Simons-matter theories with unquenched
  flavor,''
  JHEP {\bf 1107}, 099 (2011)
  [arXiv:1105.6045 [hep-th]].



\bibitem{Bigazzi:2005md}
  F.~Bigazzi, R.~Casero, A.~L.~Cotrone, E.~Kiritsis, A.~Paredes,
 ``Non-critical holography and four-dimensional CFT's with fundamentals,''
  JHEP {\bf 0510}, 012 (2005)
  [hep-th/0505140].


\bibitem{CNP}
  R.~Casero, C.~Nunez, A.~Paredes,
  ``Towards the string dual of N=1 SQCD-like theories,''
  Phys.\ Rev.\  {\bf D73}, 086005 (2006)
  [hep-th/0602027];
   R.~Casero, C.~Nunez, A.~Paredes,
  ``Elaborations on the String Dual to N=1 SQCD,''
  Phys.\ Rev.\  {\bf D77}, 046003 (2008)
  [arXiv:0709.3421 [hep-th]];
    C.~Hoyos-Badajoz, C.~Nunez and I.~Papadimitriou,
  ``Comments on the String dual to N=1 SQCD,''
  Phys.\ Rev.\ D {\bf 78}, 086005 (2008)
  [arXiv:0807.3039 [hep-th]];
E.~Conde, J.~Gaillard and A.~V.~Ramallo,
  ``On the holographic dual of N=1 SQCD with massive flavors,''
  JHEP {\bf 1110}, 023 (2011)
  [arXiv:1107.3803 [hep-th]].






\bibitem{conifold}
  F.~Benini, F.~Canoura, S.~Cremonesi, C.~Nunez, A.~V.~Ramallo,
  ``Unquenched flavors in the Klebanov-Witten model,''
  JHEP {\bf 0702}, 090 (2007)
  [hep-th/0612118];
 F.~Benini, F.~Canoura, S.~Cremonesi, C.~Nunez, A.~V.~Ramallo,
  ``Backreacting flavors in the Klebanov-Strassler background,''
  JHEP {\bf 0709 } (2007)  109
  [arXiv:0706.1238 [hep-th]];
   M.~Ihl, A.~Kundu and S.~Kundu,
  ``Back-reaction of Non-supersymmetric Probes: Phase Transition and Stability,''
  arXiv:1208.2663 [hep-th].
 



\bibitem{D3-D7}
  F.~Bigazzi, A.~L.~Cotrone, J.~Mas, A.~Paredes, A.~V.~Ramallo, J.~Tarrio,
 ``D3-D7 Quark-Gluon Plasmas,''
  JHEP {\bf 0911}, 117 (2009)
  [arXiv:0909.2865 [hep-th]];
   F.~Bigazzi, A.~L.~Cotrone, J.~Tarrio,
  ``Hydrodynamics of fundamental matter,''
  JHEP {\bf 1002}, 083 (2010)
  [arXiv:0912.3256 [hep-th]];
   F.~Bigazzi, A.~L.~Cotrone, J.~Mas, D.~Mayerson, J.~Tarrio,
  ``D3-D7 Quark-Gluon Plasmas at Finite Baryon Density,''
  JHEP {\bf 1104}, 060 (2011)
  [arXiv:1101.3560 [hep-th]];
  A.~Maga\~na, J.~Mas, L.~Mazzanti and J.~Tarrio,
  ``Probes on D3-D7 Quark-Gluon Plasmas,''
  JHEP {\bf 1207}, 058 (2012)
  [arXiv:1205.6176 [hep-th]].
  
  


\bibitem{Nunez:2010sf}
  C.~Nunez, A.~Paredes, A.~V.~Ramallo,
``Unquenched flavor in the gauge/gravity correspondence,''
  Adv.\ High Energy Phys.\  {\bf 2010}, 196714 (2010)
  [arXiv:1002.1088 [hep-th]].


\bibitem{Bianchi}
  M.~S.~Bianchi, S.~Penati, M.~Siani,
  ``Infrared stability of ABJ-like theories,''
  JHEP {\bf 1001}, 080 (2010)
  [arXiv:0910.5200 [hep-th]]; M.~S.~Bianchi, S.~Penati, M.~Siani,
  ``Infrared Stability of N = 2 Chern-Simons Matter Theories,''
  JHEP {\bf 1005}, 106 (2010)
  [arXiv:0912.4282 [hep-th]].


\bibitem{Santamaria:2010dm} 
  R.~C.~Santamaria, M.~Marino and P.~Putrov,
  ``Unquenched flavor and tropical geometry in strongly coupled Chern-Simons-matter theories,''
  JHEP {\bf 1110}, 139 (2011)
  [arXiv:1011.6281 [hep-th]].
  
  
\bibitem{Mateos:2006nu}
  D.~Mateos, R.~C.~Myers and R.~M.~Thomson,
  ``Holographic phase transitions with fundamental matter,''
  Phys.\ Rev.\ Lett.\  {\bf 97} (2006) 091601
  [hep-th/0605046].
  
  

\bibitem{Mateos:2007vn} 
  D.~Mateos, R.~C.~Myers and R.~M.~Thomson,
  ``Thermodynamics of the brane,''
  JHEP {\bf 0705}, 067 (2007)
  [hep-th/0701132].
 


\bibitem{Skenderis:2002wp} 
  K.~Skenderis,
  ``Lecture notes on holographic renormalization,''
  Class.\ Quant.\ Grav.\  {\bf 19}, 5849 (2002)
  [hep-th/0209067].

\bibitem{Karch:2005ms} 
  A.~Karch, A.~O'Bannon and K.~Skenderis,
  ``Holographic renormalization of probe D-branes in AdS/CFT,''
  JHEP {\bf 0604}, 015 (2006)
  [hep-th/0512125].


\bibitem{Kobayashi:2006sb} 
  S.~Kobayashi, D.~Mateos, S.~Matsuura, R.~C.~Myers and R.~M.~Thomson,
  ``Holographic phase transitions at finite baryon density,''
  JHEP {\bf 0702}, 016 (2007)
  [hep-th/0611099];
   D.~Mateos, S.~Matsuura, R.~C.~Myers and R.~M.~Thomson,
  ``Holographic phase transitions at finite chemical potential,''
  JHEP {\bf 0711}, 085 (2007)
  [arXiv:0709.1225 [hep-th]].
     
     
     
     
\bibitem{Ammon:2012mu}
  M.~Ammon, K.~Jensen, K.~-Y.~Kim, J.~Laia and A.~O'Bannon,
  ``Moduli Spaces of Cold Holographic Matter,''
  [arXiv:1208.3197 [hep-th]].
  
  
\bibitem{Filev:2007gb}
  V.~G.~Filev, C.~V.~Johnson, R.~C.~Rashkov and K.~S.~Viswanathan,
  ``Flavoured large N gauge theory in an external magnetic field,''
  JHEP {\bf 0710}, 019 (2007)
  [hep-th/0701001]; T.~Albash, V.~G.~Filev, C.~V.~Johnson and A.~Kundu,
  ``Finite temperature large N gauge theory with quarks in an external magnetic field,''
  JHEP {\bf 0807}, 080 (2008)
  [arXiv:0709.1547 [hep-th]];
 J.~Erdmenger, R.~Meyer and J.~P.~Shock,
  ``AdS/CFT with flavour in electric and magnetic Kalb-Ramond fields,''
  JHEP {\bf 0712} (2007) 091
  [arXiv:0709.1551 [hep-th]].

\bibitem{Filev:2011mt}
  V.~G.~Filev and D.~Zoakos,
  ``Towards Unquenched Holographic Magnetic Catalysis,''
  JHEP {\bf 1108}, 022 (2011)
  [arXiv:1106.1330 [hep-th]].

\bibitem{Erdmenger:2011bw}
  J.~Erdmenger, V.~G.~Filev and D.~Zoakos,
  ``Magnetic Catalysis with Massive Dynamical Flavours,''
  JHEP {\bf 1208}, 004 (2012)
  [arXiv:1112.4807 [hep-th]].


\bibitem{Bergman:2010gm}
  O.~Bergman, N.~Jokela, G.~Lifschytz and M.~Lippert,
  ``Quantum Hall Effect in a Holographic Model,''
  JHEP {\bf 1010} (2010) 063
  [arXiv:1003.4965 [hep-th]];
     N.~Jokela, M.~J\"arvinen and M.~Lippert,
  ``A holographic quantum Hall model at integer filling,''
  JHEP {\bf 1105} (2011) 101
  [arXiv:1101.3329 [hep-th]];
  O.~Bergman, J.~Erdmenger and G.~Lifschytz,
  ``A Review of Magnetic Phenomena in Probe-Brane Holographic Matter,''
  arXiv:1207.5953 [hep-th].

\bibitem{Karch:2008fa}
  A.~Karch, D.~T.~Son and A.~O.~Starinets,
  ``Zero Sound from Holography,''
  arXiv:0806.3796 [hep-th].


\bibitem{Bergman:2011rf}
  O.~Bergman, N.~Jokela, G.~Lifschytz and M.~Lippert,
  ``Striped instability of a holographic Fermi-like liquid,''
  JHEP {\bf 1110} (2011) 034
  [arXiv:1106.3883 [hep-th]];

\bibitem{Davison:2011ek}
  R.~A.~Davison and A.~O.~Starinets,
  ``Holographic zero sound at finite temperature,''
  Phys.\ Rev.\ D {\bf 85} (2012) 026004
  [arXiv:1109.6343 [hep-th]].

\bibitem{Jokela:2010nu}
  N.~Jokela, G.~Lifschytz and M.~Lippert,
  ``Magneto-roton excitation in a holographic quantum Hall fluid,''
  JHEP {\bf 1102} (2011) 104
  [arXiv:1012.1230 [hep-th]];
  N.~Jokela, M.~J\"arvinen and M.~Lippert,
  ``Fluctuations of a holographic quantum Hall fluid,''
  JHEP {\bf 1201} (2012) 072
  [arXiv:1107.3836 [hep-th]].

\bibitem{Jokela}
  N.~Jokela, G.~Lifschytz and M.~Lippert,
  ``Magnetic effects in a holographic Fermi-like liquid,''
  JHEP {\bf 1205} (2012) 105
  [arXiv:1204.3914 [hep-th]];
  N.~Jokela, M.~J\"arvinen and M.~Lippert,
  ``Fluctuations and instabilities of a holographic metal,''
  to appear.
 
  
  
\bibitem{Maldacena:1998im}
  J.~M.~Maldacena,
  ``Wilson loops in large N field theories,''
  Phys.\ Rev.\ Lett.\  {\bf 80}, 4859 (1998)
  [arXiv:hep-th/9803002].


\bibitem{Rey:1998ik}
  S.~J.~Rey and J.~T.~Yee,
  ``Macroscopic strings as heavy quarks in large N gauge theory and  anti-de
  Sitter supergravity,''
  Eur.\ Phys.\ J.\  C {\bf 22} (2001) 379
  [arXiv:hep-th/9803001].



\bibitem{Brandhuber:1998bs} 
  A.~Brandhuber, N.~Itzhaki, J.~Sonnenschein and S.~Yankielowicz,
  ``Wilson loops in the large N limit at finite temperature,''
  Phys.\ Lett.\ B {\bf 434}, 36 (1998)
  [hep-th/9803137].


\bibitem{Rey:1998bq} 
  S.~-J.~Rey, S.~Theisen and J.~-T.~Yee,
  ``Wilson-Polyakov loop at finite temperature in large N gauge theory and anti-de Sitter supergravity,''
  Nucl.\ Phys.\ B {\bf 527}, 171 (1998)
  [hep-th/9803135].

\bibitem{Frolov:1998td}
  V.~P.~Frolov, A.~L.~Larsen and M.~Christensen,
  ``Domain wall interacting with a black hole: A New example of critical phenomena,''
  Phys.\ Rev.\ D {\bf 59} (1999) 125008
  [hep-th/9811148].
  
  
  
\bibitem{Christensen:1998hg}
  M.~Christensen, V.~P.~Frolov and A.~L.~Larsen,
  ``Soap bubbles in outer space: Interaction of a domain wall with a black hole,''
  Phys.\ Rev.\ D {\bf 58} (1998) 085008
  [hep-th/9803158].


\bibitem{Frolov:2006tc}
  V.~P.~Frolov,
  ``Merger Transitions in Brane-Black-Hole Systems: Criticality, Scaling, and Self-Similarity,''
  Phys.\ Rev.\ D {\bf 74} (2006) 044006
  [gr-qc/0604114].
    



  
  


    
    
    
    
  
\end{thebibliography}
\end{document}